\documentclass[12pt]{iopart}
\usepackage{graphicx}
\usepackage{iopams} 
\usepackage[colorlinks=true,citecolor=blue]{hyperref}
\usepackage{ulem}
\usepackage{txfonts}
\usepackage{hyperref}
\usepackage{bbold}

\newcommand{\ket}[1]{\ensuremath{|{#1}\rangle}}

\begin{document}

\title{What can be learned from a future supernova neutrino detection?}

\author{Shunsaku Horiuchi}
\address{Center for Neutrino Physics, Department of Physics, Virginia Tech, Blacksburg, VA 24061-0435, USA}

\author{James P.\ Kneller}
\address{Department of Physics, North Carolina State University, Raleigh, NC 27695-8202, USA}

\ead{\mailto{horiuchi@vt.edu} \mailto{jim\_kneller@ncsu.edu} }

\begin{abstract}
This year marks the thirtieth anniversary of the only supernova from which we have detected neutrinos - SN 1987A. The twenty or so neutrinos that were detected were mined to great depth in order to determine the events that occurred in the explosion and to place limits upon all manner of neutrino properties. Since 1987 the scale and sensitivity of the detectors capable of identifying neutrinos from a Galactic supernova have grown considerably so that current generation detectors are capable of detecting of order ten thousand neutrinos for a supernova at the Galactic Center. Next generation detectors will increase that yield by another order of magnitude. Simultaneous with the growth of neutrino detection capability, our understanding of how massive stars explode and how the neutrino interacts with hot and dense matter has also increased by a tremendous degree. The neutrino signal will contain much information on all manner of physics of interest to a wide community. In this review we describe the expected features of the neutrino signal, the detectors which will detect it, and the signatures one might try to look for in order to get at this physics.
\end{abstract}


\maketitle


\section{Our Review}

On the 23rd of February in 1987 at UT 07:35:35, the neutrinos from a core-collapse supernova in the Large Magellanic Cloud arrived at Earth after traveling for 168,000 years. In the brief $\sim 13\;{\rm s}$ period they swept across the Earth, 25 were detected \cite{1987PhRvL..58.1490H,1987PhRvL..58.1494B,1988PhLB..205..209A,1987EL......3.1315A}.
This was a momentous event and it is tempting to regard its timing as an omen for the long period of both experimental and theoretical discoveries in neutrino physics that have occurred in the past thirty years. Over the same period there has been huge strides made in our understanding of the kind of supernova that produced the neutrino burst. In 1987 the most sophisticated hydrodynamical simulations of core-collapse supernova were still one dimensional \cite{1986ApJ...307..178B,1986NYASA.470..267W,1987ApJ...319..136A}, although it was clear conditions were such that multi-dimensional physical processes such as convection should be operating \cite{1979MNRAS.188..305E,1980ApJ...238L.139L,1987ApJ...318L..57B}. At the present time, simulations can be done in three spatial dimensions, include general relativity and much better nuclear physics, and use much more sophisticated neutrino transport, to name a few of the major improvements. 

What the current simulations have made clear is that solving the enigma of the how supernovae explode requires a better understanding of multi-dimensional effects in the collapsing core, and their interplay with the fundamental physics at play---from the nuclear equation of state and general relativity to the in-medium neutrino mixing effects. What has also become apparent is just how strongly the neutrino emission is sensitive to these multi-dimensional and fundamental physics. The corollary is that the neutrino signal, if detected with sufficient precision, holds powerful and unique information that probe the physics at the heart of the supernova phenomenon. And because of their importance to the core of the explosion, a core-collapse supernova also represents an opportunity to probe the properties of the neutrino under conditions that are impossible to achieve on Earth. Supernovae are one of the few places where neutrinos exchange energy and momentum with other neutrinos. Any difference between what we expect and what is observed may be due to neutrino physics beyond the Standard Model. If so, it would not be the first time astrophysical neutrinos have revealed that our understanding of this most ephemeral particle was incomplete. 

The goal of this review is to answer the question ``What can be learned from a future supernova neutrino detection?" In order to answer this question we have endeavored to present a comprehensive picture of supernova neutrinos at the present time rather than a history lesson. Much has changed over the period that supernovae have been studied and we hope it will become clear that understanding supernovae and their neutrino signals is still very much an on-going process. At the same time we have tried to be concise rather than undertake lengthy discussions covering all the details on every point. Forecasting all possible signal predictions and scenario tests would be near impossible. Instead, we have tried to categorically approach the task by starting from simple robust tests and progressing to more complex signals where future work would make important contributions. We hope the reader finds this review a gateway to the wider literature, particularly for those not immersed in supernova neutrinos yet interested in the wealth of physics it holds. 

Finally, we should add that other reviews of supernovae and/or supernova neutrinos have appeared in the last few years. The review by Mirrizzi \textit{et al.} \cite{Mirizzi:2015eza} goes into much more depth than we shall about the details of supernova simulations, the neutrino emission, and the physical origin of the flavor transformations. The review by Janka, Melson and Summa \cite{2016ARNPS..66..341J} focuses upon the simulations while the review by Scholberg \cite{2012ARNPS..62...81S} focuses upon detection. We cannot avoid covering some of the same material as these other reviews but our emphasis will be much more on trying to formulate a strategy for exploiting the signal from a future supernova detection. We hope the reader finds our review complimentary to the others. 

We start in Section \ref{sec:4w} with an motivational introduction to supernova neutrinos, addressing the four W's of supernova neutrinos: \textit{why}, \textit{what}, \textit{where}, and \textit{when}. In Section \ref{sec:emission}, we review the properties of neutrino emission from the pre-supernova and supernova epochs, touching on their connection to relevant supernova physics. In Section \ref{sec:oscillation}, we review processes that change the flavor composition of supernova neutrinos between their emission from the collapsed core to their arrival at Earth. We provide a summary of the expected neutrino signal in Section \ref{sec:Earth}, and review neutrino detectors and their detection channels in Section  \ref{sec:detection}. In Section \ref{sec:SNphysics}, we discuss the question of what supernova physics could be tested with future neutrino datasets. We finish with some closing remarks in Section \ref{sec:summary}. 

\section{The Four W's}\label{sec:4w}
\subsection{Why}
Put simply, the neutrino burst from a nearby core-collapse supernova would allow us to test our theories of how massive stars explode and also probe the properties of the neutrino in regimes of temperature and density not accessible here on Earth. The arrival of a Galactic supernova neutrino burst signal in the near future would be most appreciated -- if there's a shortage of one thing in this field of neutrino astrophysics it's data. While we've been waiting for a signal from a nearby supernova to arrive, the two dozen events from SN 1987A have been analyzed many times over looking for all sorts of information \cite{1987Natur.329..689K,1987PhRvL..58.2722S,1987Natur.326..135B}; for a recent review, see Ref.~\cite{Vissani:2014doa}. 

\begin{figure*}[t!] 
\includegraphics[width=0.9\linewidth]{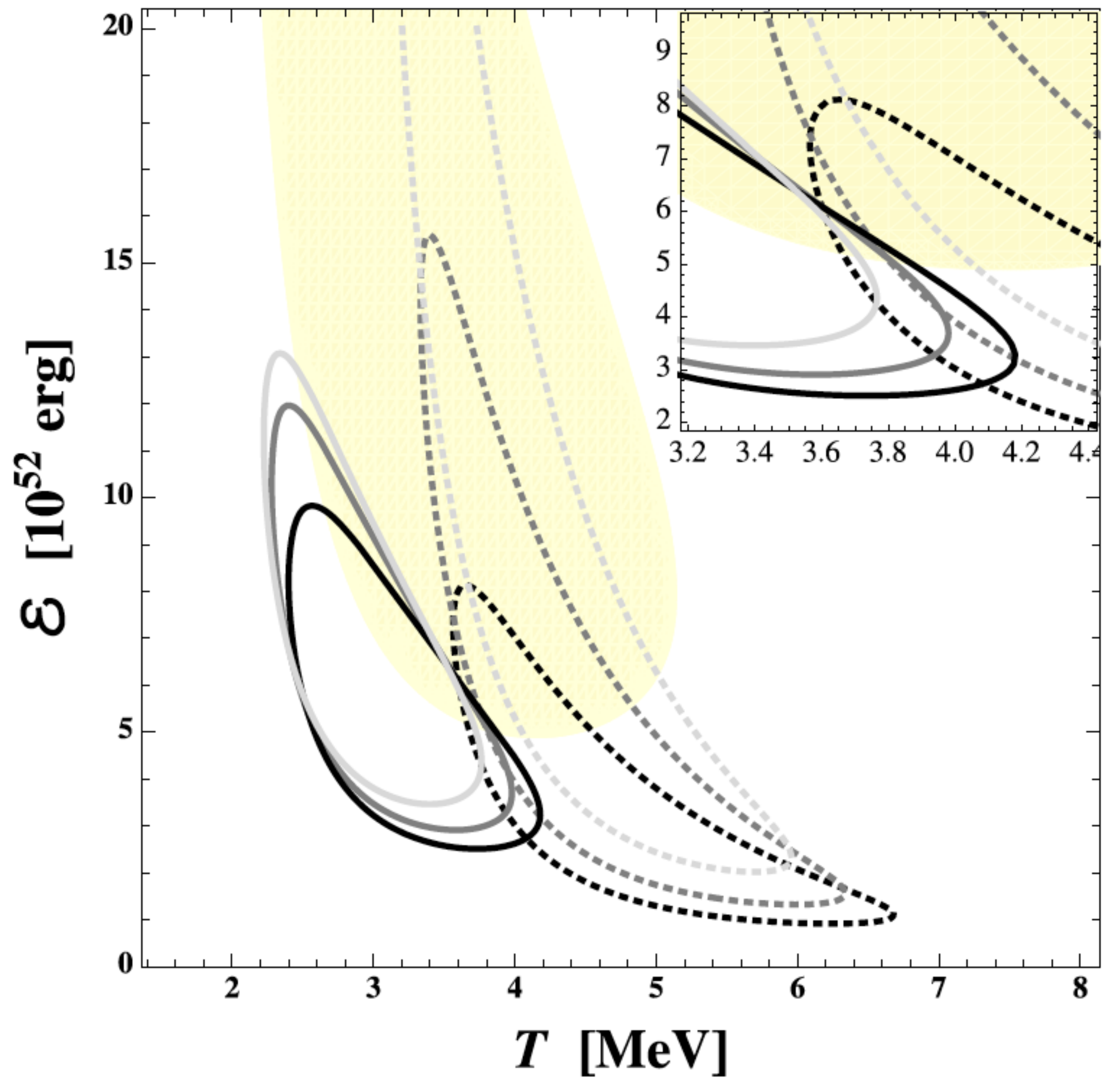}
\caption{
Regions allowed at 1$\sigma$ (68.3\% C.L..) for the energy radiated $\mathcal{E}$ and the temperature $T$ of electron antineutrinos,   as obtained by analysing separately the data  of the three relevant experiments: Kamiokande-II (3 continuous lines, corresponding to various hypotheses on the signal and background events), IMB (3 dashed lines,  corresponding to different descriptions of the efficiency function) and Baksan (shaded area). The inset zooms on the region where the contours overlap. Figure taken from Vissani ~\cite{Vissani:2014doa}. \label{fig:sn1987a} }
\end{figure*}

The most fundamental questions are the sequence of events that occurred in the core of the star and the emitted neutrino spectra. Loredo \& Lamb \cite{2002PhRvD..65f3002L} and Costantini, Ianni \& Vissani \cite{2004PhRvD..70d3006C} both find the signal from SN 1987A is best explained as a collapse, followed by an accretion phase that lasted 1-2 seconds which then transitions to a cooling phase that lasted for some ten seconds. The exploration of the neutrino spectrum has inevitably focused on the time-integrated emission given the small number of detected neutrino events. 
The events detected in Kamiokande II tended to be at lower energies while those detected by Irvine-Michigan-Brookhaven (IMB) were at noticeably higher energies. This is due to IMB's higher energy detection threshold. Figure \ref{fig:sn1987a}, from Vissani \cite{Vissani:2014doa}, shows the 68\% allowed regions separately for IMB (dotted), Kamiokande II (solid), and Baksan (shaded), for a two parameter (temperature and total energetics) anti-neutrino thermal spectrum analysis. The three IMB contours employ different low energy detector efficiencies: a functional form with thresholds of 13, 15, and 17 MeV (from dark to thin shade). The three Kamiokande II regions employ a 7.5 MeV detection cutoff but different time window selections (15 seconds or 30 seconds) and background considerations (including or neglecting). The Baksan region is large due to the small event number statistics. As can be seen, the Kamiokande II region shows the least variation due to detector considerations. The 68\% contours of all three detectors overlap with each other in a region around temperature 3.7 MeV, provided the lower threshold is adopted for IMB. The neutrino emission parameters are, within their errors, in broad agreement with what is predicted by recent core-collapse simulations~\cite{Pagliaroli:2008ur}. Going beyond the thermal spectrum, Mirizzi \& Raffelt \cite{2005PhRvD..72f3001M} considered a parametric non-thermal spectral shape that includes a spectral pinch, and Costantini \textit{et al.} \cite{Costantini:2006xd} and Y\"uksel \& Beacom \cite{2007PhRvD..76h3007Y} considered a nonparametric approach that makes no \textit{a priori} assumptions of the neutrino spectrum. These studies find the combined data is compatible with a thermal spectrum with a modest pinch to the peak. The neutrino signal from SN 1987A has also been frequently used to derive limits on various neutrino properties such as their mass, charge, magnetic moment, lifetime, mixing with heavy sterile neutrinos \cite{1987PhRvL..58.1906A,1987Natur.329...21B,1988PhRvL..60Q1789G,1988PhRvL..61...27B,1989PhRvL..62..505C,1989PhRvL..62..509K,2000NuPhB.590..562D}, and much more. 

The surprisingly large amount that was learned from such a pauce signal whets one's appetite for what might be learned from the next. The prospects for learning much more about supernovae and the neutrino from a similar event in the future have grown considerably. First, we shall collect much more data. In the thirty years since the detection of neutrinos from SN 1987A, the scale and sensitivity of neutrino detectors have increased tremendously. If we shift SN 1987A to the `standard' distance of a Milky Way supernova, $d=10\;{\rm kpc}$, the number of events we would have detected in IMB, Kamiokande-II and Baksan would have been 650 total. In comparison, we expect to detect with current and very near future detectors $\sim 12,000$ events or more \cite{Mirizzi:2015eza}.  Second, what has also become apparent over the past thirty years is that the neutrino signal contains much more needed information, because (i) we have learned core-collapse supernovae do not explode as spherically \cite{2001PhRvD..63j3004L,2001PhRvL..86.1935M,2003ApJ...584..971B,2006A&A...447.1049B,2006A&A...453..661K,2007PhR...442...23B,2008ApJ...678.1207I,2013ApJ...775...35C,2014ApJ...785..123C}, (ii) the neutrino spectra emitted at the neutrinosphere are sensitive to fundamental physics of the nuclear equation-of-state \cite{1994ApJ...425..195S,2005ApJ...629..922S,2007ApJ...667..382S,2009A&A...496..475M,2011ApJ...730...70O,2013ApJ...765...29C,Suwa:2012xd,2013ApJ...774...17S,2014EPJA...50...46F}, general relativity \cite{2006A&A...447.1049B,2016ApJ...831...98R}, and in-medium nuclear effects \cite{1999PhRvC..59.2888R,2012PhRvC..86f5803R,2012PhRvC..86f5806H}, and (iii) we have also learned that the neutrino flavor composition is not fixed at the neutrinosphere but rather evolves through the mantle of the star, decoheres as it propagates through the vacuum to Earth, and may oscillate again if it travels through Earth before reaching a detector \cite{2000PhRvD..62c3007D}.  These would allow scrutiny of much more core-collapse theory than was possible with SN 1987A, by using the structures observed in time, neutrino energies, neutrino flavor, neutrino energetics, and even spatial distribution \cite{2017PhRvL.119e1101W}. Decoding the complex signal will be a real challenge but, as we're trying to argue, it's also a golden opportunity to extract a lot of valuable information for a wide range of physics. For that reason, the difficulty of decoding the signal is a challenge worth rising to. Similar issues of interpreting data that are related to the processes one wishes to understand via a complicated path occur in many areas of physics and every time the issues have been confronted and overcome because the physics was worth it. Supernova neutrinos are the same. 

\subsection{What}

Our next question is what kind of star explodes as a core collapse supernova? Theory indicates that core-collapse supernovae can be divided into two categories according to the progenitor. The first are the collapse of stars with cores made up of a mixture of oxygen-neon-magnesium---so called ONeMg supernovae (also known as `electron capture supernovae')---and the second are the collapse of stars with cores made up of iron. According to stellar evolution theory, both supernova classes are the endpoints of stars that start their lives with masses much greater than the Sun. However, over their lifetimes such stars will lose a substantial amount of their atmospheres and will be considerably less massive at the point of collapse. The amount of atmosphere lost strongly depends upon the amount of metals in the atmosphere of the star---`metals' here meaning anything heavier than helium---which drives line-driven winds. Many massive stars also evolve with companion stars---indeed, the progenitor of SN 1987A was part of a triple star system \cite{1987ApJ...321L..41W,1987ApJ...323L..35S}---allowing a history of mass transfer. The exact structure of the star at the point of collapse also depends upon a lot of other factors, such as the nuclear burning, the energy transport mechanisms (i.e., whether it is convection or radiation) through the star, rotation (i.e., the distribution of angular momentum), internal magnetic field and, if the star has a companion, the orbital parameters. Not all these details are well understood or, if understood, then not well implemented in stellar evolution models. Simulations of the core-collapse supernovae indicate the internal structure of the progenitor is crucial for determining what happens when the star eventually collapses. Small changes in the progenitor---due to, say, convection in a layer which is undergoing nuclear burning---can lead to substantial differences in the supernova dynamics and can even be the difference between whether or not a star explodes \cite{2013ApJ...778L...7C,2015MNRAS.448.2141M,2016ApJ...833..124M}. 

\begin{center}
\begin{table}
\begin{tabular}{lll}
 & ONeMg & Fe core \\
\hline
$M_{\rm ZAMS}$ & $\sim 8$--$11 M_\odot$ & $\gtrsim 11 M_\odot$ \\
Compactness $\xi_{1.0}$ & 2.5 & 1.0--1.7 \\
Compactness $\xi_{2.5}$ & $5 \times 10^{-6}$ & 0.003--0.4 \\
\end{tabular}
\caption{Comparison of ONeMg and Fe core progenitors. The first row is the zero-age main sequence mass. The mass at which the transition from ONeMg to Fe core occurs is highly uncertain. The second and third rows show compactness, defined in Eq.~(\ref{eq:compactness}). For ONeMg the $8.8 M_\odot$ progenitor of \cite{Nomoto:1984,Nomoto:1987} is adopted. For Fe core, we show the range in the 101 solar metal progenitors of \cite{Woosley:2002zz}.} \label{tab:progenitors}
\end{table}
\end{center}

At the present time theory indicates that for single stars with the same initial metallicity as the Sun, ONeMg supernovae occur in stars with initial masses---the so called the zero-age main sequence (ZAMS) mass---in the range $8\;{\rm M_{\odot}}$ to $11\;{\rm M_{\odot}}$ \cite{Jones:2013wda}, while iron core supernovae occur in stars above $11\;{\rm M_{\odot}}$. A short summary is given in Table \ref{tab:progenitors}, where we also show representative values of the core compactness, $\xi$, which is related to how fast or show the mass density falls away from the core (its explicit definition is given in Eq.~\ref{eq:compactness}). Nevertheless, there is significant uncertainty due to the microphysics and macrophysics mentioned in the preceding paragraph. Testing this theoretical prediction is not easy: there are not a lot of known progenitors of core-collapse supernova, and estimating the ZAMS mass is non-trivial. One way is to see what kinds of stars don't explode as supernovae. The compilation of white dwarf masses by Ref.~\cite{Dobbie:2006rr} suggests that stars up to 6.8--$8.6 M_\odot$ produce white dwarfs and thus beyond this is where core collapse starts. The direct method to finding supernova progenitors is to wait for the supernova to be discovered and then re-examine previous images of the same area of the sky. This was the approach used to identify the progenitor of SN 1987A \cite{1987ApJ...321L..41W,1987ApJ...323L..35S}, and dozens of other progenitors in nearby galaxies have been found by this method since \cite{2009MNRAS.395.1409S,2013MNRAS.436..774E,2012A&A...538L...8G,2016MNRAS.457.1107H}; see the review by Smartt \cite{2009ARA&A..47...63S}. Armed with pre-explosion images, one can try to link the properties of the progenitor to the properties of the supernova. In the case of SN 1987A, the progenitor was a blue supergiant. That it was not a red supergiant was a big surprise at the time \cite{1992PASP..104..717P} and required a re-examination of stellar evolution theory. More recently, observations of supernova progenitors in nearby galaxies have found red supergiant and also yellow supergiants undergoing supernovae. Smartt \textit{et al.} \cite{2009MNRAS.395.1409S} combined confirmed Type IIP supernova progenitors and upper limits placed by deep pre-images without progenitor confirmations, and derived statistically the lower limit of the mass of stars which explode as Type IIP supernova of $M_{min} = 8.5^{+1}_{-1.5}\;{\rm M_{\odot}}$, consistent with theoretical predictions.

In the past decade observers have taken less biased approaches to look for progenitors of supernovae such as ``A Survey About Nothing" \cite{2008ApJ...684.1336K,2015MNRAS.453.2885R} which aims to find massive stars undergoing core collapse regardless of whether they explode or not, i.e., even those that fail, presumably by forming a black hole. Although the survey strategy is such that no electromagnetic signal of failed explosions are required, we note that failed supernovae cannot be hidden entirely. For example, failed explosions have to emit almost as many neutrinos as a successful supernova and also, if the star has a sufficiently tenuous envelope, the mass-energy carried away by the neutrinos is enough to unbind the outermost layer of the atmosphere \cite{2013ApJ...769..109L,2013ApJ...768L..14P}. Two candidates for failed supernova have been reported \cite{2015MNRAS.453.2885R,2015MNRAS.450.3289G,2017MNRAS.468.4968A}. The upper limit on the fraction $f$ of stars which fail to explode is $f = 0.14^{+0.33}_{-0.10}$ at 90\% confidence \cite{2017MNRAS.469.1445A} if the failed candidate found by Gerke \textit{et al.} \cite{2015MNRAS.450.3289G} is included, and $f < 0.35$ at 90\% confidence if it is excluded; hardly insignificant fractions.

Before we dive further into the neutrinos from core-collapse supernova, we should mention that neutrinos are also emitted from the other common form of supernovae - thermonulcear or Type Ia. Type Ia supernova are thought to occur due to the explosive nuclear burning of a carbon-oxygen white dwarf when its mass crosses over the maximum mass which can be supported by electron degeneracy pressure -- the Chandrasekhar mass. The nuclear burning produces elements in the so-called `iron group' and some of these capture electrons emitting neutrinos. Also, the temperature in the supernova becomes so hot that the `thermal' production of neutrino-antineutrino pairs also occurs. The signal from these supernovae has been calculated  \cite{Odrzywolek2011a,2016PhRvD..94b5026W,2016arXiv160907403W} and it is found the flux of neutrinos from this kind of supernova are approximately four orders of magnitude lower than the flux from a core-collapse at the same distance. In addition the spectrum peaks at lower energy. Together, the lower flux and less-energetic spectrum makes detecting these kind of supernovae more difficult, nevertheless, it has been shown that a next-generation neutrino detector on the scale of Hyper-Kamiokande will detect a few events from a Type Ia supernova at the Galactic center and can distinguish between explosion mechanisms if the supernova is closer than $\sim 3\;{\rm kpc}$. 

Recently another type of supernova known as a pair-instability supernova (PISN) has also been considered \cite{PhysRevD.96.103008}. A PISN occurs when the carbon-oxygen core of a very massive star starts to produce electron-positron pairs. The pair creation softens the Equation of State (EOS) causing a contraction which, in turn, leads to explosive nuclear burning of the oxygen. The energy released is enough to unbind the star and leaves no remnant -- just like a Type Ia. In simulations of PISN huge quantities of $^{56}$Ni and other iron-group elements are produced thus the neutrino emission from a PISN is similar to the origin of neutrinos in a Type Ia. But because the mass of the cores of these stars is so much bigger, the luminosity is about two orders of magnitude higher than a Type Ia though the spectrum of neutrino emission is similar. 

\subsection{Where and When}

\begin{figure*}[t!] 
\includegraphics[width=0.9\linewidth]{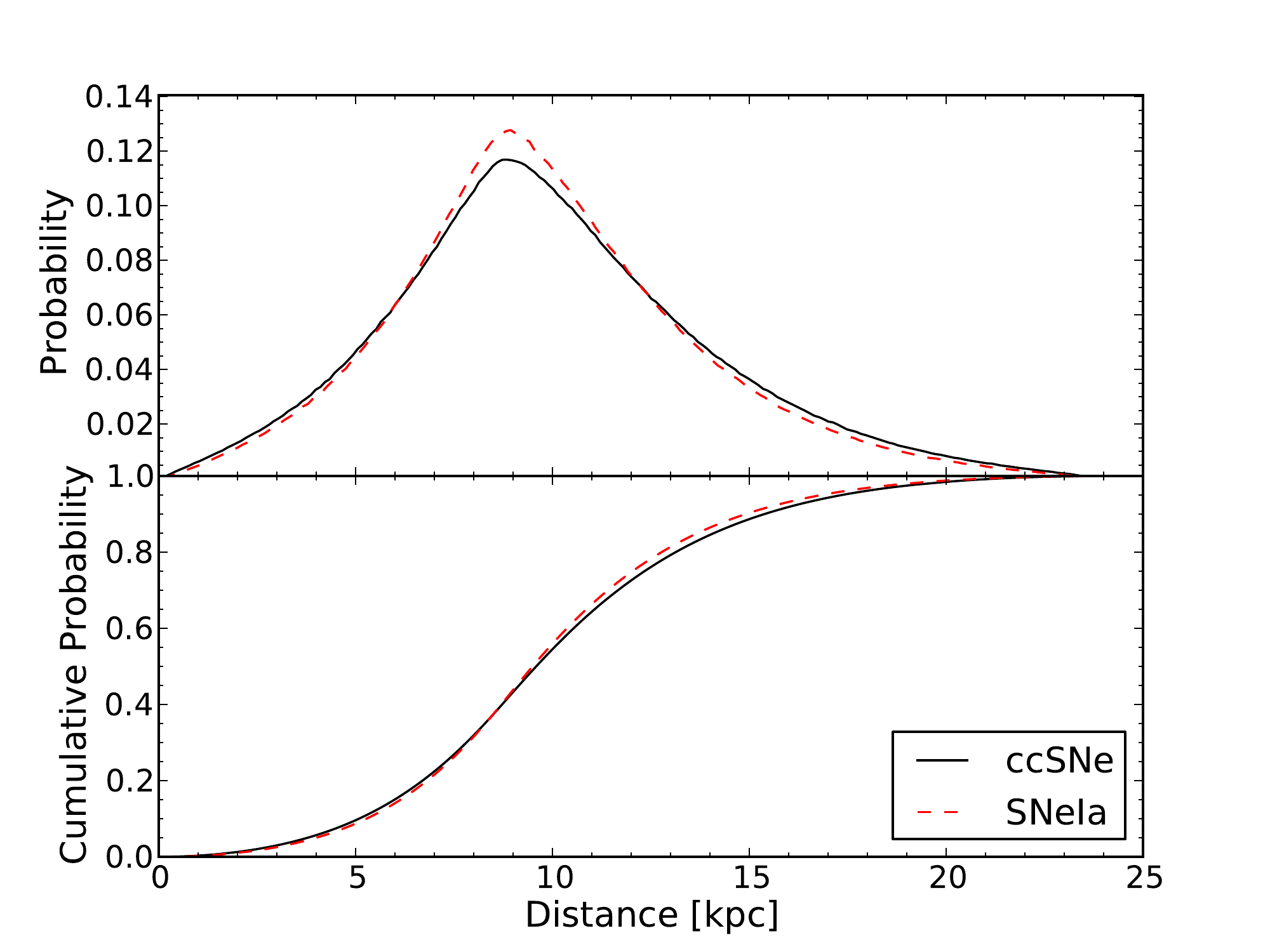}
\caption{The differential (top panel) and cumulative probability (bottom panel) of the distance to supernova in the Milky Way from the model by Adams \textit{et al.} \cite{2013ApJ...778..164A}. \copyright AAS. Reproduced with permission. \label{fig:SNdistribution} }
\end{figure*}

The \$64,000 question is when will the next Galactic supernova occur? Multiple approaches are viable, and many studies have estimated the core-collapse rate of the Milky Way.
Using a model for the distance, extinction, and magnitude probability distribution of supernovae in the Milky Way galaxy, Adams \textit{et al.} \cite{2013ApJ...778..164A} find that the historical rate of core-collapse supernovae in the Galaxy is $3.2^{+7.3}_{-2.6}$ per century. This represents ${\sim}70\%$ of the total supernova rate, the remainder being Type Ia supernovae. Scaling from supernova rates in other galaxies yields an estimate of $2.8 \pm 0.6$ per century \cite{1991ARA&A..29..363V,Li:2010kd}. Other methods yield similar rates: from the massive stellar birthrate, 1--2 supernova per century \cite{Reed:2005en}; from radioactive aluminum emission measurements, $1.9 \pm 1.1$ per century \cite{Diehl:2006cf}; and from the pulsar birthrate, $2.8 \pm 0.1$ per century \cite{Keane:2008jj}. 

The distribution of supernova distances have been investigated by several authors \cite{2013ApJ...778..164A,Mirizzi:2006xx}. They typically predict a peak near or just beyond the Galactic Center of the Milky Way. The results of the model by Adams \textit{et al.} is shown in figure (\ref{fig:SNdistribution}). They find that the most probable distance to the next core-collapse supernova in the Milky Way will be just shy of $d\sim 10\;{\rm kpc}$ \cite{2013ApJ...778..164A}. The lower panel shows the cumulative probability. Up to $d\sim 10\;{\rm kpc}$, the cumulative probability that a supernova occurs at distance $d$ or smaller follows $P(<d) \propto d^{2.5}$. The probability the next supernova occurs within a distance of $1\;{\rm kpc}$---an extremely nearby explosion with qualitatively new physics potential---is 0.2\%. Upgrades to neutrino detectors are in place to register neutrino events even in these very nearby cases with high event rates \cite{Orii:2015khh}. The probability within $5\;{\rm kpc}$ is 10\%. 


\section{The Neutrino Emission}\label{sec:emission}

Our understanding of the neutrino emission from core-collapse supernovae has improved tremendously in recent years due to the growing sophistication and duration of simulations. For both ONeMg and Fe-core supernova the neutrino emission can be divided into four `epochs': the presupernova phase, the collapse phase, the accretion phase, and finally the cooling phase. These epochs and their approximate duration are listed in table (\ref{tab:epochs}). In some literature, the presupernova, collapse and accretion phase are collectively referred to as the ``pre-explosion phase''. Let us consider them sequentially.
\begin{center}
\begin{table}
\begin{tabular}{lll}
 Epoch & Duration & Dominant neutrino \\
 \hline
 Presupernova & $\sim$ days & $\nu_e$ and $\bar{\nu}_e$ \\
 Collapse & $\lesssim 50$ ms & initially mostly $\nu_e$, later all flavors \\
 Accretion & $\sim 100$ ms for ONeMg core & $\nu_x < \bar{\nu}_e < \nu_e$ \\
 & $\sim 200-700 \;{\rm ms}$ for Iron core & $\nu_x < \bar{\nu}_e < \nu_e$ \\ 
 Cooling & $\sim 10 \;{\rm s}$ & $\nu_x \sim \bar{\nu}_e \sim \nu_e$
\end{tabular}
\caption{The four epochs of neutrino emission for a core-collapse supernova.} \label{tab:epochs}
\end{table}
\end{center}

\subsection{The presupernova epoch}

All stars emit neutrinos with the amount of neutrino emission dependent on the mass of the star, it's evolutionary state, and the reaction chains by which it is burning its nuclear fuel \cite{2004APh....21..303O,2004AcPPB..35.1981O,2015ApJ...808..168K,2015arXiv151102820P,2016PhRvD..93l3012Y}. The neutrino emission occurs via two processes: `weak' and `thermal' \cite{1964ApJS....9..201F,1980ApJS...42..447F,1985ApJ...296..197M,1996ApJS..102..411I,2016PhRvD..94d3005G}. The names are a little misleading: neutrinos can only be produced by weak reactions and in a star, the rate of emission is always very sensitive to the temperature. The origin of the names is that the neutrinos from weak processes emerge from reactions where a neutron is converted to a proton or vice-versa whereas thermal neutrinos do not. While on the main sequence, massive stars burn their hydrogen fuel via the CNO cycle emitting two neutrinos carrying on average a combined energy of $E\sim 1.5\;{\rm MeV}$ for every helium nucleus formed. These neutrinos are from weak processes. The number of neutrinos emitted from thermal processes during the main sequence is small but once triple-alpha begins the thermal processes increase significantly. At the point where carbon burning begins a star begins to emit more energy in neutrinos than it does in photons \cite{Woosley:2002zz}. The neutrino emission continues to accelerate until, at the point of collapse, the neutrino luminosity may be of order $\sim 10^6$ \cite{2005NatPh...1..147W} times larger than the photon luminosity. The calculations by Kato \textit{et al.} \cite{2015ApJ...808..168K} included just the thermal processes and found that there is a strong dependence on the progenitor, with many more neutrinos emitted by Fe-core progenitors than ONeMg progenitors. In a following study, Yoshida \textit{et al.} \cite{2016PhRvD..93l3012Y} considered the time variation of the neutrino emission and showed the emission decreases somewhat during the oxygen and silicon shell burning phases. When only thermal processes are included, one finds the number of electron neutrinos and electron antineutrinos emitted from a pre-supernova core is a factor of a few times larger than that of heavy leptons and all the spectra tend to peak at energies of $E_{\nu}\sim 1 \;{\rm MeV}$ which means low-threshold detectors are needed to see them. The weak processes are subordinate in terms of the amount of energy they carry away but weak processes are capable of emitting neutrinos up to $E_{\nu}\sim 10\;{\rm MeV}$ which makes them much easier to detect \footnote{Wright \textit{et al.} \cite{2016PhRvD..94b5026W,2016arXiv160907403W} found a similar result for the neutrinos from Type Ia supernovae} \cite{2015arXiv151102820P}. At the end of core silicon burning, the emission due to weak processes begins to overtake those of thermal processes so becomes predominantly electron type. Note that neutrino flavor transformation in the envelope of the star will introduce $\mu$ and $\tau$ flavor content as these neutrinos emerge from the star. 
	
\subsection{The collapse phase}

The cores of these stars are supported against gravity by electron degeneracy pressure with subdominant thermal pressure support. Eventually however, the Fermi energy of electrons increases enough that electron captures on the nuclei present in the core become energetically feasible, e.g.,\ ${\rm Fe} + e^- \to {\rm Mn}+\nu_e$. The neutrinos escape and thus such reactions reduce the pressure gradient which is holding up the star. Contributing to the pressure loss for more massive Fe-core progenitors is the photo-disintegration of the nuclei \cite{Woosley:2002zz}. Initially, for Fe-core progenitors, as the electron fraction $Y_e$---the ratio of the number density of electrons (or protons) to the total nucleon number density---decreases due to the electron captures, nuclei heavier than iron are formed \cite{2007PhR...442...38J}. The electron captures and photo-disintegration cause the average adiabatic index to fall below the equilibrium condition leading to instability and collapse. The collapse is swift, occurring on a free-fall timescale $\sim 0.04 {\rm \, s \, } (\bar{\rho}/10^{10} {\rm \, g \,cm^{-3}})^{-1/2} $, where $\bar{\rho} \sim 10^{10} {\rm \, g \,cm^{-3}}$ is the average stellar core mass density. This infall phase leads to the first important physical effect for the neutrino emission: neutrino trapping \cite{Freedman:1973yd,Sato:1975vu}. As the core collapses its temperature and density increases. Neutrinos that are initially freely streaming out of the core eventually start to diffusively escape when their mean free path becomes shorter than the core radius. This leads to the formation of a so-called \textit{neutrinosphere}, which defines the characteristic surface determining neutrino escape, much like the photospheres of stars for photons. Since the diffusion time scale, $\sim 0.2  {\rm \, s \, } (\bar{\rho}/10^{12} {\rm \, g \,cm^{-3}})$, is longer than the free-fall time scale of the core, $\sim 0.004 {\rm \, s \, } (\bar{\rho}/10^{12} {\rm \, g \,cm^{-3}})^{-1/2} $, the neutrinos that are generated are trapped in the core. Within the neutrinosphere, neutrinos are thermalized and diffuse from hot to cold regions. Eventually convective cells within the outer layers of the protoneutron star form and the neutrinos are carried to the neutrinosphere in the same way photons are carried along to the photosphere of the Sun in convective cells \cite{1986NYASA.470..267W}. 

The core collapse continues until it is halted by the repulsive part of the nucleon-nucleon potential. This occurs when nuclear densities of $\sim 3 \times 10^{14} {\rm \, g \,cm^{-3}}$ are reached. The sudden deceleration of the material in the core drives a pressure wave outwards through the core that eventually steepens into a shock wave. This shock would produce an optical explosion should it reach the photosphere of the progenitor. As simple and elegant as this mechanism may appear, the energetic costs of the propagating shock are unfortunately immense. First, the shock dissociates nuclei into free nucleons which then rapidly generate neutrinos via charge-current $e^- + p \to n + \nu_e$. When the shock is inside the neutrinosphere these neutrinos are trapped, but when the shock crosses the $\nu_e$ neutrinosphere, the neutrinos are free to escape. The passing of the shock through the neutrinosphere produces a huge spike in the neutrino emission called the \textit{neutronization burst}. The $\nu_e$ luminosity reaches a staggering $\sim 10^{53} {\rm \, erg/s}$ for $O$(10) ms, robbing the shock of $\sim 10^{51} {\rm \, erg}$ of precious energy. For Fe-core supernovae the energy costs makes the shock eventually stall at a distance of some 100--200 km from the core which we define as the end of the collapse phase. For ONeMg supernovae the shock velocity decreases but the shock never stalls \cite{2006A&A...450..345K}. This does not mean the star is exploding: the velocity of the material that passes through the shock is still negative, i.e., the material is falling towards the protoneutron star (PNS). For ONeMg supernovae we define the end of the collapse phase as being the time when the neutrino heating timescale becomes longer than the advection timescale. This is around $60-80$ ms depending upon the EOS \cite{2006A&A...450..345K}.

When all the input physics is held fixed, it is found the neutrino emission during the collapse phase depends upon whether the core was made up of ONeMg or iron but does not depend upon much else. 2D and 3D simulations give very similar results to spherically symmetric simulations. Also the neutrino emission for Fe-core supernovae up to the peak of the neutronization burst is very similar for all progenitor masses \cite{2013ApJ...767L...6B}, only after the peak do differences between progenitor masses emerge. These features make the neutronization burst a standard candle with unique potential to probe progenitor, core-collapse, and neutrino physics. However there is some degree of uncertainty in the nuclear physics which is relevant for this phase. Sullivan \textit{et al.} \cite{2016ApJ...816...44S} have undertaken a large set of simulations of numerous progenitors with various EOS and found changes of $+16/-4\%$ in the mass of the inner core at the point of shock formation and a range of $\pm 20\%$ of the peak $\nu_e$ luminosity during the deleptonization burst. Similarly there has been some recent re-assessment of the importance of nuclear excited states during this epoch \cite{2013PhRvC..88a5807M,2015PrPNP..85...33B,2016PhRvC..94e5808M}. While Fischer, Langanke and {Mart{\'{i}}nez-Pinedo, \cite{2013PhRvC..88f5804F} found no effect of `hot nuclei' upon the dynamics, they did change change the emerging neutrino spectra. 

The neutronization burst is followed by the rise of emission of the $\mu$ and $\tau$ neutrino and antineutrino flavors (often called the heavy lepton flavors and denoted as $\nu_x$ though the heavy lepton antineutrinos are occasionally treated as separate). The electron degeneracy of the post-bounce region is not high, and relativistic positrons are created thermally. This leads to the production of $\bar{\nu}_e$ via $e^+ n \to p + \bar{\nu}_e$, as well as all neutrinos by pair-annihilation $e^+  e^- \to \nu + \bar{\nu}$, nucleon-nucleon bremsstrahlung $NN' \to NN'+\nu\bar{\nu}$, and neutrino/anti-neutrino annihilation $\nu_i+\bar{\nu}_i \to \nu_j + \bar{\nu}_j$. The rise of these contributions to the neutrino emission means the luminosities of all flavors become rather similar with the luminosity of the electron neutrinos almost equal to the luminosity of the electron antineutrinos and both are $\sim 10-50\%$ higher than the luminosity of the heavy lepton flavors. It is after the neutronization burst peak that the mean energies of the different flavors forms the well-known hierarchy $\langle E_{\nu_e}\rangle < \langle E_{\bar{\nu}_e}\rangle < \langle E_{\nu_x}\rangle$. Less well known is that the pinch parameters $\alpha_\nu$ (which describes the spectral deviation from Fermi-Dirac, explained in the next subsection), which lie in the range of $3\lesssim \alpha_{\nu} \lesssim 4.5$ at this time \cite{Tamborra:2014hga}, also form a hierarchy with the heavy lepton flavors less pinched (lower $\alpha_\nu$) than the electron antineutrinos which are less pinched than the electron neutrinos.

\subsection{The energy spectrum of supernova neutrinos}

The literature usually characterizes the neutrino emission in terms of the luminosity of a particular flavor $L_{\nu}$, and its mean energy $\langle E_{\nu} \rangle$. However, care should be taken since other definitions are also used in the literature, e.g., the root-mean-squared mean energy $\langle E_{\nu}^2 \rangle^{1/2}$ and also the root-mean-squared deviation $(\langle E_{\nu} \rangle - \langle E_{\nu} \rangle^2)^{1/2}$.
The ratio $L_{\nu} / \langle E_{\nu} \rangle$ gives the overall number of neutrinos emitted but not the spectrum $\Phi_{i}(E)$. For simplicity the spectrum can be taken to be a Fermi-Dirac distribution, but more recently the spectrum was studied by Keil, Raffelt and Janka \cite{2003ApJ...590..971K} who found it more closely followed a Fermi-Dirac distribution with a pinched peak (which is also what is reconstructed from SN 1987A neutrinos with nonparametric inferential statistical methods \cite{2007PhRvD..76h3007Y}). They suggested the spectral functional form $\propto E^\alpha_\nu e^{-(\alpha_\nu+1)E/\langle E_\nu \rangle}$, where $\alpha_{\nu}$ is a numerical parameter describing the amount of spectral peak pinching. Normalizing, this yields the commonly adopted spectrum
\begin{equation}
\Phi_{i}(E) = \frac{L_{\nu}}{\langle E_{\nu} \rangle}\;
\frac { (\alpha_{\nu} + 1)^{\alpha_{\nu} + 1} }{ \langle E_{\nu} \rangle\,\Gamma (\alpha_{\nu} + 1) } \,\left( \frac{E}{\langle E_{\nu} \rangle}\right)^{\alpha_{\nu}} \exp \left(  - \frac{(\alpha_{\nu} + 1)\,E }{\langle E_{\nu} \rangle } \right) \label{eq:Phi}. 
\end{equation}
A value of $\alpha_\nu =2.0$ corresponds to a Maxwell-Boltzmann spectrum, while $\alpha_\nu =2.3$ to a Fermi-Dirac spectrum. The $\alpha_\nu$ parameter can be computed from the energy moments $\langle E_{\nu}^k \rangle$ by
\begin{equation}
\frac{\langle E_{\nu}^k \rangle}{\langle E_{\nu}^{k-1} \rangle} = \frac{k+\alpha_\nu}{1+\alpha_\nu}\langle E_{\nu} \rangle.
\end{equation}
Commonly it is sufficient to use $k=2$ which results in
\begin{equation}
\alpha_{\nu} = \frac{2\langle E_{\nu} \rangle^2 - \langle E^2_\nu \rangle}{\langle E^2_\nu \rangle-\langle E_{\nu} \rangle^2}.
\end{equation}
All of the neutrino spectral parameters---$L_\nu$, $\langle E_{\nu} \rangle$, and $\alpha$---are different for different neutrino flavors, vary considerably in time, and moreover strongly depend on the progenitor. The following sections focus on these dependencies. Bearing in mind the strong caveat of time evolution, representative values for ${\nu}_e$ are $L_\nu \Delta t \sim 5 \times 10^{52}$ erg, $\langle E_{\nu} \rangle \sim 12$ MeV, and $\alpha \sim 3$; for $\bar{\nu}_e$ they are $L_\nu \Delta t \sim 5 \times 10^{52}$ erg, $\langle E_{\nu} \rangle \sim 15$ MeV, and $\alpha \sim 3$; and for ${\nu}_x$ they are $L_\nu \Delta t \sim 5 \times 10^{52}$ erg, $\langle E_{\nu} \rangle \sim 16$ MeV, and $\alpha \sim 2$. In other words, the energetics is approximately equipartitioned and there is an energy hierarchy. 

\subsection{The accretion phase}

The next phase is critical for triggering an explosion, i.e., the shock must be energetically revived in order to reach the progenitor photosphere and gravitationally unbind the progenitor envelope. The mechanism by which this is achieved constitutes the explosion mechanism. Many ideas of how this accomplished have been explored. The common traits of all explosion mechanisms are the identification of new energy reservoirs and the mechanisms to channel this energy to the region immediately below the stalled shock. Some examples include magneto-rotational mechanisms \cite{1970ApJ...161..541L}, using the gravitational binding energy released from further collapse \cite{1993ApJ...414..701G}, as well as new physics such as sterile neutrinos \cite{Hidaka:2006sg}. Jet-induced collapse, in particular in the context of gamma-ray bursts, is another alternative. In these mechanisms, the explosion mechanism and the thermal neutrino emission may not be intimately linked. On the other hand, the thermal neutrinos play an integral part in the neutrino-driven delayed explosion mechanism. Here, the stalled shock is revived via neutrino heating, which opens a unique opportunity to test the explosion mechanism by future neutrino detections. Here, we focus on the neutrino mechanism. 

Interior to the stalled shock, the collapsed core by now has formed a hot and dense protoneutron star with a radius of $\lesssim 50$ km, which continues to emit neutrinos of all flavors. The total energetics of the neutrino emission is closely described by the gravitational binding energy liberated,
\begin{equation}
\Delta V_{\rm grav} \simeq  \left( \frac{3}{5} \frac{G_N M^2}{R} \right)_{\rm PNS} - \left( \frac{3}{5} \frac{G_N M^2}{R} \right)_{\rm Fe \, core} \sim 3 \times 10^{53} {\rm \, erg},
\end{equation}
which significantly exceeds the energy needed by the stalled shock to trigger an explosion, $E_{\rm kin} \sim 10^{51} {\rm \, erg}$. To obtain an order of magnitude picture of how the neutrino mechanism works, consider a mass shell at radius $R$ that is interior to the stalled shock but outside the neutrinosphere, i.e., $R_\nu \ll R \ll R_s$. It experiences heating by the neutrino luminosity $L_\nu$ emitted from the neutrinosphere, with a heating rate $Q_+ \sim L_\nu \sigma(T_\nu) / (4 \pi R^2)$, where $\sigma(T_\nu)$ is the neutrino absorption cross section on free nucleons and $T_\nu$ is the neutrinosphere temperature. On the other hand, the shell is subject to neutrino cooling at a rate $Q_- \sim - \sigma(T) a c T^4$, where $\sigma$ is the absorption cross section, $a$ is the radiation constant, and $T$ is the temperature of the mass shell. Since $\sigma \propto T^2$, and expressing $L_\nu \simeq \pi R_\nu^2 a c T_\nu^4$, the net heating can be written
\begin{equation}
Q_{\rm tot} = Q_+ + Q_- = Q_+ \left[ 1 - \left( \frac{2R}{R_\nu} \right)^2 \left( \frac{T}{T_\nu} \right)^6 \right].
\end{equation}
In the radiation dominated PNS atmosphere $R_\nu < R < R_s$, the temperature falls with radius $T \propto R^{-1}$, so one can obtain the radius where there is net energy gain, 
\begin{equation}
r_g = \sqrt{\frac{2 R_s^3}{R_\nu} \left( \frac{T_s}{T_\nu} \right)^3},
\end{equation}
the so-called gain radius. For typical parameters found in simulations $T_\nu = 4.2$ MeV, $T_s = 1.5$ MeV, $R_\nu = 80$ km, and $R_s = 200$ km, one obtains $R_g \approx 95$ km, indicating that there is indeed a region of net energy gain interior to the stalled shock.
The success of the neutrino mechanism to trigger an explosion depends crucially on the net neutrino heating, which in turn depends on the neutrino emission from the neutrinospheres and their semi-transparent coupling with the post shocked matter including potential novel neutrino flavor mixing effects. Thus, accurate neutrino transport becomes necessary, leading to the development of full Boltzmann transport equation solvers to various approximation schemes, e.g., see Sumiyoshi \textit{et al.} \cite{2015ApJS..216....5S} for a comparison of methods. Simultaneously, hydrodynamical instabilities in the collapsing core have been shown to grow from small perturbations to impact the supernova mechanism. Due to the complex nature of the phenomena that occur during the accretion phase, numerical simulations are absolutely critical for realistic investigations of the core collapse and its neutrino signal during this epoch. Whats is found is that during the accretion phase the luminosities of the different flavors are more-or-less constant with $L_{\nu_e} \approx L_{\bar{\nu}_e}$ and each is approximately $50\%$ higher than $L_{\nu_x}$. The mean energies generally increase during the accretion phase and it is often seen in simulations that $\langle E_{\bar{\nu}_e}\rangle \approx \langle E_{\nu_x}\rangle$ as the cooling epoch is approached. At the transition from the collapse to the accretion phase the hierarchy in the pinch parameters changed so that the electron antineutrinos are now more pinched than the electron neutrinos but the heavy lepton flavors remaining the least pinched. Over the course of the accretion phase the amount of pinching of each spectra decreases \cite{Tamborra:2014hga}. 

Eventually, the net neutrino heating must overcome the impinging ram pressure due to the accretion of mass, which depends on the stellar structure. For the case of a star with an ONeMg core, where the progenitor density sharply declines outside the core, the significant reduction of ram pressure due to the lower mass accretion ram pressure does lead to supernovae even in one-dimensional, spherically symmetric, simulations. Indeed, as previously mentioned, simulations of ONeMg supernovae \cite{2006A&A...450..345K,2010PhRvL.104y1101H,2017arXiv170203927R} find the shock velocity never reaches zero. Once the neutrino heating timescale becomes shorter than the advection timescale, the velocity of the material that has immediately passed through the shock climbs from very negative to zero and then eventually becomes positive. This takes $\sim 100$ ms to occur.

For an Fe-core supernova, the matter accreted at the start of the accretion phase is silicon but this can later switch to oxygen before the shock is revived. The change in the composition of the accreting mass shell from silicon to oxygen reduces the ram pressure and causes the shock to expand \cite{Tamborra:2014hga}. Even so, there is now consensus within the modeling community that one-dimensional spherically symmetric simulations of Fe core collapse do not explode \cite{2001PhRvD..63j3004L,2001PhRvL..86.1935M,2002A&A...396..361R,2003ApJ...592..434T,2005ApJ...620..840L}. 
Thus, solving the mystery of the explosion mechanism of Fe core supernovae requires a better understanding of multi-dimensional phenomena in the collapsing core. The failure of one dimensional simulation to explode indicates the spherically symmetry has to be broken and from two and three dimensional simulations it is found this occurs during the accretion phase. A number of multi-dimensional physics effects are seen to emerge. 

\begin{itemize}

\item \textit{Convection}: The first is convection which has long been recognized as an important ingredient in supernova studies \cite{1979MNRAS.188..305E,1987ApJ...318L..57B,1990RvMP...62..801B,1994ApJ...435..339H,1995PhR...256..135J}. The origin, duration and location of the convection has changed over time with the current understanding that convection develops in two regions in a supernova: the first is inside the PNS below the neutrinosphere driven by a negative lepton number gradient, and the second is in the gain region---the region immediately below the shock where neutrino heating dominates over neutrino cooling---driven by a negative entropy gradient \cite{1994ApJ...435..339H,1996A&A...306..167J,2006ApJ...645..534D}. The convection in the outer region is often referred to as being neutrino driven. The effect of convection in both regions is to transport heat outwards with the convection inside the PNS also transporting lepton number \cite{1979MNRAS.188..305E}. Convection in the inner region develops within $\sim 30- 40\;{\rm ms}$ after bounce \cite{2006A&A...447.1049B} (which means it begins towards the end of the collapse phase) and leads to a boost in the luminosity of the $\bar{\nu}_e$ and heavy lepton flavors by $\sim 15\%$ and $\sim 30\%$, respectively \cite{2006ApJ...645..534D}, but reduces the mean energies \cite{2006A&A...447.1049B}. Interestingly it was shown by Roberts \textit{et al.} \cite{2012PhRvL.108f1103R} that the convection in the PNS is sensitive to the nuclear symmetry energy. The prompt convection also manifests in the gravitational wave signal as well, opening promising multi-messenger possibilities \cite{Mueller:2012sv}. Convection in the outer region affects many aspects of the explosion. First it transports heat from the region close to the PNS out to the stalled shock \cite{1990RvMP...62..801B,1994RvMA....7..103M}. Convection changes the accretion rate 
onto the PNS and also the pattern of accretion \cite{1992ApJ...395..642H,1994ApJ...435..339H,1996A&A...306..167J,2006ApJ...652.1436F}. 
With convection, accretion is directed into `downflows,' i.e., narrow channels of downward moving material surrounded by hot, lower density bubbles, see, e.g., figure (9) in Bruenn \textit{et al.} \cite{2016ApJ...818..123B}. At the locations where the downflows meet the PNS, hotspots of neutrino emission occur \cite{2015ApJS..216....5S,2016ApJ...818..123B}. A rotating PNS can cause a `lighthouse' effect.

\item \textit{SASI}: The Standing Accretion Shock Instability (SASI) \cite{2003ApJ...584..971B} is a very interesting phenomenon in which small perturbations of a spherical accretion shock---which is the situation when the forward shock stalls---grow rapidly in amplitude until the shock is seen to `slosh' back and forth with a very pronounced periodicity. In three dimensions the SASI can also develop a spiral mode which is seen to aid the explosion \cite{2013ApJ...770...66H,2015MNRAS.452.2071F}. The SASI affects the mass accretion onto the proto-neutron star such that the neutrino luminosity also begins to exhibit periodic fluctuations. The SASI and neutrino driven convection are often regarded as being in competition with each other and which dominates is seen to be related to the accretion rate \cite{2012ApJ...761...72M,2014MNRAS.440.2763F}.

\item \textit{Turbulence}: More recently the attention of some in the simulation community has begun to focus upon the small scale physics and in particular, turbulence  \cite{2011ApJ...742...74M,2012ApJ...749..142F,2013ApJ...765..110D,2013PhST..155a4022E,2014ApJ...783..125H,2015ApJ...808...70A,2015ApJ...799....5C,2015ApJ...808...70A,2016ApJ...820...76R}. What turbulence does is effectively stiffen the EOS of the material behind the shock so that it exhibits a greater pressure for a given density. This causes the shock to be pushed outwards compared to 1D simulations increasing the size of the gain region \cite{2015ApJ...799....5C}. 

\item \textit{LESA}: Finally we mention a new phenomenon that has been discovered in some simulations known as the Lepton-Emission Self-sustained Asymmetry (LESA) \cite{2014ApJ...792...96T,Tamborra:2014hga,2017ApJ...839..132T}. 
The LESA is a large scale asymmetry where the proto-neutron star / accreting matter system during the accretion phase develops and maintains a significant, quasi-stable dipole moment in the electron neutrino/antineutrino emission. The difference in the flux can be as large as a factor of $\sim 2$ in certain directions. The dipole is constrained to just the emission of electron lepton number: there is essentially no asphericity in the emission of the heavy lepton flavors nor in the overall luminosity \cite{2014ApJ...792...96T}. 

\end{itemize}

What could be called `criticality studies' have examined the effectiveness of various dimensionality effects upon the explosion. All indicate that a star is easier to explode in two dimensions than in one, i.e., the critical neutrino luminosity is lower when the same star is modeled in 2D than 1D, and while most find it more difficult to explode a star in 3D than 2D  \cite{2013ApJ...775...35C,2014ApJ...786...83T,2015ApJ...799....5C,2015MNRAS.452.2071F}, others find the opposite \cite{2013ApJ...765..110D}. Fernandez \cite{2015MNRAS.452.2071F} found the difference between 2D and 3D was larger when the dynamics was dominated by the SASI; when the dynamics were convection dominated the difference between 2D and 3D was smaller. The difference between 2D and 3D was also dependent upon the spatial resolution of the simulation.The general trend of greater difficulty of exploding a star in 3D compared to 2D (and also that the difference depends upon whether dynamics are convection or SASI dominated) is also seen in the more sophisticated simulations including neutrino transport \cite{2012ApJ...755..138H,2014ApJ...785..123C,2015ApJ...807L..31L} but, in one case, once the star was exploding, M\"uller \cite{2015MNRAS.453..287M}, found the explosion energy in 3D was larger than for the same progenitor in 2D. The reason for the difference is an interesting one with explanations focusing upon the very different behavior of turbulence in 2D than in 3D \cite{2012ApJ...755..138H,2015ApJ...799....5C}, the greater buoyancy compared to the drag of plumes in 2D than 3D \cite{2013ApJ...775...35C}, and the difference between the SASI in 3D---when it can posses spiral modes---compared to 2D \cite{2015MNRAS.452.2071F}. 

\subsubsection{Progenitor dependence}\label{sec:accretion:progenitor}

In parallel to the rapid progress in multi-dimensional simulation efforts, spherically symmetric studies continue to be valuable. They are relatively computational inexpensive and it is now feasible to perform hundreds of simulations using reliable neutrino transport schemes. This provides an efficient way to test the impacts of a variety of input physics. Since spherically symmetric simulations do not typically lead to explosions, explosions of Fe-core progenitors are artificially induced. The prescriptions for how to make a one-dimensional simulation explode vary depending upon the details of the simulation, see, e.g., Ref.~\cite{2015ApJ...806..275P} for a summary of the various techniques. In the case of the simulations by Fischer et al \cite{2010A&A...517A..80F}, the electron neutrino /antineutrino - nucleon absorption cross sections and emissivities were multiplied by a factor of 5--7: the PUSH method \cite{2015ApJ...806..275P} adds heat in the gain region proportional to the $\mu$ and $\tau$ flavor neutrino fluxes without changing the charged current reaction rates. The advantage of the the latter approach is that it does not keep the neutron-proton ratio in equilibrium beyond where it would naturally depart thus producing more realistic electron fraction profiles.

\begin{figure*}[t] 
\includegraphics[width=0.8\linewidth]{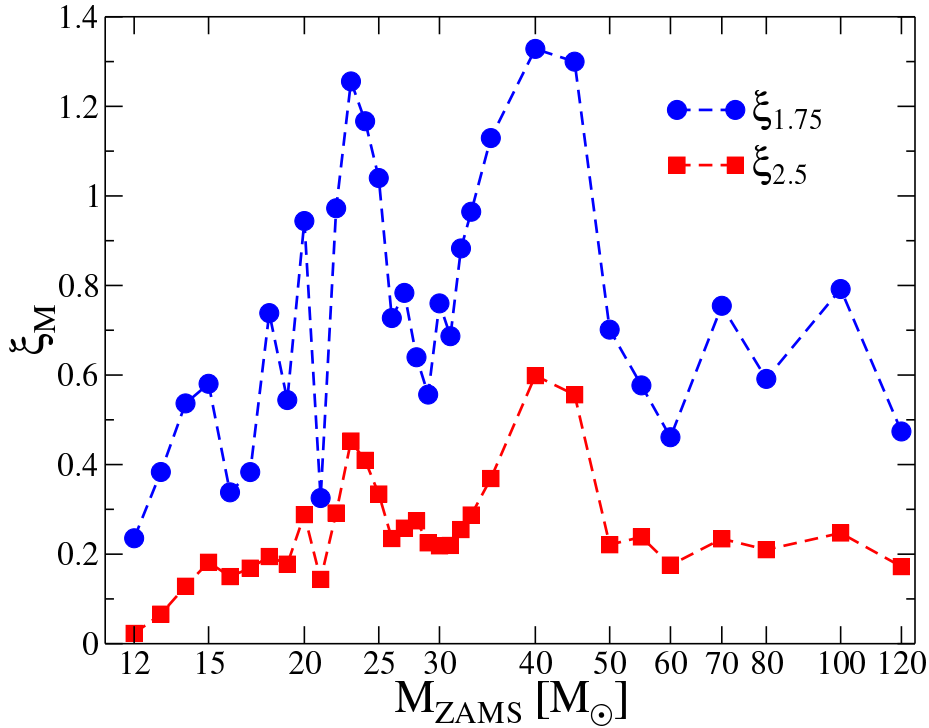}
\caption{The compactness parameters for the 32 presupernova models of Woosley \& Heger \cite{2007PhR...442..269W} vs. ZAMS mass as evaluated from collapse simulations with the Lattimer and Swesty \cite{1991NuPhA.535..331L} LS220 EOS. The figure shows both $\xi_{1.75}$ and $\xi_{2.5}$. The mapping between ZAMS mass and precollapse structure is highly non-monotonic, making the former an ill-suited parameter for describing progenitor structure in core collapse simulations. Figure taken from O'Connor \& Ott \cite{2013ApJ...762..126O} \copyright AAS. Reproduced with permission. \label{fig:O&OFig1} }
\end{figure*}

One of the insights gained from recent simulation suites in 1D is the quantitative importance of the progenitor properties for the explosion mechanism. The neutrino emission up to a postbounce time of $t=440\;{\rm ms}$, i.e., up to a point during accretion phase, was studied by O'Connor \& Ott \cite{2013ApJ...762..126O} using spherically symmetric full general relativistic simulations in a wide range of progenitors. They found the emission closely followed the compactness $\xi_{M}$ of the progenitor, not the ZAMS mass of the progenitor. The compactness is defined by the same authors to be \cite{2011ApJ...730...70O}
\begin{equation}\label{eq:compactness}
\xi_M = \left. \frac{M/ M_{\odot} }{ R(M_{bary}=M) / 1000\;{\rm km} } \right|_{t=t_{bounce}}
\end{equation}
where $M$ is some chosen mass and $R(M_{bary}=M)$ is the radial coordinate which encloses that (baryonic) mass at the time of core bounce. The plot of the compactness $\xi_{1.75}$ and $\xi_{2.5}$ versus ZAMS mass from O'Connor \& Ott is shown in figure (\ref{fig:O&OFig1}). The neutrino luminosities and mean energies O'Connor \& Ott found in their simulations for two different EOS are shown in figure (\ref{fig:O&OFig3}). The figure shows how these quantities correlate very closely with the compactness and this correlation is independent of the EOS. Nakamura \textit{et al.}~\cite{2015PASJ...67..107N} also found close correlations of many quantities with the compactness in a suite of 378 two-dimensional axis-symmetric simulations. For example, quantities such as the mass accretion rate, remnant mass, neutrino luminosities, and synthesized nickel mass all scale with the progenitor compactness, as shown in Figure \ref{fig:NakamuraFig14}. In particular, the mass accretion rate has been argued to be an important driver for the competition between the growth of SASI versus convection in the neutrino mechanism \cite{2012ApJ...761...72M}. Progenitors with high compactness lead to high accretion rates and when the simulations are undertaken it is seen the dynamics are SASI dominated; if the progenitor has low compactness then the simulations are convection dominated. 

\begin{figure*}[p] 
\includegraphics[width=0.9\linewidth]{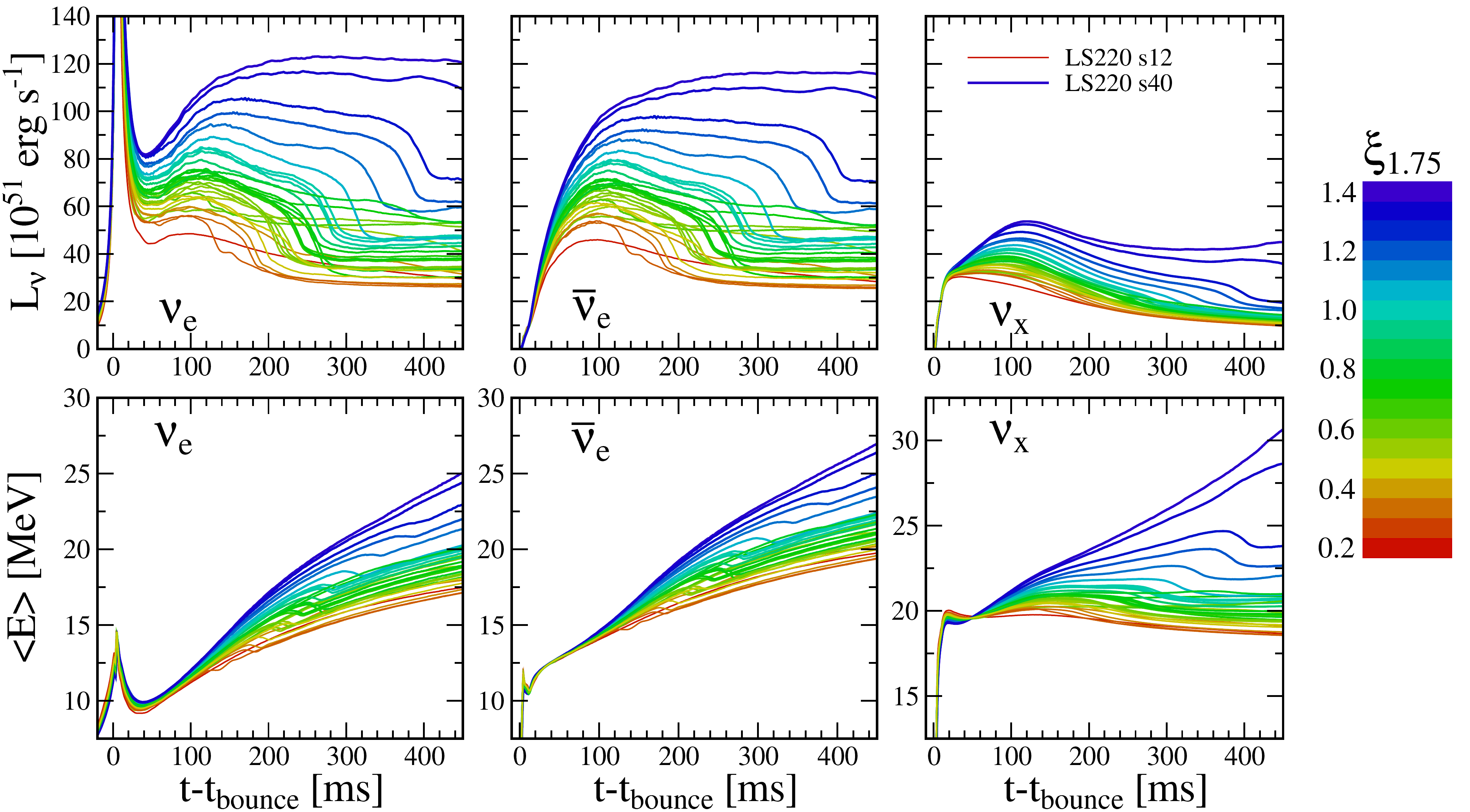}
\includegraphics[width=0.9\linewidth]{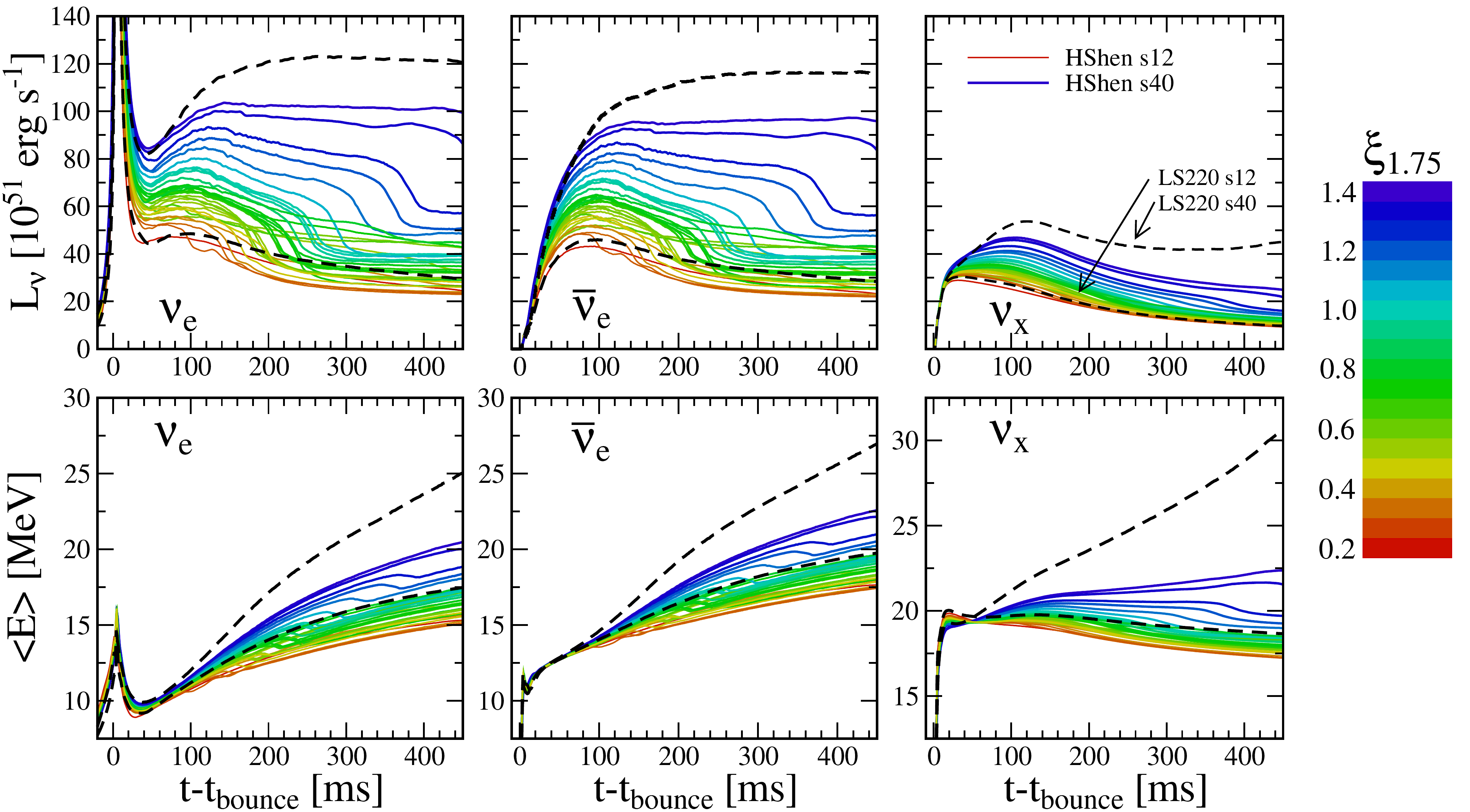}
\caption{Neutrino luminosities (top panels) and average energies (bottom panels) plotted as a function of postbounce time for all 32 models of Woosley \& Heger \cite{2007PhR...442..269W}. The top set of panels shows results obtained with the LS220 EOS. The bottom panel shows the same for the HShen EOS, but includes, for reference, two Lattimer and Swesty \cite{1991NuPhA.535..331L} LS220 models: s12WH07 and s40WH07. The left, center, and right panels show results for $\nu_{e}$, $\bar{\nu}_{e}$ and $\nu_{x}$ respectively. The curves are color- and line-weight-coded with increasing compactness $\xi_{1.75}$ with the mapping from color to compactness parameter shown on the right. There is a clear trend in all luminosities and average energies with compactness parameter. The progenitor with the highest compactness, s40WH07, forms a black hole at 503 ms after bounce. None of these models explode, but the onset of an explosion in any of these models may lead to a sudden deep drop (strongest for $\nu_{e}$, $\bar{\nu}_{e}$) in the luminosities and average energies (Fischer et al. \cite{2010A&A...517A..80F}), although this is likely suppressed by multidimensional effects. The smaller drop observed for most models here is due to the sudden decrease of the accretion rate when the silicon-oxygen interface reaches the stalled shock. Figure from O'Connor \& Ott \cite{2013ApJ...762..126O} \copyright AAS. Reproduced with permission. \label{fig:O&OFig3} }
\end{figure*}

Criticality studies by a number of authors \cite{1993ApJ...416L..75B,2005ApJ...623.1000Y,2006ApJ...650..291Y,2008ApJ...688.1159M,2009ApJ...703.1464F,2012PhRvL.108y1101K,2012ApJ...746..106P,2013ApJ...775...35C,2016ApJ...816...43S,2016ApJ...818..165Y} have examined the minimum neutrino luminosity needed to overcome the ram pressure from a given mass accretion rate. The accretion rate is related to the compactness---the more compact the star then the higher the accretion rate---so stars with high compactness should take longer to explode or, maybe don't explode at all. Various authors have confirmed this prediction, sometimes refining / replacing the criteria for explodability by replacing compactness with another quantity which measures something similar \cite{2012ApJ...757...69U,2015PASJ...67..107N,2015ApJ...801...90P,2016ApJ...816...43S,2016ApJ...818..124E,2016ApJ...821...38S}. Whatever is used, the basic result is that the explodability of a star is not correlated with the ZAMS mass. Instead one finds islands of ZAMS masses which produce failed supernova amid regions of ZAMS masses where the star does explode. One of these regions that leads to failed supernova is found between $\sim 20\;{\rm M_{\odot}}$ to $\sim 25\;{\rm M_{\odot}}$ with the exact range depending how the simulation was done \cite{Sukhbold:2013yca}. This island of ZAMS masses which produce failed supernovae appears to match the conclusion from progenitor surveys made by Kochanek \textit{et al.} \cite{2008ApJ...684.1336K} that there are a lack of Type IIP supernova progenitors with masses above $\sim 18 \;{\rm M_{\odot}}$  (see also Smartt \textit{et al.} \cite{2009MNRAS.395.1409S} for updates). Failure to explode is not the only explanation for this observation, e.g., it may be that stars with ZAMS masses above $\sim 18 \;{\rm M_{\odot}}$ evolve to a state where they are too dim to observe before the star explodes \cite{2012A&A...544L..11Y} or they may be dust obscured \cite{2012MNRAS.419.2054W}. However, it is intriguing that a failed explosion explanation is also able to explain the mass function of compact objects \cite{Kochanek:2013yca,Kochanek:2014mwa}, as well as the dearth of optically luminous supernova compared with the birth rate of massive stars \cite{2011ApJ...738..154H,2013ApJ...769..113H,2014MNRAS.445L..99H}. The scenario also implies $3/4$ of Type Ibc supernovae arise from binary stripped progenitors of mass $<16.5 M_\odot$, while $1/4$ arise from single stars of ZAMS mass $30$-$40 M_\odot$ \cite{2014MNRAS.445L..99H}--- and such high binary fractions are consistent with recent reports of binarity in massive O stars \cite{Sana:2012px}. It may also be consistent with systematic three-dimensional simulations \cite{2014MNRAS.445L..99H}, although more simulations are required. 

\begin{figure*}[t!] 
\includegraphics[width=0.9\linewidth]{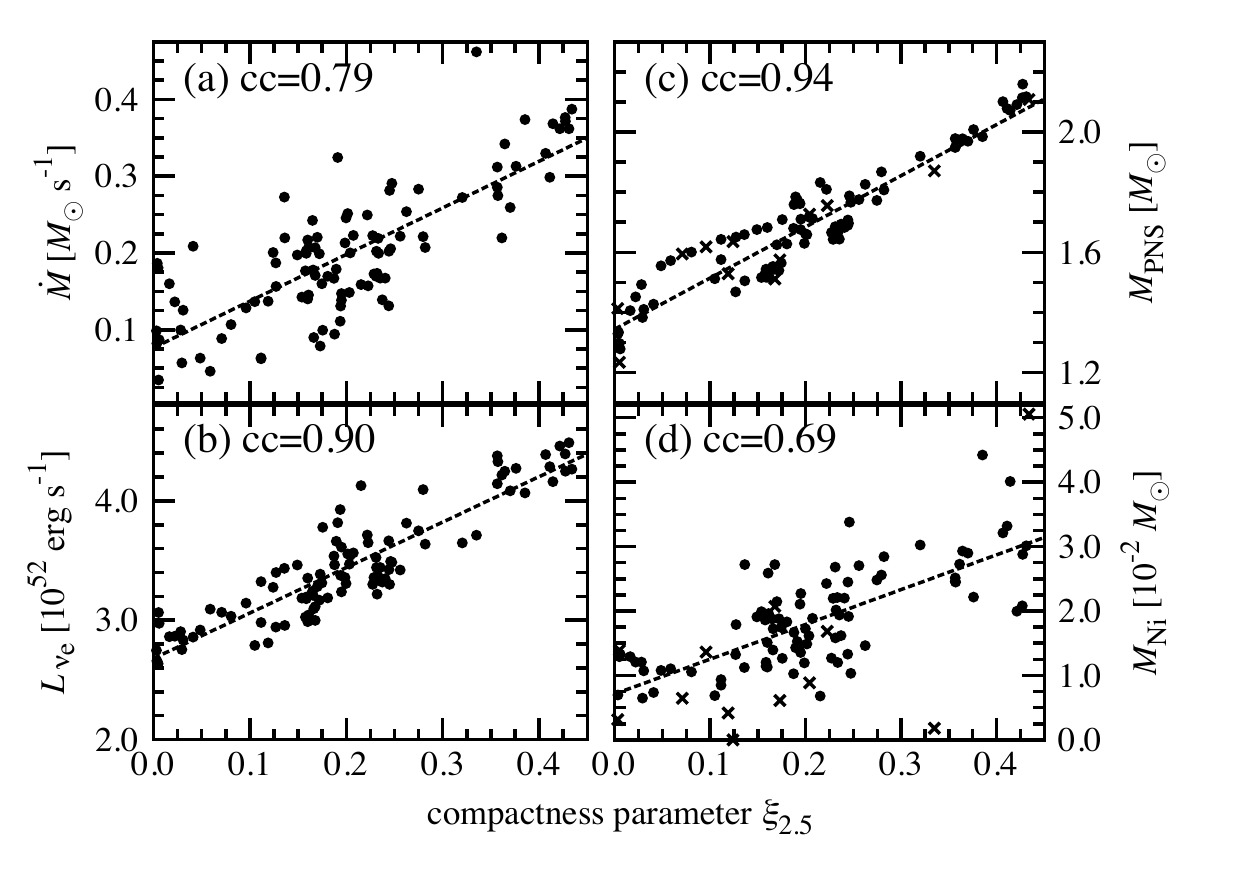}
\caption{Core-collapse properties based on a section of 101 solar metallicity progenitors as functions of the compactness parameter $\xi_{2.5}$: (a) mass accretion rate, (b) electron neutrino luminosity, (c) protoneutron star mass, and (d) mass of nickel in the outgoing unbound material. All quantities are the final time of the simulations. Dashed lines are linear fits to the simulation results. Figure from Nakamura \textit{et al.}~ \cite{2015PASJ...67..107N} \label{fig:NakamuraFig14} }
\end{figure*}

\subsection{The cooling phase}

The accretion phase is seen to last at most a second in simulations. The data from SN 1987A also indicate the existence of an accretion phase of similar duration. However long it lasts, once the shock is revived and the shock wave is moving outwards again, the mass accretion onto the proto-neutron star drops and the neutrino emission changes. This is the end of the accretion phase and the beginning of the cooling phase of the neutrino signal. In one-dimensional simulations the transition to the cooling phase is quite abrupt but in two and three dimensional simulations the transition is more smooth with continued accretion from downflows even when the shock has begun to move outwards. Because multi-dimensional simulations are so computationally expensive, few extend beyond $t_{pb} \sim 1\;{\rm s}$. Most of our theoretical understanding of the neutrino emission during the cooling phase comes from one-dimensional simulations.
\begin{figure*}[t!] 
\includegraphics[width=0.9\linewidth]{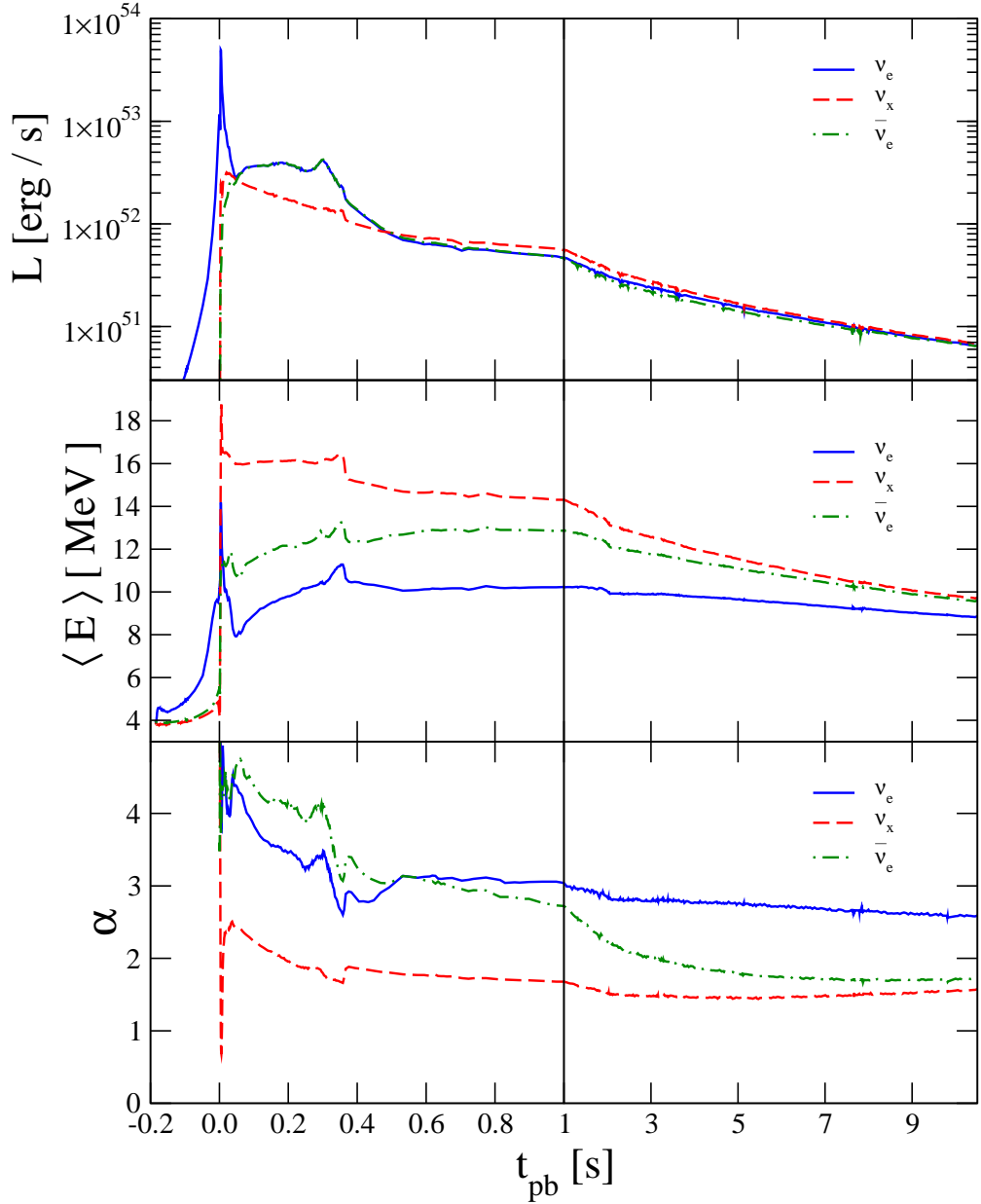}
\caption{The luminosity (top panel), mean energy (middle panel), and pinch paramter as a function of postbounce time from the same simulation of a $10.8\;{\rm M_{\odot}}$ progenitor by Fischer \textit{et al.} \cite{2010A&A...517A..80F}. \label{fig:L,meanE} }
\end{figure*}
In particular a set of three, long duration one-dimensional simulations were undertaken by Fischer \textit{et al.}~\cite{2010A&A...517A..80F} for a $8.8\;{\rm M_{\odot}}$ ONeMg supernova, and two Fe-core supernova with masses of $10.8\;{\rm M_{\odot}}$ and $18\;{\rm M_{\odot}}$. At the present time these are the best long duration simulations available and have become a de facto standard set of neutrino spectra. The neutrino luminosity $L_{\nu}$, mean energies $\langle E_{\nu} \rangle$ , and post-bounce pinch parameters $\alpha$ for the simulation of the $10.8\;{\rm M_{\odot}}$ progenitor from their study are shown in are shown in figure (\ref{fig:L,meanE}). The spike in the electron neutrino luminosity close to $t_{pb}= 0\;{\rm s}$ is the neutronization burst. In this simulation the accretion phase lasts until $t_{pb}\approx 0.3\;{\rm s}$ and then the shock is revived. The general features of the luminosities and mean energies of the $18\;{\rm M_{\odot}}$ progenitor they studied are similar. The simulation of the $8.8\;{\rm M_{\odot}}$, ONeMg progenitor has essentially no accretion phase. The trends in the luminosities and mean energies during the collapse and accretion phases of the different flavors, described previously,  are seen in the figure and we see that as the cooling phase is entered, the luminosities of the electron neutrino and antineutrino drop suddenly and become essentially the same as the heavy lepton flavors. However the hierarchy of mean energies established after the neutronization burst remains and we also observe how the hierarchy in the pinch parameters switches back to electron neutrinos being more pinched than the electron antineutrinos which are more pinched than the heavy lepton flavors. 

Finally, the convection inside the PNS that was established towards the end of the collapse phase continues through accretion and into the cooling phase. Its duration depends upon the neutrino opacities and EOS. When this convection terminates there is a change in the neutrino luminosity and cooling timescale \cite{2012PhRvL.108f1103R}. At these same late times Horowitz \textit{et al.} \cite{2016arXiv161110226H} found that nuclear pasta phases alter the neutrino emission slowing the neutrino diffusion and greatly increasing the length of the neutrino signal.

\subsection{Phase Transitions and Black hole formation}

Depending upon the EOS of dense matter, it is possible a newly formed PNS becomes unstable as it grows and collapses again to form an object supported not by nucleon-nucleon repulsion but some other source of pressure. Many scenarios have been considered with the most popular the phase transition from hadronic to quark matter. For a long time the conventional wisdom was that the maximum mass of the exotic star that would form was less than the maximum mass of cold neutron stars \cite{2001ApJ...550..426L,2002PhLB..526...19B,2004PhRvD..70d3010M} and so exotic stars could be identified by their different mass-radius relationship \cite{2000NuPhA.677..463S}. That appears not to be the case with more recent EOS's leading to exotic stars which can look, from the outside, very similar to neutron stars and have masses larger than $\sim 2\;{\rm M_{\odot}}$ \cite{2005ApJ...629..969A,2013A&A...551A..61Z}, compatible with the observations \cite{2016ARA&A..54..401O}.

Simulations \cite{1988ApJ...335..301T,1993ApJ...414..701G,2009PhRvL.102h1101S,2008PhRvD..77j3006N,2011ApJS..194...39F} indicate the sequence of events when a phase change occurs is very similar to the collapse of the original Fe or ONeMg core. These secondary collapses have been named quark-novae \cite{2002A&A...390L..39O} even though the cause may not be a phase transition to quark matter. Whatever the origin of the instability, the consequence of the second collapse is yet more release of gravitational binding energy which has to emerge in the form of neutrinos. The inner most region of the PNS collapses and rebounds; the pressure wave generated by the bounce is transformed into a shock wave and when the shock passes through the neutrinospheres, there is a large burst of neutrino emission and the energy of the emitted neutrinos changes. But unlike the neutronization burst which is principally the emission of electron neutrinos only, the second neutrino burst is in all flavors though dominated by the emission of electron antineutrinos \cite{2009PhRvL.102h1101S}, and easily detectable \cite{Dasgupta:2009yj}. As a consequence of the second collapse some of the outer layers of the neutron-rich pre-collapse PNS are ejected and later a proton-rich neutrino driven wind forms above the newly created proto-exotic star. The nucleosynthesis that occurs in the ejected material was considered by Nishimura \textit{et al.} \cite{2012ApJ...758....9N} who found it could produce nuclei of the so-called weak r-process and, when uncertainties in the explosion dynamics were taken into account, could just about make nuclei associated with the strong r-process. 

If the collapse of the PNS is not halted by an exotic phase or the proto-exotic star itself becomes unstable, a black hole can form. Simulations of black hole formation in supernovae \cite{1995ApJ...443..717B,1996ApJ...468..823B,2008ApJ...688.1176S,2009A&A...499....1F,2011ApJ...730...70O} indicate the collapsing PNS releases more gravitational binding energy and with the increasing density, for a brief period, the luminosity and the mean energies of the $\mu$ and $\tau$ flavour neutrinos actually increase \cite{2004ApJS..150..263L}. Both Fischer \textit{et al.} \cite{2009A&A...499....1F} and Sumiyoshi, Yamada and Suzuki \cite{2008ApJ...688.1176S} found a strong dependence of the neutrino spectra upon the progenitor, and several studies \cite{2006PhRvL..97i1101S,2007ApJ...667..382S,2008PhRvD..78h3014N} observed a strong dependence of the number of detected neutrino events upon the EOS they used in the computations. When the neutrinospheres pass below the event horizon the neutrino emission is switched off promptly within $\sim 0.5\;{\rm ms}$ \cite{2001PhRvD..63g3011B}. Interestingly general relativistic effects prevent the spike in luminosity from being as large as it could be: when the neutrinosphere is below the photon sphere---the radius where massless particles can orbit a gravitational source---only neutrinos emitted within a narrow cone of the radial direction are able to escape to infinity \cite{2003gieg.book.....H}. The trajectories of neutrinos emitted outside this cone turn around and return to the neutrinosphere \cite{2017PhRvD..96b3009Y}.

\subsection{Simulation code comparison}

As summarized in the above sections, there is considerable variation in the predicted neutrino spectra from core-collapse supernovae. Comparing the simulations in figures (\ref{fig:O&OFig3}) and (\ref{fig:L,meanE}) at the same 300 millesecond post bounce time, we see the $\bar{\nu}_e$ luminosity varies by a factor of $\sim 5$ and $\bar{\nu}_e$ mean energies by some $\sim 40$\%. Some of this variation is due to real differences between the progenitors, some is due to different choices for the physical inputs (for example the equation of state or the nuclear reaction network) and the rest is due to different codes using different schemes to implement the neutrino transport and the numerical algorithms used to solve the equations. Quantifying the variation due to the first two is relatively straightforward: one can change the progenitors and rerun the simulation with the same code. This is what O'Connor \& Ott \cite{2013ApJ...762..126O} did to produce figure (\ref{fig:O&OFig3}), and similarly differences due to the Equation of State are routinely determined by using multiple different forms \cite{2005ApJ...629..922S}. 
The last source of variation is more difficult to quantify. Code comparison is really the only way to validate a simulation code: as Liebendoerfer \textit{et al.} \cite{Liebendoerfer:2003es} point out, comparing the output of a simulation code with observations is meant to determine our understanding of supernovae, and cannot simultaneously also be used to determine whether a code is accurate. In Yamada, Janka \& Suzuki \cite{1999A&A...344..533Y} the authors compared two different neutrino transport methods and found broad overall agreement but with notable important differences in some regions of the simulation volume. Liebendoerfer \textit{et al.} ~\cite{Liebendoerfer:2003es} made a detailed comparison of the results of spherical symmetric simulations for two different progenitors from two different codes with different implementations of Boltzmann neutrino transport. Differences in luminosities and mean energies were at a few tens of percent or less. The results from this study have often been used by other authors to gauge the accuracy of their codes \cite{2010ApJS..189..104M,2015ApJS..219...24O,2015MNRAS.453.3386J}.
More recently the results from two codes using different multidimensional Boltzmann neutrino transport methods were compared by Richers \textit{et al.} \cite{2017ApJ...847..133R} who found good agreement in all spectral, angular, and fluid interaction quantities. 

Comparisons of the time-integrated neutrino emission parameters have been performed by a number of authors in the context of the diffuse supernova neutrino background. For example, Table III of Ref.~\cite{Horiuchi:2008jz} lists the Fermi-Dirac temperatures for almost a dozen simulations. A more recent and comprehensive collection can be found in Tables 2--4 of Ref.~\cite{Mathews:2014qba}. However, it should be cautioned that such brute force compilations come with caveats: for example, many older results are now known to be unphysical due to their lack of important microphysics or have artificial macroscopic effects. In Ref.~\cite{Horiuchi:2017qja}, a comparison of the time-integrated neutrino spectral parameters of 101 progenitor models was explored based on the axisymmetric simulations of Ref.~\cite{2015PASJ...67..107N}. The $\bar{\nu}_e$ total energetics varied from $(3$--$10) \times 10^{52}$ erg, mean energy from 14.2--16.5 MeV, and alpha from 3.4--4.6, with visible correlations with the progenitor compactness. The authors also compared spectral parameters with the axisymmetric simulations of the Garching group \cite{Summa:2015nyk}. They found consistent total energetics and mean energies with minimal differences between simulations. However, $\alpha$ differed systematically by $\Delta \alpha \sim 1$, being attributed to the fact that $\alpha$ quantifies second order spectral shape and is thus more sensitive to detailed microphysics implementations than total energetics or mean energy. 


\section{Flavor Changing Processes}\label{sec:oscillation}


The neutrino spectra at Earth are not what was emitted from the PNS. The flavor composition changes due to neutrino flavor mixing and a review focused upon just this aspect of supernova neutrinos can be found in \cite{2009JPhG...36k3201D}. At the present time calculations of the neutrino flavor evolution are computed by post-processing stellar models or supernova hydro simulations so are not entirely self-consistent (unless no flavor change is found). We describe the flavor transformation processes before indicating which of them are expected to occur during the four epochs. 

\subsection{Formalism}

The flavour evolution of the neutrinos is a quantum mechanical problem whose most general formulation has been described by a set of Quantum Kinetic Equations \cite{2013PhRvD..87k3010V,2014PhRvD..89j5004V}. When the transfer of momentum and energy with the matter is negligible one finds the equations reduce to a Schr\"odinger equation that has been presented in the literature in a number of different formulations. 

The energy of the neutrinos are much larger than their masses so their velocity is very close to the speed of light. This allows us to replace the time variable in the Schr\"odinger equation with the spatial coordinate along the trajectory - $\lambda$. For a neutrino traveling along a pure radial direction the trajectory coordinate could be the radial position $r$. The neutrino state $\ket{\phi(\lambda_0)}$ at some initial location $\lambda_0$ is related to the state $\ket{\phi(\lambda)}$ at location $\lambda$ by an evolution matrix $S(\lambda,\lambda_0)$ which is the solution of the equation
\begin{equation}
\imath \frac{dS}{d\lambda} = H\,S. \label{eq:Schrodinger}
\end{equation}
In this equation $H$ is the Hamiltonian. This equation needs further qualification before it is well defined which we now go through.


\subsubsection{The evolution matrix}

First, both the Hamiltonian and the evolution matrix are specified relative to basis states. 
So let us be specific and denote the evolution of a neutrino in a basis $(X)$ at initial position $\lambda_1$ to a possibly different basis $(Y)$ at position $\lambda_2$ by $S^{(YX)}(\lambda_2,\lambda_1)$. There are three bases which the reader will usually encounter in the literature: the flavour basis, the matter basis and the mass basis. The flavour basis states are the neutrino states that participate in the weak interaction so these are the neutrino states produced in the core of the supernova and detected here on Earth. The definition of the mass states are those states which which diagonalize the free Hamiltonian in the vacuum and the matter basis is defined as being the basis where the eigenvalues of the Hamiltonian appear on the diagonal. Note this does not mean the off-diagonal elements of the Hamiltonian in the matter basis are zero. The reason the matter basis is useful is because when a neutrino evolves adiabatically, the evolution matrix in the matter basis is diagonal. The matter basis will become the mass basis in the vacuum. Quantities in the flavor basis are labeled with one of three indicii $e, \mu$ and $\tau$. When we refer to a generic index we shall use Greek letters $\alpha, \beta$. Quantities in the mass/matter basis are given numeric indicii $1,2$ or $3$ and we shall use Latin letters $i, j$ to refer to a generic index. For a generic index in either basis we shall use $x$ and $y$. 

While $S$ is the quantity we compute, what is often reported are known as either survival or transition probabilities The transition probabilities are the set of probabilities that the neutrino in a given initial state $x$ of $(X)$ at $\lambda_1$ is detected in the state $y$ of $(Y)$ at $\lambda_2$ and are denoted by $P_{yx}(\lambda_2,\lambda_1)$. These probabilities can be arranged into a matrix $P^{(YX)}$ and are related to the elements of $S^{(YX)}$ by $P_{yx} = |S_{yx}|^2$. So, for example, the electron neutrino survival probability $P_{ee}$ is the square amplitude of the $ee$ element of $S^{(ff)}$ and the transition probability for a neutrino in matter state $3$ at $\lambda_1$ to matter state $2$ at $\lambda_2$ is $P_{23} = |S_{23}|^{2}$. One can even talk about the probability that an initial flavor state is later found in a given matter state. These flavor to matter state probabilities are computed from the evolution matrix $S^{(mf)}(\lambda_2,\lambda_1)$ and turn out to be essential to calculating the neutrino flux from supernovae. It is the probabilities which are usually reported in the literature but it is the evolution matrix which is more fundamental. Note that due to unitarity we have many similar relationships between the elements of $P^{(YX)}$ that can all be succinctly summarized as $\sum_x P_{yx} = \sum_y P_{yx} = 1$. These relationships allow us to construct `missing' probabilities from just a handful of given results. In fact in any given $3\times 3$ transition matrix $P^{(YX)}$ there are just four independent elements of the matrix, the other five are contingent. 


\subsubsection{The Hamiltonian}

Using Standard Model physics, in vacuum the Hamiltonian comprises a single term $H_V$ and the phenomenon of neutrino oscillations arises because the neutrino states that participate in weak reaction Feynman diagrams - so called `flavour' states - are not eigenstates of this Hamiltonian i.e. $H_V$ is not diagonal in the `flavour basis'. The neutrino states which are eigenstates of the vacuum Hamiltonian have definite mass so are called `mass states'. If the neutrino energy $E$ is much larger than the rest masses and we consider three neutrino flavours, the vacuum Hamiltonian in the flavour basis is 
\begin{equation}
 H^{(f)}_V = E + \frac{1}{2E}\,U_V\left( \begin{array}{*{20}{c}} m_1^2 & 0 & 0  \\ 0 & m_2^2 & 0  \\ 0 & 0 & m_3^2 \end{array} \right) U_V^{\dagger}
\end{equation}
where $m_i$ are the neutrino masses. The leading term proportional to the neutrino energy $E$ is often omitted because its effect is to multiply the neutrino state by an overall phase which is not observable. In this equation $U_V$ is a unitary matrix often known by the alternative moniker of the `mixing matrix'. $U_V$ can be parameterized in multiple ways but the most common parameterization is that of the Particle Data Group \cite{Olive:2016xmw} which uses three mixing angles $\theta_{12}$, $\theta_{13}$ and $\theta_{23}$, a CP-violating phase $\delta$, and two Majoranna phases $\alpha_21$ and $\alpha_32$. The structure of $U_V$ is
\begin{equation}
\fl
U_V = \left( \begin{array}{ccc} c_{12}\,c_{13} & s_{12}\,c_{13} & s_{13}\,\rme^{-\rmi\delta} \\   -s_{12}\,c_{23} -c_{12}\,s_{13}\,s_{23}\,\rme^{\rmi\delta} & c_{12}\,c_{23}-s_{12}\,s_{13}\,s_{23}\,\rme^{\rmi\delta} & c_{13}\,s_{23} \\ s_{12}\,s_{23} -s_{12}\,s_{13}\,c_{23}\,\rme^{\rmi\delta} & c_{12}\,s_{23}-s_{12}\,s_{13}\,c_{23}\,\rme^{\rmi\delta} & c_{13}\,c_{23} \end{array}  \right)
\,\left( \begin{array}{ccc} 1 & 0 & 0 \\ 0 & \rme^{\rmi\alpha_{21}/2} & 0 \\ 0 & 0 & \rme^{\rmi\alpha_{31}/2} \end{array} \right)  \label{eq:UV}
\end{equation}
where $c_{ij} = \cos\theta_{ij}$ and $s_{ij} = \sin\theta_{ij}$. In principle three more phases are necessary to make $U_V$ completely general but these phases can be absorbed by redefining the charged fermion states or are not observable. The mixing matrix for the antineutrinos, $\bar{U}_V$, is the complex conjugate of ${U}_V$ so that the vacuum Hamiltonian for the antineutrinos, $\bar{H}^{(f)}_V$, is the complex conjugate of $H^{(f)}_V$.

In matter one needs to add to $H$ an extra term $H_M$ called the `matter potential'. In contrast to $H_V$, the matter potential is diagonal in the flavour basis. For three neutrino flavuors and at temperatures where the density of charged muons and taus is negligible, the matter potential has the form 
\begin{equation}
 H^{(f)}_M(\lambda) = \left( \begin{array}{*{20}{c}} V_{CC} + V_{NC} & 0 & 0  \\ 0 & V_{NC} & 0  \\ 0 & 0 & V_{NC} \end{array} \right)
\end{equation}
Here $V_{CC}$ is the charged current potential $V_{CC}(\lambda) = \sqrt{2} G_F n_e(\lambda)$ with $G_F$ Fermi's constant and $n_e(\lambda)$ the net electron density at the point $\lambda$ along the trajectory, while $V_{NC}(\lambda) = - G_{F} n_n(\lambda) / \sqrt{2}$ with $n_n(\lambda)$ the neutron density at $\lambda$. Again, one often sees the matter potential written without $V_{NC}$ because for three flavours, it's effect is to multiply the neutrino state by a global, and unobservable, phase. For the antineutrinos the matter potential, $\bar{H}_M$ is $\bar{H}_M = - H_M$. Small corrections to $H^{(f)}_M$ appear when one considers `radiative effects' \cite{2008PhRvD..77f5024E}. These lead to a small difference in the coupling of $\mu$ and $\tau$ neutrino flavors to ordinary matter (i.e.\ just electrons, protons and neutrons) that can be represented by adding an additional potential $V_{\mu\tau}$ in the $\tau\tau$ element of $H^{(f)}_M$. This potential is approximately a factor of $10^{-5}$ compared to the charged current and neutral current potentials $V_{CC}$ and $V_{NC}$.

The density of neutrinos is a core collapse supernova is so large that one needs to add another term to $H$ known as the `self-interaction' Hamiltonian $H_{SI}$. The form of the self-interaction at a given spacetime point $t,{\bf r}$ (which we omit for the sake of clarity) can be found in Duan \textit{et al.} \cite{Duan:2006an} and is given by
\begin{equation}
\label{eqn:HSI}
H^{(f)}_{SI}\left( {\bf{q}} \right) = \sqrt 2 {G_F} \int \left( 1 - {\bf{\hat q}} \cdot {\bf{\hat q}}' \right)\,\left[ \rho\left({\bf q}' \right) - \rho^*\left( {\bf q}' \right) \right]\,\frac{ d^{3}{\bf q}' }{(2\pi)^{3}}
\end{equation}
where $\rho({\bf q})$ and $\bar{\rho}({\bf q})$ are the differential density matrix of the neutrinos and antineutrinos at spacetime position $t,{\bf r}$ with energy $\left|{\bf q}\right|$ propagating in the directions between ${\bf{\hat q}}$ and ${\bf {\hat q}}+d{\bf{\hat q}}$, per unit energy (the hats on ${\bf q}$ and ${\bf q}'$ indicate unit vectors). If we ignore scattering, the density matrices evolve according to $\rho(\lambda,{\bf q}) = S(\lambda,{\bf q};\lambda_0,{\bf q_0})\,\rho(\lambda_0,{\bf q_0})\,S^{\dagger}(\lambda,{\bf q};\lambda_0,{\bf q_0})$. For antineutrinos the form of the self-interaction Hamiltonian, $\bar{H}_{SI}$ is $\bar{H}_{SI} = - H^{\ast}_{SI}$. Note the presence of the $1 - {\bf{\hat q}} \cdot {\bf{\hat q}}'$ term in the integral which is equivalent to $1-\cos\theta$ where $\theta$ is the angle between the neutrino trajectories. At large radii, when the all neutrino trajectories approach colinearity, this term strongly suppresses the self-interaction. 

If one considers physics Beyond the Standard Model (BSM) then either these three Hamiltonians are modified or additional terms are added. For example: if one considers additional, `sterile' flavours of neutrino then the size of the Hamiltonian has to grow to accommodate the new flavors which introduces new mixing angles and masses. In addition, the sterile states do not feel any effect from the matter, including the neutral current potential, so this potential can no longer be dropped from $H_M$ \cite{1997PhRvD..56.1704N,1999PhRvC..59.2873M,2001NuPhB.599....3P,2006PhRvD..73i3007B,2012JCAP...01..013T,2014PhRvD..90j3007W,2014PhRvD..89f1303W,2014PhRvD..90c3013E}. Any physics which can lead to neutrino - antineutrino conversion - such as neutrino magnetic moments - also mean one has to enlarge the Hamiltonian because now neutrinos and antineutrinos must be considered simultaneously, not separately.  \cite{1988PhRvL..61...27B,1999PhLB..470..157M,1999APh....11..317N,2007JCAP...09..016B,2012JCAP...10..027D,2016arXiv161008586T}. Finally any new neutrino interactions with matter \cite{1987PhLB..199..432V,1996PhRvD..54.4356N,1996NuPhB.482..481N,1998PhRvD..58a3012M,2002PhRvD..66a3009F,PhysRevD.76.053001,2010PhRvD..81f3003E,2016PhRvD..94i3007S}, or among themselves 
\cite{2008PhRvD..78k3004B,2017JCAP...05..051D}, will also modify the Hamiltonian. For a recent review of non-standard neutrino interactions the reader is referred to Ohlsson \cite{2013RPPh...76d4201O}. Typically the new interactions are expressed relative to the charged current potential $V_{CC}$ and a matrix $\epsilon$ that may be a function of the composition of the matter. Alternatively, it has been shown that Super Symmetry can also modify the matter Hamiltonian by adding new radiative corrections \cite{2010PhRvD..81a3003G,2010JCAP...05..029G}. 


\subsubsection{Resonances and the Matter Basis}

We have defined two different bases - the flavor and matter basis - and of course the two bases are related. If we are given a Hamiltonian in the flavor basis, $H^{(f)}$, then we can compute it's eigenvalues - $k_1$, $k_2$ and $k_3$ - and put those eigenvalues along the diagonal of a matrix $K$. The eigenvalues are placed so that their order reflects the same ordering as the masses in the vacuum Hamiltonian $H_V$. The matrix $H^{(f)}$ is related to $K$ by the `matter mixing matrix' $U$ by the equation
\begin{equation}
H^{(f)} = U K U^{\dagger}
\end{equation}
Note that if $H^{(f)}$ is a function of position then both $K$ and $U$ are also functions of position. The same procedure can be repeated for the antineutrinos by starting with $\bar{H}^{(f)}$, computing $\bar{K}$ and introducing an antineutrino matter mixing $\bar{U}$. In the vacuum $U$ becomes the vacuum mixing matrix $U_V$ and $\bar{U}$ becomes $U_V^{\star}$. The matter and flavor basis are related via $U$: if $\psi^{(f)}$ is a neutrino wavefunction in the flavor basis the the same wavefuncton in the matter basis, $\psi^{(m)}$, is related to $\psi^{(f)}$ by $\psi^{(f)} = U \psi^{(m)}$. For antineutrino we use $\bar{U}$ to relate the two bases. The transformation of $S$ from one basis to another differs from the transformation of a wavefunction because the evolution matrix is a function of two positions. The flavor basis evolution matrix $S^{(ff)}(\lambda_2,\lambda_1)$ is related to the matter basis evolution matrix $S^{(mm)}(\lambda_2,\lambda_1)$ by 
\begin{equation}
S^{(ff)}(\lambda_2,\lambda_1) = U(\lambda_{2})\, S^{(mm)}(\lambda_2,\lambda_1)\, U^{\dagger}(\lambda_1).
\end{equation}
Similarly we have  
\begin{equation}
S^{(mf)}(\lambda_2,\lambda_1) = S^{(mm)}(\lambda_2,\lambda_1)\, U^{\dagger}(\lambda_1) = U^{\dagger}(\lambda_{2})\, S^{(ff)}(\lambda_2,\lambda_1)
\end{equation}
The structure of $U(\lambda_{2})$ may be very different from the structure of $U(\lambda_{1})$ and this difference is the origin of the MSW effect. Exactly how different depends upon the mass ordering and the structure of the total Hamiltonian. If we restrict ourselves to Standard Model physics then for a normal mass ordering, at high densities such as at the neutrinosphere $R_{\nu}$,
\begin{equation}
\fl
\hspace{1cm}
U(R_{\nu}) \approx \left( \begin{array}{ccc} 0 & 0 & 1 \\ 1/\sqrt{2} & 1/\sqrt{2} & 0 \\ -1/\sqrt{2} & 1/\sqrt{2}  & 0 \end{array} \right),
\hspace{1cm}
\bar{U}(R_{\nu}) \approx \left( \begin{array}{ccc} 1 & 0 & 0 \\ 0 & 1/\sqrt{2} & 1/\sqrt{2} \\ 0 & -1/\sqrt{2} & 1/\sqrt{2}  \end{array} \right) \label{eq:UNMO}
\end{equation}
while for inverted mass ordering at the same high densities,
\begin{equation}
\fl
\hspace{1cm}
U(R_{\nu}) \approx \left( \begin{array}{ccc} 0 & 1 & 0 \\ 1/\sqrt{2} & 0 & 1/\sqrt{2} \\ -1/\sqrt{2} & 0 & 1/\sqrt{2} \end{array} \right), 
\hspace{1cm}
\bar{U}(R_{\nu}) \approx \left( \begin{array}{ccc} 0 & 0 & 1 \\ 1/\sqrt{2} & 1/\sqrt{2} & 0 \\ -1/\sqrt{2} & 1/\sqrt{2}  & 0 \end{array} \right) \label{eq:UIMO}
\end{equation}
Beyond the neutrinosphere the matter mixing matrices will remain very close to those at $R_{\nu}$ until they reach the neutrino resonances whereupon they transform to a different structure over a relatively narrow spatial window - the `resonance width'. Resonances are those locations where two diagonal elements of the total flavour basis Hamiltonian become equal. Denoting two different generic flavour states by $\alpha$ and $\beta$, a neutrino resonance occurs whenever one $H^{(f)}_{\alpha\alpha} = H^{(f)}_{\beta\beta}$ or $\bar{H}^{(f)}_{\alpha\alpha} = \bar{H}^{(f)}_{\beta\beta}$. If we restrict ourselves to Standard Model physics, the equality occurs because the charged current potential and the self-interaction potential contributions to a given pair of diagonal element offsets the difference between the vacuum difference of the same two elements. In practice, with Standard Model physics only, the self-interaction potential is usually not important for determining the location of the resonances for supernova neutrinos so we usually just have to deal with the charge current potential which only contributes to the $ee$ element. Thus resonances occur when any of the equalities $H_{V,ee}+V_{CC} = H_{V,\mu\mu}$, $H_{V,ee}+V_{CC} = H_{V,\tau\tau}$ for the neutrinos, or and $\bar{H}_{V,ee}-V_{CC} = \bar{H}_{V,\mu\mu}$, $\bar{H}_{V,ee}-V_{CC} = \bar{H}_{V,\tau\tau}$ for the antineutrinos, are satisfied. Although we list four possible equalities, only two can be satisfied once the mass ordering is given. To denote the two resonances, the one at higher density is appropriately known as the high (H) density resonance while the one at lower density is the low (L) density resonances \cite{2000PhRvD..62c3007D}. If the mass ordering is normal both the H and L resonances occur for the neutrino only: the H resonance leads to mixing between matter neutrino states $\nu_2$ and $\nu_3$ and the L resonance mixes $\nu_1$ and $\nu_2$. If the mass ordering is inverted the H resonance switches to the antineutrinos and the mixing is between states $\bar{\nu}_1$ and $\bar{\nu}_3$, the L resonance remains in the neutrinos and continues to mix $\nu_1$ and $\nu_2$. We also note that the resonance locations depend upon the neutrino energy because the vacuum Hamiltonian depends upon the neutrino energy as $1/E$. This means the charged current potential $V_{CC}$ needed to satisfy a given resonance condition is smaller for higher energy neutrinos. If we go Beyond the Standard Model new resonances arise \cite{1999PhRvC..59.2873M,PhysRevD.76.053001,2010PhRvD..81f3003E,2016PhRvD..94i3007S} but in an effort to be concise, we refer the reader to the literature.

The degree of mixing at a resonance is determined by the diabaticity $\Gamma_{ij}$ (the inverse of the adiabaticity $\gamma_{ij}$ i.e.\ $\Gamma_{ij} = 1/\gamma_{ij}$. General expressions for the diabaticity are known \cite{2012JPhG...39c5201G} but are not very transparent. 
The diabaticity is largest at the resonances and using a two flavour approximation the adiabaticity at the resonance is found to be 
\begin{equation}
\Gamma_{ij} = \left( \frac{\pi}{2 |H_{e\alpha}|^2} \frac{dV_{CC}}{d\lambda} \right)_{res}
\end{equation}
where flavour state $\alpha$ and the states $i$ and $j$ are determined by the resonance and mass ordering : e.g.\ $\alpha = \tau$, $i=2$, $j=3$ for the H resonance and a NMO. One can similarly define a diabaticity for the antineutrino states. When the diabaticity is large $|\Gamma_{ij}| \gg 1$, the neutrino hops from state $i$ to state $j$ (and vice versa); when $|\Gamma_{ij}| \ll 1$ a neutrino in matter state $i$ remains in matter state $i$ as it traverses the resonance.


\subsection{Flavor Changing Processes in Supernova}

With the overview of different flavor changing processes complete, we now describe what has been found when they have been applied to the case of supernova neutrinos using just Standard Model physics.  

\subsubsection{Matter Effects}

During all four epochs listed in table (\ref{tab:epochs}) the potential $V_{CC}$ dominates over the vacuum Hamiltonian at the point of neutrino emission for typical neutrino energies. This means there will be MSW effects due to the different structure of the matter mixing matrices at the point of emission and where the neutrino or antineutrino exits the star. The physics of the MSW effect in supernovae is very similar to that of matter effects in the Sun but with three complications: first the electron density in the core of the supernova means that the charged-current potential is much larger than the differences between all the diagonal elements of the vacuum Hamiltonian which are $\sim ( m_{i}^{2} - m_j^2 ) / E$; second, when the star begins to explode the electron density profile of the supernova evolves with time; and third the density profile is not spherically symmetric. 

The first complication means we have to use a full three-flavour treatment of flavour transformation rather than two (though it is usually possible to factorize the flavor transformation into a sequence of two-flavor mixings \cite{2000PhRvD..62c3007D}). 
This leads to the second complication - the time dependence of the density profile. 
\begin{figure*}[t!] 
\includegraphics[width=0.9\linewidth]{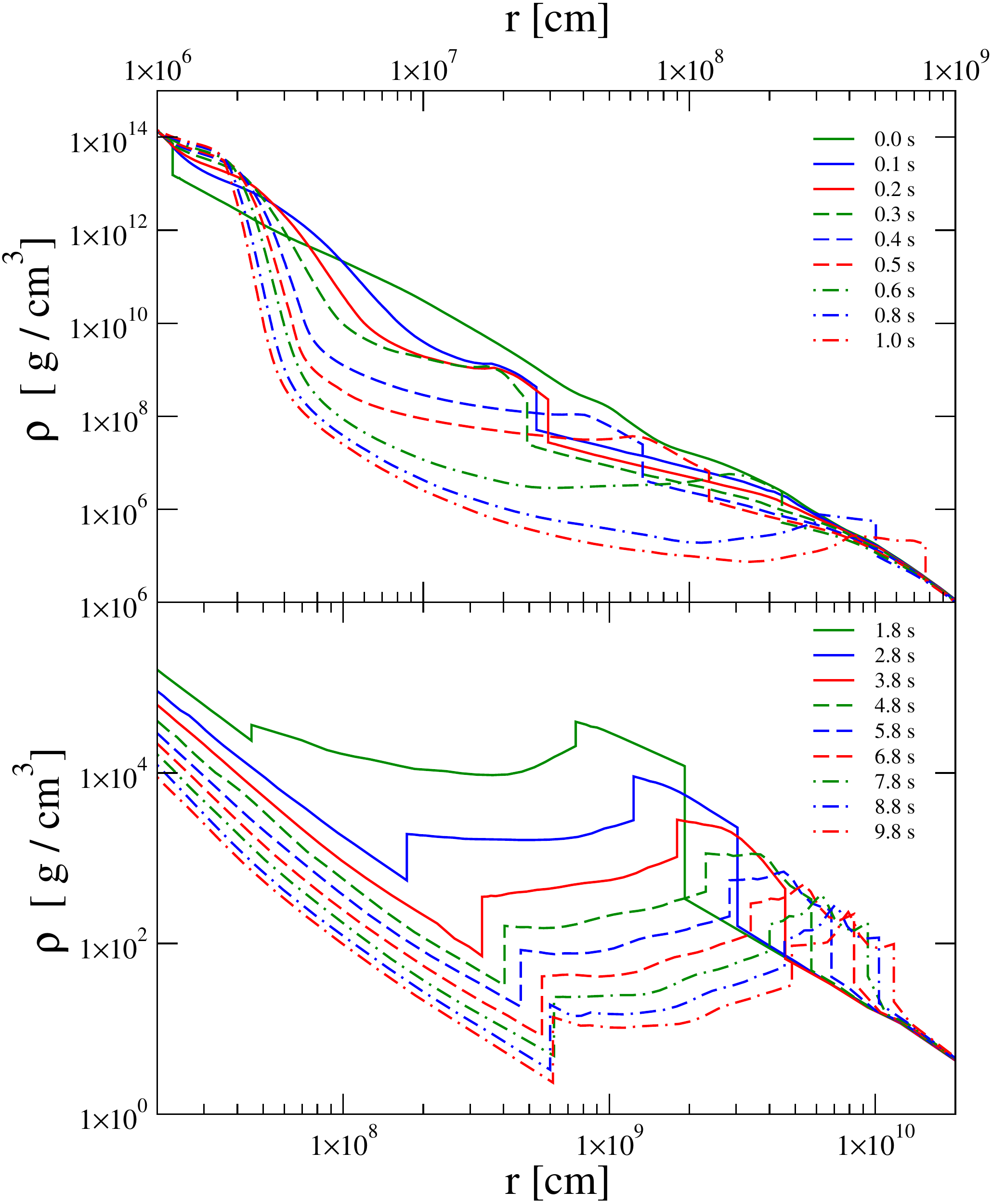}
\caption{The density as a function of the radial distance $r$ at various postbounce times. The data is for the spherically symmetric explosion of a $10.8\;{\rm M_{\odot}}$ progenitor. The data is taken from the simulation by Fischer \textit{et al.} \cite{2010A&A...517A..80F} \label{fig:rho} }
\end{figure*}
The density $\rho$ as a function of the radius $r$ - the density profile - at various postbounce epochs of the explosion of the $10.8\;{\rm M_{\odot}}$ progenitor by by Fischer \textit{et al.} \cite{2010A&A...517A..80F} are shown in figure (\ref{fig:rho}). The reader will observe several features in these curves. The one that is always present is the forward shock - the increase in density (if we are looking towards the core). As the reader can see, the shock is at $r\sim 12\;{\rm km}$ at $t_{pb} = 0.0\;{\rm s}$ and has propagated to $r\sim 200\;{\rm km}$ at $t_{pb} = 0.1\;{\rm s}$ whereupon it stalls at that radius until $t_{pb} = 0.3\;{\rm s}$. At $t_{pb} = 0.3\;{\rm s}$ the shock begins to move outwards again. Notice how the fractional density jump across the forward shock remains more-or-less the same for all the times shown. The region below the stalled shock is where the neutrino driven convection, SASI etc.\ occur during the accretion phase. The matter fluctuation due to these effects do not appear in these density profiles because the simulations by Fischer \textit{et al.} are one dimensional. Once the shock is revived and accretion is choked off, a neutrino driven wind forms above the proto neutron star. In multi-dimensional simulations it is seen the density profile of the wind is close to spherically symmetric \cite{1996A&A...306..167J} - see, for example, figure (6) from Arcones \& Janka \cite{2011A&A...526A.160A} or figures (10) and (11) in \cite{Kneller:2008PhRvD..77d5023K}. The forward shock is not the only discontinuity in the density profiles. In the bottom panel one sees how the density profile develops two more discontinuities during the cooling phase: a reverse shock due to the neutrino driven wind colliding with slower moving ejecta above it \cite{1996A&A...306..167J}, and a contact discontinuity. The reverse shock is the first jump up in density (looking from the core outwards) and the contact discontinuity is the second jump. Notice how the density contrast across these two discontinuities changes with time. Also, at very late time, $t_{pb} \sim 7\;{\rm s}$, the reverse shock ceases to move outwards and even the contact discontinuity appears to be stalling at $t\sim 10\;{\rm s}$. The forward shock, the contact discontinuity and the reverse shock and their general behavior are common to other simulations \cite{Kneller:2008PhRvD..77d5023K,2011A&A...526A.160A} however a counter example where the reverse shock continues to advance can be found in Arcones, Janka \& Scheck \cite{2007AA...467.1227A}. What changes between simulations is the postbounce time for the forward shock to reach various densities due to differences in the size of the envelopes of the star, and the epoch when the reverse shock stalls due to differences in the amount of neutrino heating in the driven wind \cite{Kneller:2008PhRvD..77d5023K}. A word of caution: one has to be careful about the use of density profiles from simulations to study matter effects upon neutrinos. Due to the numerical integration schemes employed in the codes, the shocks in the simulations are not discontinuities, they are simply sections of larger density gradient than the surroundings spread over several zones. This becomes a problem typically at large radii because the gradients in the `shock' zones are not sufficiently steep to cause the neutrino to jump between the matter eigenstates \cite{2013PhRvD..88b3008L}. Before one uses the profile the shock - and any other discontinuity - must be steepened. The density profiles shown in figure (\ref{fig:rho}) were steepened from the original simulation data. In multi-dimensional simulations the forward shock, the reverse shock and contact discontinuity are seen to be aspherical  \cite{1996A&A...306..167J,Kneller:2008PhRvD..77d5023K,2011A&A...526A.160A}. As the neutrino driven wind passes through the reverse shock from below, and the forward shock propagates through the progenitor profile, fluid elements acquire non-radial velocities leading to turbulence in the region between the reverse and forward shock. As with turbulence below the stalled shock during the accretion phase, to correctly capture the turbulence in the envelope during the cooling phase requires a high resolution, preferably three-dimensional, simulation. Unfortunately these do not yet exist. 

The dynamics of the density profile will leave an imprint in the flavor transformation. During the accretion phase no effect upon the transition probabilities is seen because the shock is close to the proto neutron star at densities far above the MSW resonances. What one expects is that at early times - the presupernova epoch, the accretion phase and the beginning of the cooling phase - the matter effects are essentially static and the neutrino evolution through the envelope of the star or the supernova is adiabatic. When the mixing angle $\theta_{13}$ was unknown there was a possibility that the neutrino for a NMO (antineutrino for the IMO) might 'hop' from one matter eigenstate to another at the H resonance but now this angle is known, it is clear no hopping occurs. It is not until the shock has been revived and propagated out to the relevant densities of the stellar envelope - which occurs during the cooling phase - that one sees any dynamism in the matter effects. The first to appreciate the shock could reach the relevant densities while the proto-neutron star was emitting neutrinos were Schirato \& Fuller \cite{2002astro.ph..5390S}. Follow-up studies by numerous authors improved upon the calculations by Schirato \& Fuller by using more realistic density profiles and by solving the Schr\"odinger equation numerically rather than making use of analytic formulae \cite{2004JCAP...09..015T,Kneller:2008PhRvD..77d5023K,2013PhRvD..88h3012N}. While these improvements did lead to quantitative changes, qualitatively the phenomenology found by Schirato \& Fuller still holds. The shock is created at high densities / small radii and moves outwards to larger radii, lower densities. As the shock approaches the H resonance densities the transition probabilities of the lowest energy neutrinos (NMO) or antineutrinos (IMO) begin to change first followed by the higher energies. The energy dependence arises from the $1/E$ term in the MSW resonance condition i.e. the density at which the resonance condition is satisfied for, say, a $5\;{\rm MeV}$ neutrino is higher than the density at which the condition is satisfied for, say, a $50\;{\rm MeV}$ neutrino. Since the density profile of the supernova is (to first order) a monotonically decreasing function of the radius $r$, the H resonance for a $50\;{\rm MeV}$ neutrino will be at larger radius than the H resonance for a $5\;{\rm MeV}$ neutrino. Exactly when the transition probabilities begin to change depends very much on the progenitor. For the ONeMg, $8.8\;{\rm M_{\odot}}$ simulation by the Basel group the shock reaches the H resonance within 0.1 second \cite{2013PhRvD..88b3008L} - see also \cite{PhysRevD.78.023016}. For the $10.8\;{\rm M_{\odot}}$ model - shown in figure(\ref{fig:rho}) - it takes 1 second, and for the $18\;{\rm M_{\odot}}$ model it takes 3 seconds \cite{2013PhRvD..88b3008L}. 

\begin{figure*}[t!] 
\includegraphics[width=0.5\linewidth]{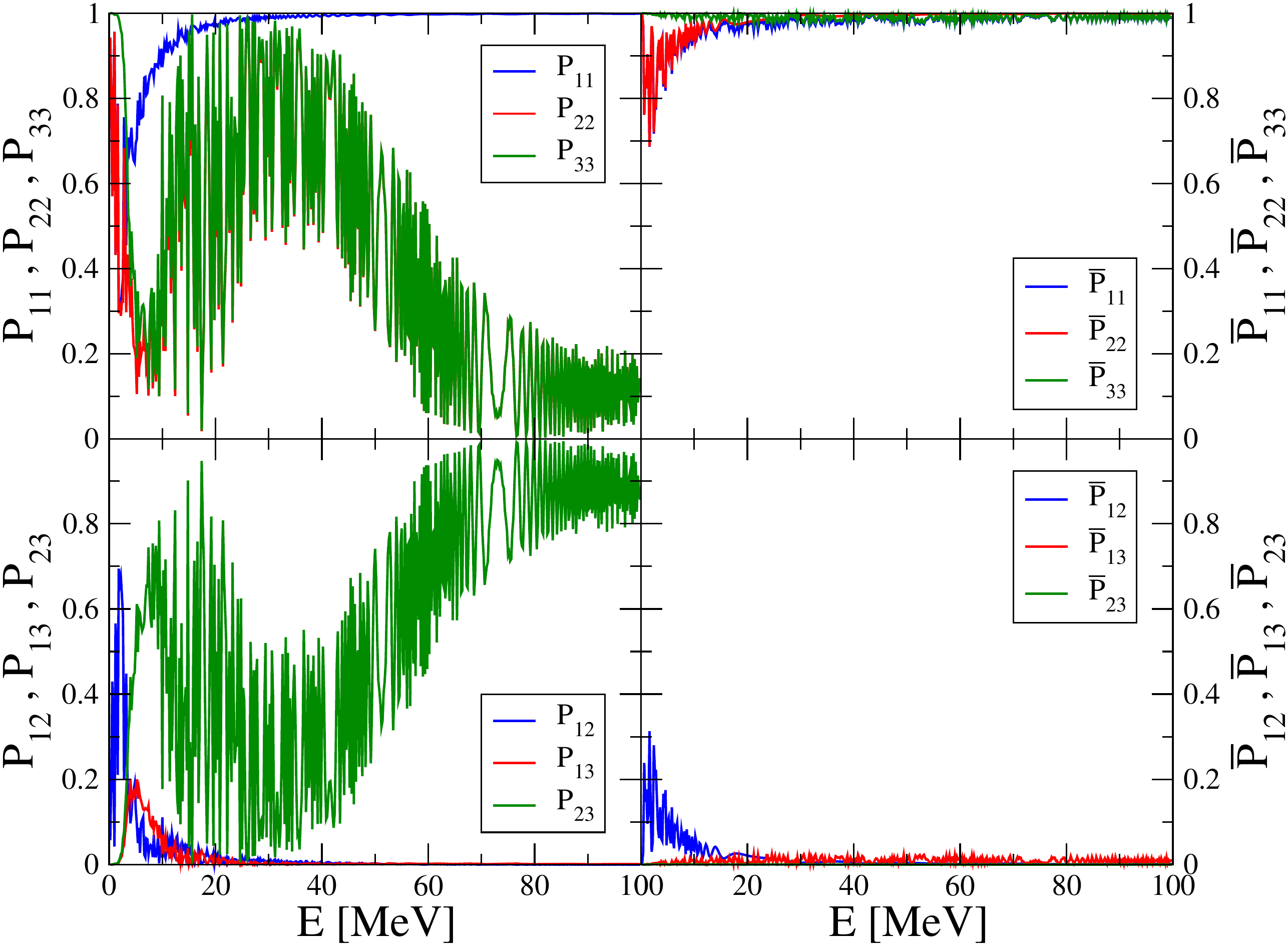}
\includegraphics[width=0.5\linewidth]{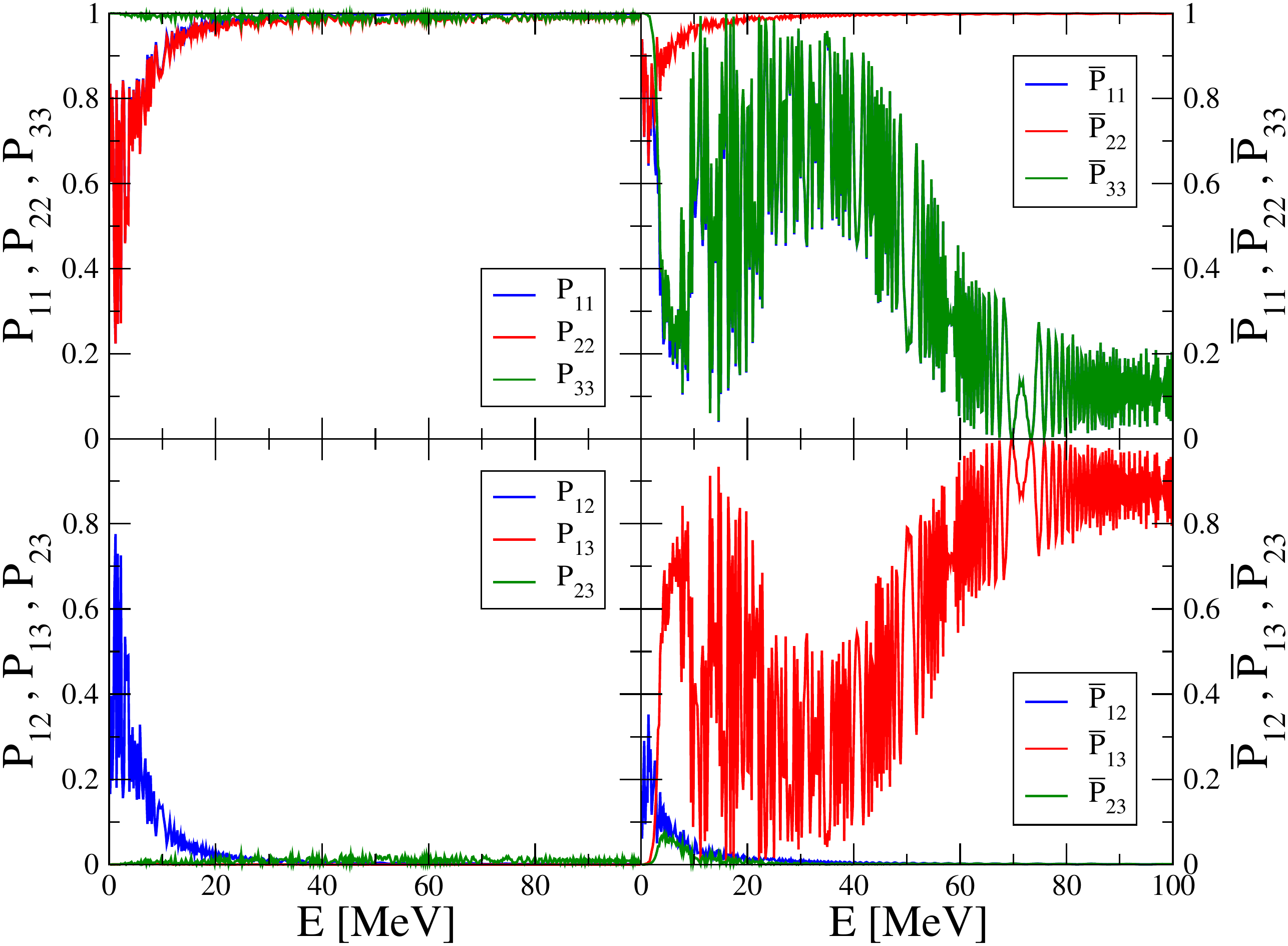}
\caption{The matter basis survival and transition probabilities at the edge of the supernova for neutrinos (left panels) and antineutrinos (right panels) in the NMO (left quartet) and IMO (right quartet). The density profile used was the $t_{pb}=2.8\;{\rm s}$ snapshot shown in figuure (\ref{fig:rho}) and the calculation starts at $r=1000\;{\rm km}$. The figure is adapted from figures (8) and (9) in Lund \& Kneller \cite{2013PhRvD..88b3008L}. \label{fig:matter effects}  }
\end{figure*}
What the studies that came after Schirato \& Fuller revealed is that there could be additional effects due to the presence of the other discontinuities in the profile. The presence of these additional discontinuities leads to the phenomenon of Phase Effects  \cite{2006PhRvD..73e6003K,2007PhRvD..75i3002D}. As the name implies, Phase Effects are due to a phase which is 
\begin{equation}
\Phi_{ij} = \int_{\lambda_1}^{\lambda_2} d\lambda \left( k_i - k_j \right) \label{eq:phaseeffects}
\end{equation}
where $\lambda_a$ are the location of discontinuities and $k_i$ are the instantaneous eigenvalues of the Hamiltonian $H$. The Phase Effects appear in the transition probabilities as a term proportional to $\cos\Phi_{ij}$ and lead to oscillations of the transition probabilities as a function of energy because the eigenvalues are energy dependent. The effect is similar to the change in the intensity of light reflected from a thin film as a function of the wavelength. 

An example of a recent calculation of the neutrino transition probabilities in the matter basis at the edge of the supernova as a function of neutrino energy is shown in figure (\ref{fig:matter effects}). The calculation uses the $t_{pb}=2.8\;{\rm s}$ snapshot of the $10.8\;{\rm M_{\odot}}$ simulation from Fischer \textit{et al.} \cite{2010A&A...517A..80F} shown in figure (\ref{fig:rho}). To isolate the effects of matter only the calculation was started at $r=1000\;{\rm km}$. At this snapshot time the reverse shock, contact discontinuity and forward shock are located at the H resonance density for neutrino energies above $E\gtrsim 5\;{\rm MeV}$ and the forward shock is located at the L resonance density for neutrino energies below $E\lesssim 3\;{\rm MeV}$. However the L resonance is quite broad - because $\theta_{12}$ is large - so neutrinos up to $E\sim 20\;{\rm MeV}$ are beginning to feel the effects of the forward shock in the $\nu_1 - \nu_2$ mixing channel. The phase effects can be seen clearly in many mixing channels, e.g.\ $\bar{P}_{23}$ for the IMO. The phase effects look random in this figure but that is due to the energy resolution of the calculation which was $200\;{\rm keV}$. A very high energy resolution calculation - see the bottom panel of figure (4) from Kneller \& Volpe \cite{2010PhRvD..82l3004K} for example - shows the oscillations are smooth and have a `period' of $\sim 50\;{\rm keV}$. 

\subsubsection{Turbulence Effects}

Finally, we must address the third complication - the asphericity of the explosion. We have already mentioned that simulations indicate the forward shock propagates at different speeds for different radial directions. The location and motion of the contact discontinuity and reverse shock are even more variable \cite{Kneller:2008PhRvD..77d5023K,2011AA...526A.160A} such that the reverse shock stalls much earlier in some directions than along others. 
But perhaps an even more significant effect from the asphericity of the explosion is the turbulence generated in the fluid. Turbulence is known to affect neutrino evolution  \cite{2013PhRvD..88b3008L,1990PhRvD..42.3908S,Loreti:1995ae,1996PhRvD..54.3941B,Friedland:2006ta,Fogli:2006JCAP...06..012F,Choubey:2007PhRvD..76g3013C,2010PhRvD..82l3004K,2011PhRvD..84h5023R,2013PhRvD..88d5020K,2013PhRvD..88b5004K,2015PhRvD..92a3009K}. In order to study the turbulence effects upon the neutrinos the ideal would be to take density profiles from multi-dimensional simulations of supernovae, construct the charged-current potential $V_{CC}$ from them, insert the potential into the Schr\"odinger equation, and solve it. However this ideal approach cannot yet be adopted without further modeling: the only study which has come close to the ideal is that by Borriello \textit{et al.} \cite{2014JCAP...11..030B}. The problem is that the spatial resolution of the simulations is not yet sufficient to capture the turbulence on scales where it might affect neutrinos. Ideally the simulation should be three-dimensional and not two is because turbulence in two dimensions is very different from that in three. 

In order to circumvent these short comings the usual approach is to adopt the density profile from a one-dimensional (turbulence free) simulation and insert turbulence into it via a model. Typically the model is to multiply the charged-current potential derived from the 1D simulation by $1+F(r)$ where $F$ is a random field. The statistical properties of the field are that it has a root-mean square amplitude $C_{\star}$ and a power spectrum $E(q)$ - where $q$ is the wavenumber - which is usually taken to be an inverse power law with exponent $\alpha$ - also known as the `spectral index'. If needed, realizations of $F$ can be generated with a Fourier series i.e.
\begin{equation}
F(r) \propto C_{\star} \sum_a^{N_q} \left\{ G_a \cos \left( q_a r + \eta_a \right) \right\}
\end{equation}
where $N_q$ is some carefully chosen integer for the number of modes, $G_a$ the amplitude of the mode, $q_a$ the wavenumber and $\eta_a$ a phase. Algorithms for assigning the amplitudes, wavenumbers, and phases can be found in the literature. Note the equation for $F(r)$ omits additional terms which are used to force $F(r)$ to vanish at the boundaries of the turbulent region. For each realization one constructs the neutrino Hamiltonian and solves for the evolution matrix. Three examples are shown in figure (\ref{fig:turbulenceeffects}) taken from \cite{2015PhRvD..92a3009K}. Notice how the difference from the turbulence free calculation is of roughly equal order for all three cases even though the turbulence amplitude differs by two orders of magnitude between them.
\begin{figure*}[t!] 
\includegraphics[width=0.9\linewidth]{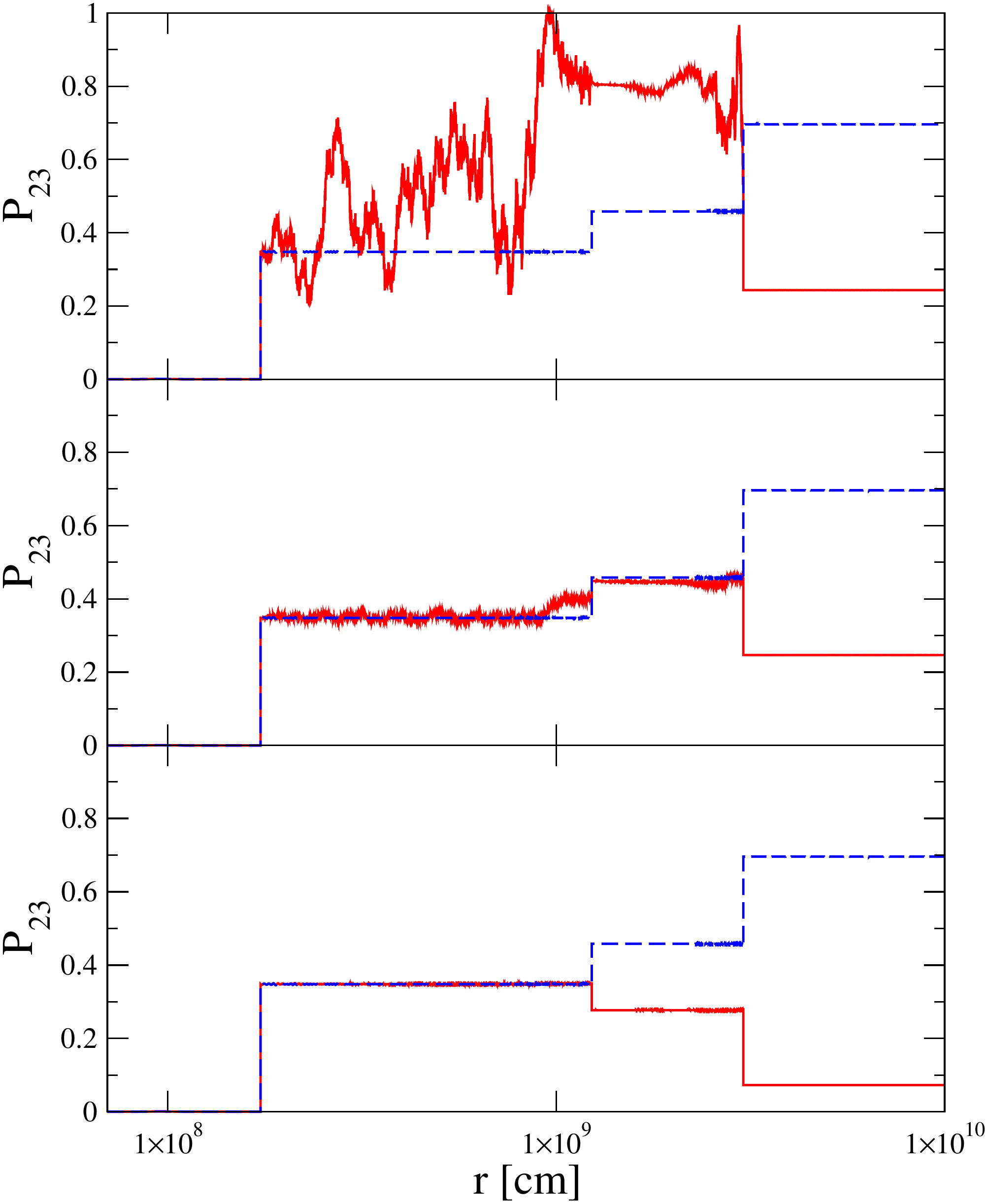}
\caption{The two effects of turbulence. The evolution of the matter basis transition probability $P_{23}$ as a function of distance $r$ for a $15\;{\rm MeV}$ neutrino and a normal mass ordering. The dashed line is the evolution through the underlying profile and the solid line is the evolution with a single realization of turbulence with spectral index $\alpha=5/3$. In the top panel the rms amplitude of the turbulence is set to $C_{\star}=10\%$, in the middle $C_{\star}=1\%$ and in the bottom $C_{\star}=0.1\%$\label{fig:turbulenceeffects} }
\end{figure*}
One may try to understand such results as being due to a more complicated MSW effect but an alternative perspective that works very well is to use time-dependent perturbation theory with the turbulence as an external perturbing potential \cite{Patton2015a}. Via this approach it has been shown the turbulence can effect the neutrinos via two different mechanisms which are shown in figure (\ref{fig:turbulenceeffects}). In the top panel we see what have been called `Stimulated Transitions' while in the bottom we see `Distorted Phase Effects'. The middle panel has some Stimulated Transition but the Distorted Phase is larger. Whether the Stimulated Transition effects and the Distorted Phase effects occur can be established by a set of criteria among a set of six lengthscales \cite{2017arXiv170206951K}. From these criteria is has been found that no turbulence effects should occur during the accretion phase of the supernova. 
\begin{figure*}[t!] 
\includegraphics[width=0.5\linewidth]{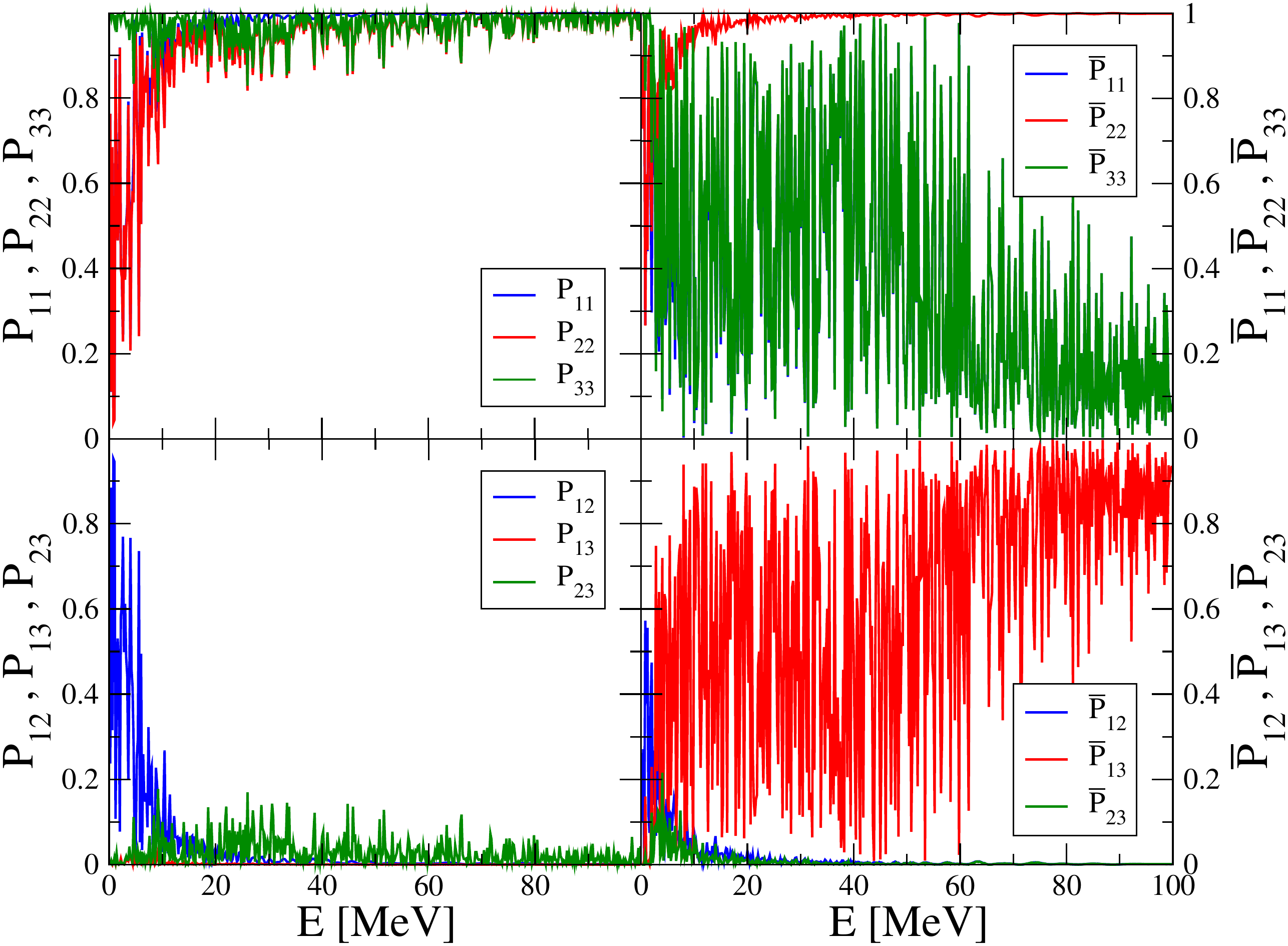}
\includegraphics[width=0.5\linewidth]{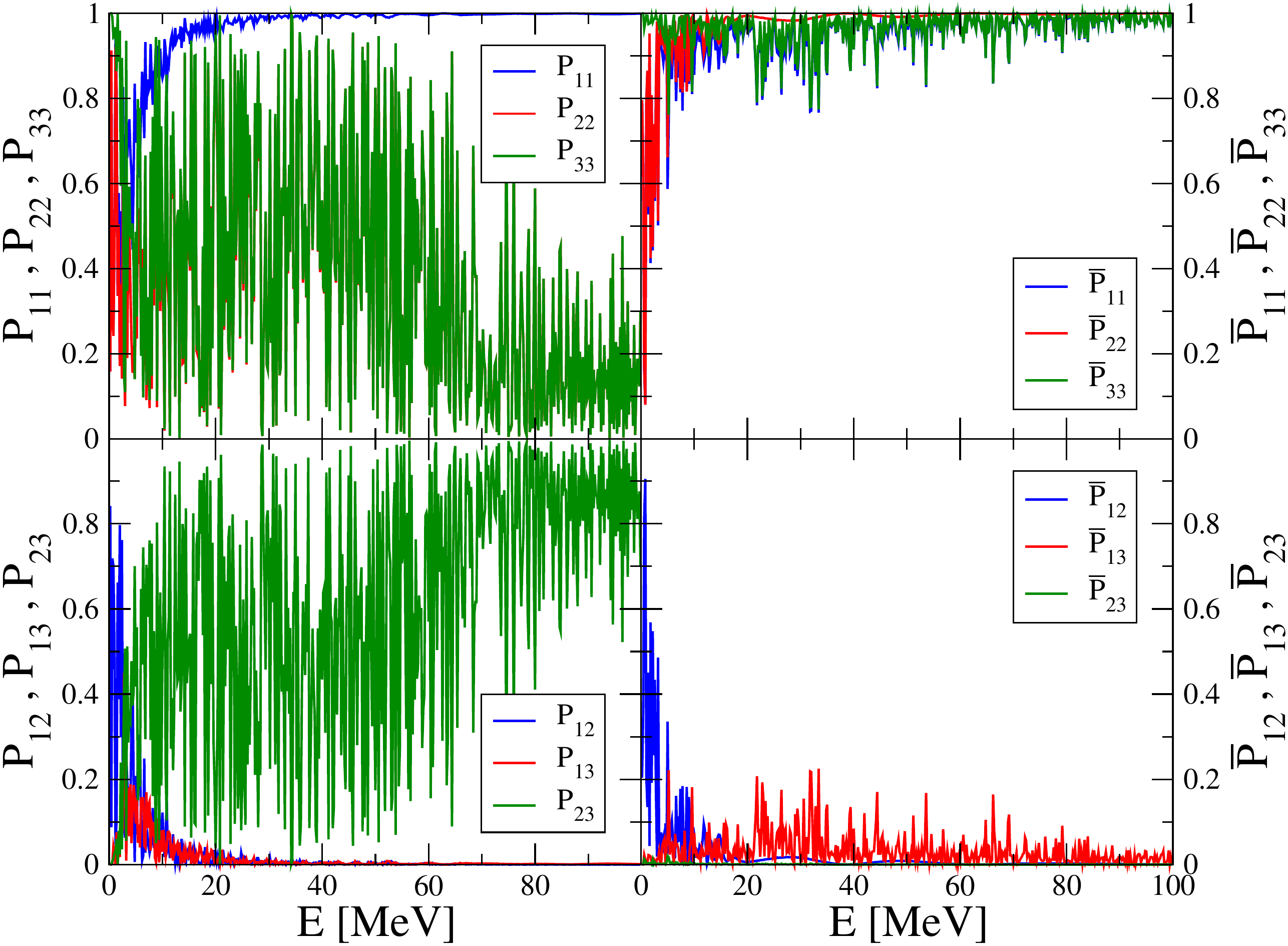}
\caption{The matter basis survival and transition probabilities at the edge of the supernova for neutrinos (left panels) and antineutrinos (right panels) in the NMO (left quartet) and IMO (right quartet). The density profile used was the $t_{pb}=2.8\;{\rm s}$ snapshot shown in figure (\ref{fig:rho}) and turbulence was added at the level of 30\% and the calculation starts at $r=1000\;{\rm km}$. The figure is adapted from figure (14) in Lund \& Kneller \cite{2013PhRvD..88b3008L}. \label{fig:matterplusturbulence}  }
\end{figure*}
During the cooling phase the turbulence moves farther out into the star to lower densities and when the turbulence is in the vicinity of a MSW resonance, turbulence effects appear in the transition probabilities \emph{for both the neutrino and antineutrino} \cite{2015PhRvD..92a3009K}. If the MSW resonance is in the neutrino sector then the turbulence effects are larger in the neutrino transition probabilities than antineutrino: if the MSW resonance occurs in the antineutrinos then the opposite mass ordering of turbulence effects is seen.
As a function of time and fixed energy, it is the Distorted Phase Effects which first alter the transition probability with the Stimulated Transitions taking over later if the amplitude of the turbulence $C_{\star}$ is larger than $C_{\star} \gtrsim 1\%$ (which is likely but not known for certain). As the turbulence passes through the MSW resonance for that energy the Stimulated Transitions die away leaving just the Distorted Phase Effects which themselves eventually fade. During the period of strongest turbulence effects depolarization occurs. What this means is that the transition probabilities become essentially random. In practice depolarization would be seen either by looking at neutrinos within a small energy window or for a fixed energy over a small time window. When averaged, say over an energy window of order $\sim 1\;{\rm MeV}$, the average transition probability, $\langle P_{ij}\rangle$, would approach the average of the distribution of $P_{ij}$ for each energy individually. Interestingly the depolarization limit depends upon the number of neutrino flavors that are mixing due to the turbulence. If the turbulence amplitude is smaller than $C_{\star} \lesssim 30\%$ then turbulence affects just two states so the average transition probabilities approach $\langle P_{ij}\rangle \approx 1/2$. If the turbulence amplitude is larger than $C_{\star} \gtrsim 30\%$ one finds all transition probabilities approach $\langle P_{ij}\rangle \approx 1/3$ because the turbulence affects all mixing all the states simultaneously \cite{2015PhRvD..92a3009K}. There is some sensitivity to the power spectrum of the turbulence with harder spectra producing depolarized transition probabilities at smaller turbulence amplitudes. 

For a given turbulence profile at a fixed time, the transition probability in the resonance channel oscillates very rapidly with the neutrino energy. An example of a calculation showing turbulence effects is shown in figure (\ref{fig:matterplusturbulence}) which is adapted from figure (14) in Lund \& Kneller \cite{2013PhRvD..88b3008L}. For this calculation 30\% turbulence was added with a Kolmogorov power spectrum. This figure should be compared with figure (\ref{fig:matter effects}) which uses the same density profile and neutrino mixing parameters and starts from the same location. One sees the amplitude of the fluctuations in the probabilities have grown now that turbulence has been added and in some mixing channels for some energies - e.g.\ $P_{23}$ in the range from $10\;{\rm MeV}$ to $40\;{\rm MeV}$ - the range of fluctuations strongly hints the probability is two-flavor depolarized at this time. In a very high energy resolution calculation - e.g.\ see the other panels of figure (4) from Kneller \& Volpe \cite{2010PhRvD..82l3004K} - the `period' of the oscillations when turbulence is added becomes much smaller, of order $\sim 10\;{\rm keV}$. There are also changes seen in other mixing channels: for example the transition probability $P_{13}$ in the NMO was essentially zero for all energies in figure (\ref{fig:matter effects}) but in figure (\ref{fig:matterplusturbulence}) we see it can reach $\sim 20\%$ for some energies.

\subsubsection{Self-Interaction Effects}

The subject of neutrino self-interaction effects in supernovae is still an area of intense interest and, unfortunately, uncertainty. The addition of $H_{SI}$ to the neutrino Hamiltonian makes the evolution of one neutrino dependent upon every other neutrino emitted. In principle the evolution matrix - or whatever quantity one is following - would have to be a seven dimensional matrix dependent upon time, three spatial coordinates and three momenta. Calculating such a quantity is beyond the capability of present codes. In order to make progress a number of symmetries have to be imposed to reduce the dimensionality. At the present time the only self-consistent calculations which include $H_{SI}$ are those which assume a steady-state (stationary in time) and a spherically symmetric  neutrino source at some fixed radius $R_{\nu}$. This model is known as the `bulb' model. In the bulb model the evolution matrix is a function of just three quantities: the radius from the center of the star (spherical symmetry), the neutrino energy, and the angle between the trajectory of a neutrino and the radial direction at a given radius (axial symmetry) e.g.\ at the neutrinosphere. Calculations of $S$ with these three degrees of freedom are known as `multi-angle' calculations. The first multi-angle calculations were performed by Duan \textit{et al.} \cite{Duan:2006jv,Duan:2006an} and are still not very common because they are computationally expensive. The most comprehensive set of multi-angle calculations to date are those by Wu \textit{et al.} \cite{PhysRevD.91.065016} which processed the $18\;{\rm M_{\odot}}$ simulation by Fischer \textit{et al.} \cite{2010A&A...517A..80F} for the IMO. The results of their multi-angle calculation are shown in figure (\ref{fig:WuMA}). 
\begin{figure*}[t!] 
\includegraphics[width=\linewidth]{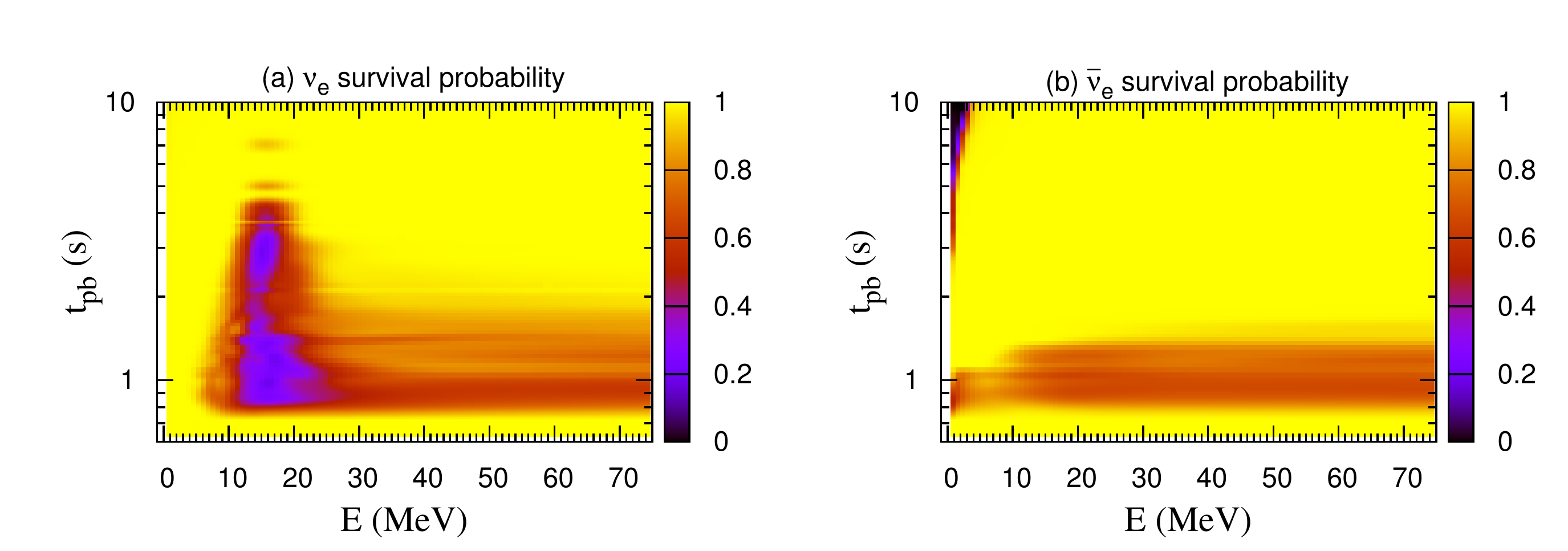}
\caption{The angle average electron neutrino (left panel) and antineutrino (right panel) survival probability at $r=500\;{\rm km}$ as a function of the neutrino energy and postbounce time. The figure is from Wu \textit{et al.} \cite{PhysRevD.91.065016} and is reprinted courtesy of APS. \label{fig:WuMA}}
\end{figure*}
The figure shows that up to the accretion phase, there are no affects from self-interactions. Then, starting at $t_{pb}=0.7\;{\rm s}$, the electron neutrino survival probability for energies greater than $E\sim 5\;{\rm MeV}$ dips blow unity and the same for all antineutrinos. For the neutrinos in the energy window $10\;{\rm MeV} \lesssim E \lesssim 20\;{\rm MeV}$ the survival probability is of order $\sim 30\%$. The self-interaction affects for the antineutrinos are only brief because they are gone by $t_{pb} \sim 1.3\;{\rm s}$. In the neutrinos they last a bit longer: up to $t_{pb} \sim 2\;{\rm s}$ for all energies above $E\sim 5\;{\rm MeV}$ and then just in a window $10\;{\rm MeV} \lesssim E \lesssim 20\;{\rm MeV}$ up to $t_{pb} \sim 5\;{\rm s}$.

\begin{figure*}[t!] 
\includegraphics[width=\linewidth]{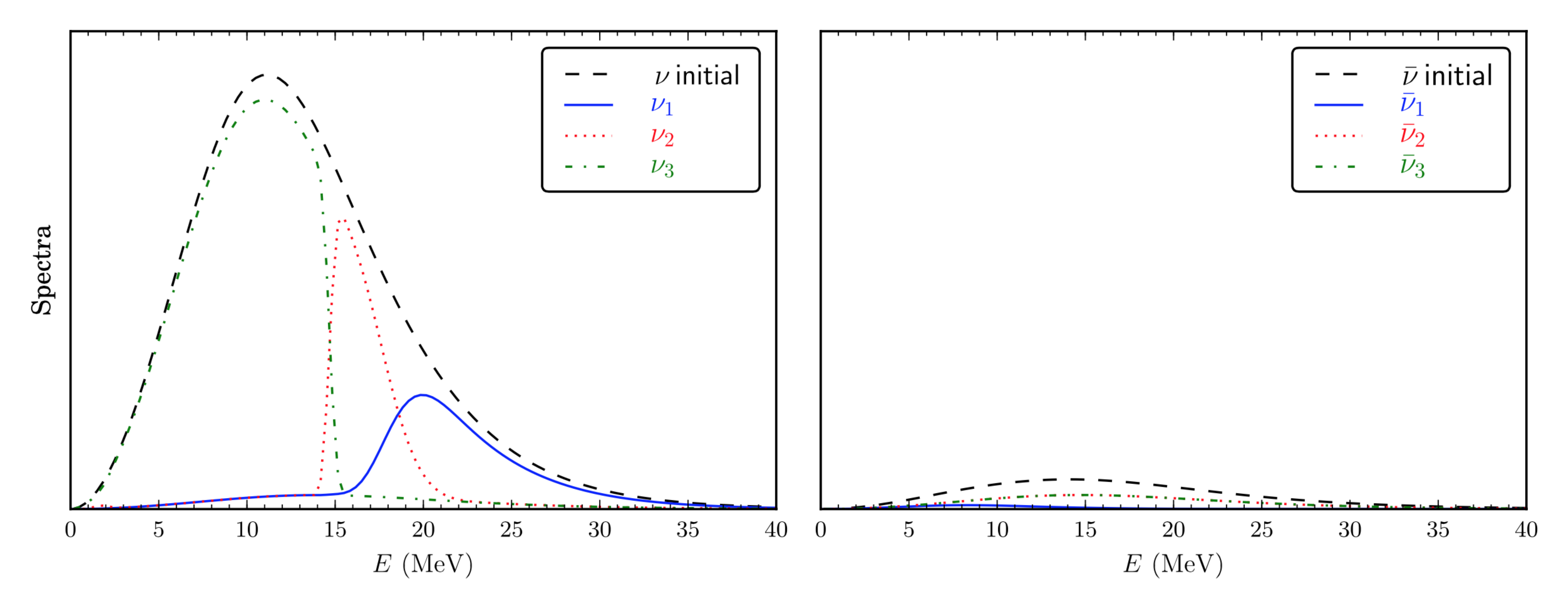}
\includegraphics[width=\linewidth]{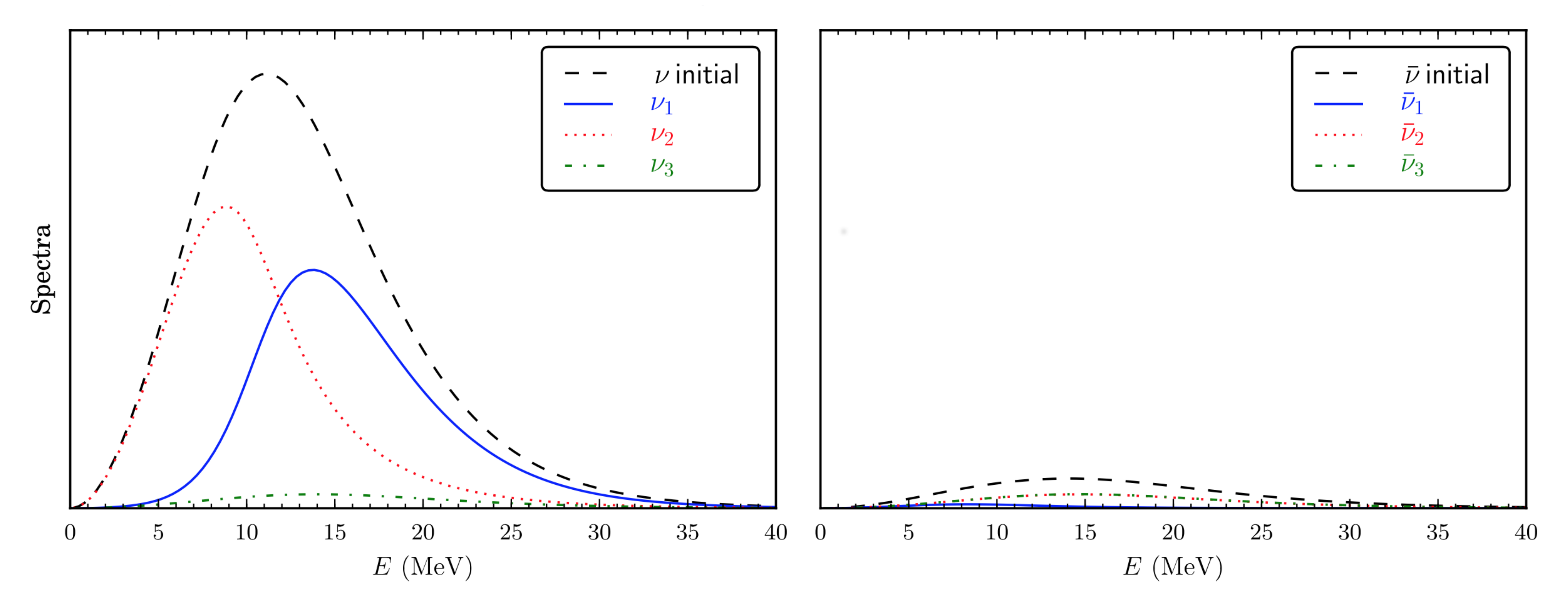}
\caption{ The expected signal from the neutronization burst of the ONeMg simulation by Fischer \textit{et al.} \cite{2010A&A...517A..80F}. The top panels are for the NMO, the bottom pair are for the IMO. The left panel in each pair is the scaled neutrino number flux, summed over all neutrino flavors, in the vacuum mass basis created by integrating the final emission angle averaged neutrino spectral energy distribution and fluxes over the first 30 ms of the neutrino burst signal. The right panel in each pair is for the antineutrino number flux again in the vacuum mass basis. The figure is taken from Cherry \textit{et al.} \cite{PhysRevD.85.125010} and is reprinted courtesy of APS. \label{fig:multiangleONeMg}  }
\end{figure*}
The neutrino flavor transformation in ONeMg supernovae is not as well known. 
To date there have been just a few multi-angle calculations by Cherry \textit{et al.} \cite{2010PhRvD..82h5025C,PhysRevD.84.105034,PhysRevD.85.125010} which focus upon the collapse epoch, particularly the neutronization burst. The results of Cherry \textit{et al.} \cite{PhysRevD.85.125010} for the multi-angle calculation of the neutronization burst in the $M=8.8\;{\rm M_{\odot}}$ ONeMg simulation by Fischer \textit{et al.} \cite{2010A&A...517A..80F} are shown in figure (\ref{fig:multiangleONeMg}). The NMO case has an interesting sequence of  spectral splits in the neutrinos around $E=15\;{\rm MeV}$ where the initial electron type neutrinos emerge mostly as mass state $\nu_{3}$ below this energy, mostly as mass state $\nu_2$ between $15\;{\rm MeV}$ and $25\;{\rm MeV}$, and then as mass state $\nu_1$ above $25\;{\rm MeV}$. In the IMO the electron neutrinos emerge as $\nu_{2}$ below $13\;{\rm MeV}$ and as mass state $\nu_1$ above $13\;{\rm MeV}$. Given the different proportions of electron neutrino in each mass state, the neutronization burst would exhibit two steps in the event rate as a function of energy if the ordering were NMO and just one if the ordering were an IMO. 

More commonly one sees reference in the literature describing flavor transformation calculations to a `single-angle' approximation. The approximation reduces the number of degrees of freedom of $S$ to two by assuming the evolution of one neutrino at a given energy emitted at some angle with respect to the radial direction is representative of all neutrinos with that energy. Often the results from the single-angle calculations match closely the multi-angle but the agreement is not always true \cite{2010PhRvD..82h5025C,2011PhRvL.106i1101D}.
\begin{figure*}[b!] 
\includegraphics[width=0.45\linewidth]{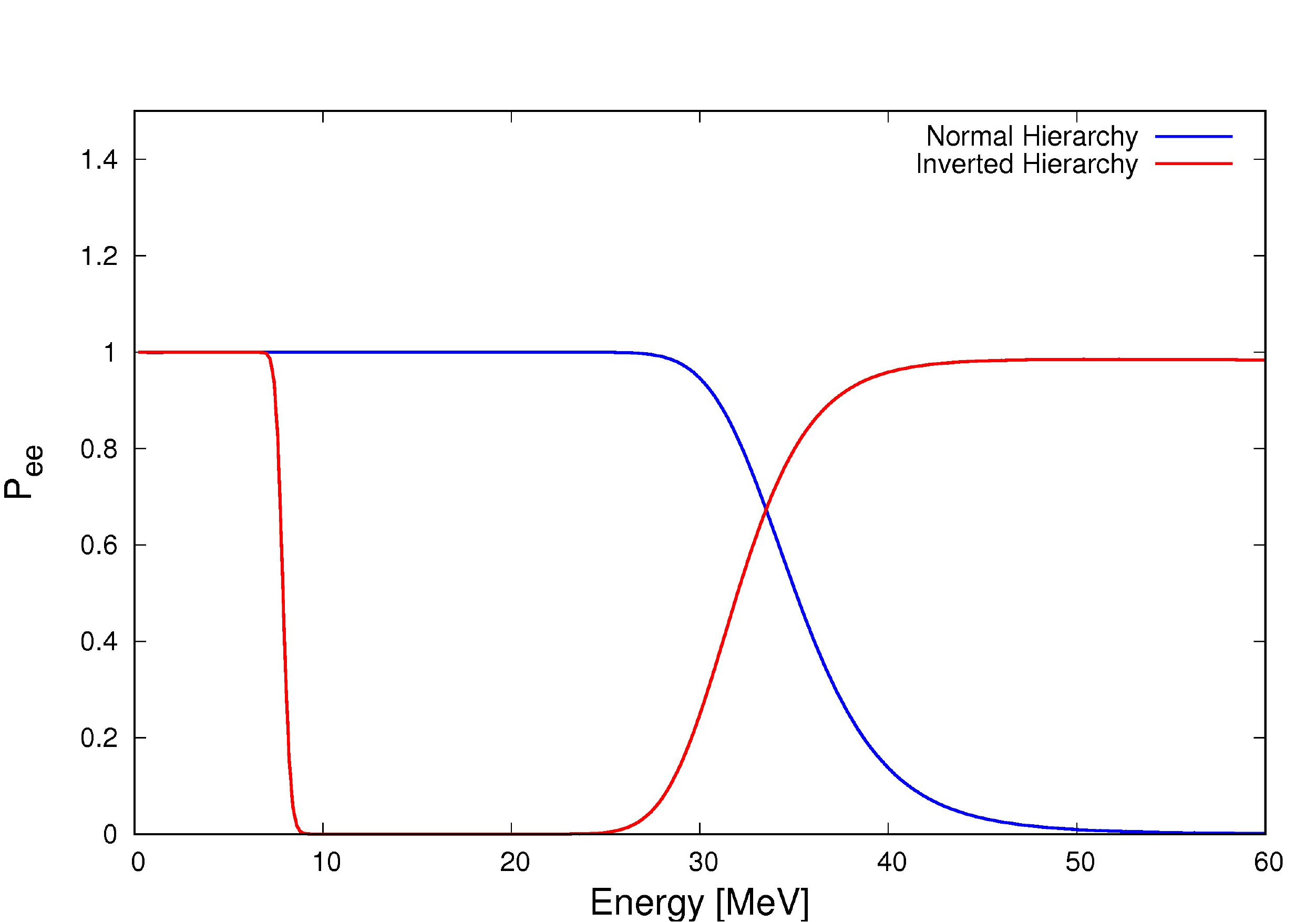}
\includegraphics[width=0.45\linewidth]{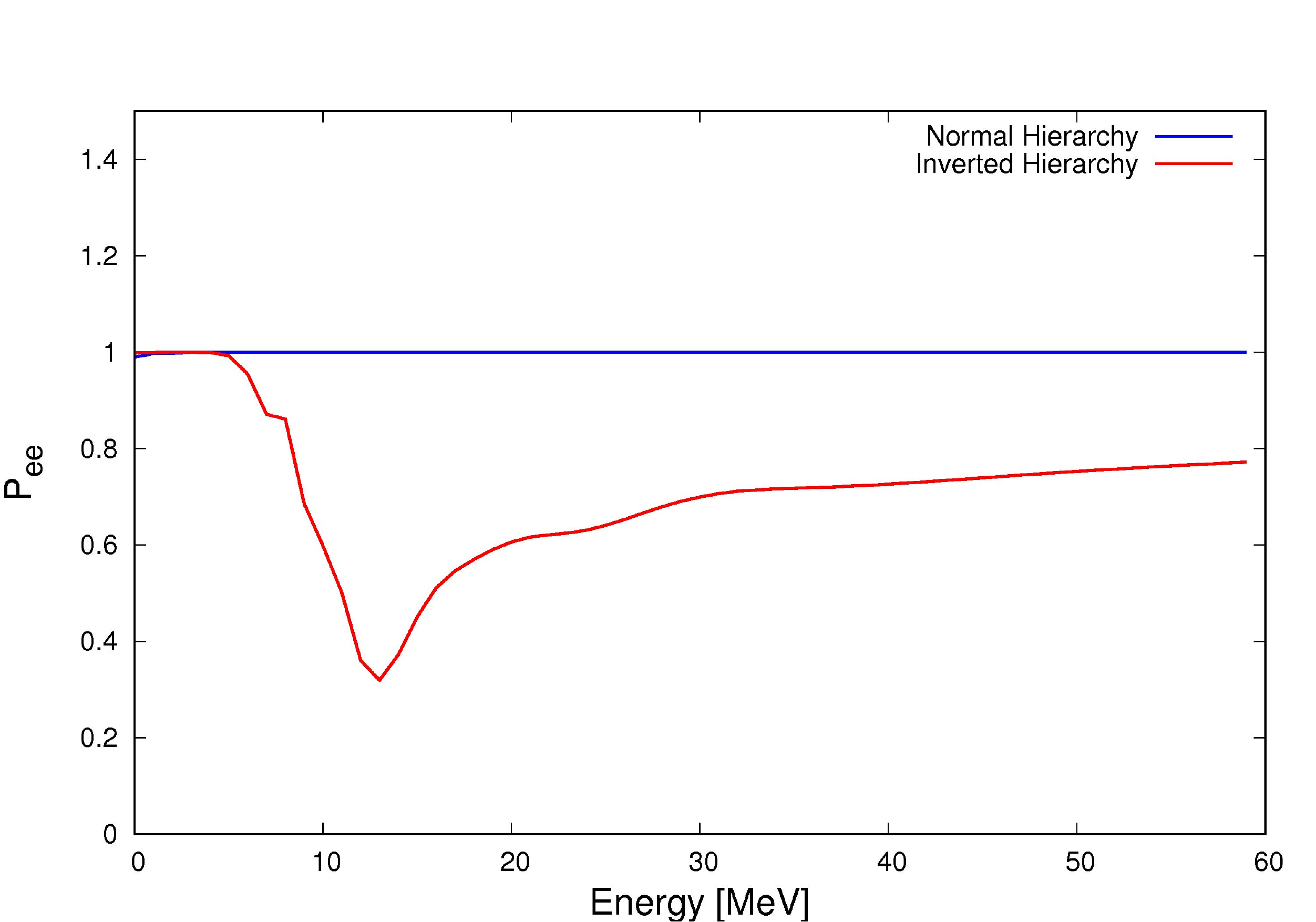}
\caption{The electron neutrino survival probability at $r=1000\;{\rm km}$ as a function of the neutrino energy. The calculation uses the luminosities, mean energies, mean square energies, and density profile from the $t=1\;{\rm s}$ snapshot of the $10.8\;{\rm M_{\odot}}$ simulation by Fischer et al . \cite{2010A&A...517A..80F} shown in figures (\ref{fig:rho}) and (\ref{fig:L,meanE}). On the left we show the results from a single angle calculation, on the right the angle-averaged result of a multi-angle calculation. The figure is courtesy of Yue Yang \cite{Yangprivate}. \label{fig:SAvsMA}}
\end{figure*}
For example, in figure (\ref{fig:SAvsMA}) we see a comparison of a single-angle and a multi-angle calculations for both normal and inverted mass ordering. The figure is courtesy of Yue Yang \cite{Yangprivate} and the calculation is described in the caption. In the single-angle calculation we see nice clean spectral splits where the probability changes from unity to zero (or vice versa). In the inverted mass ordering the transition at $E\sim 8\;{\rm MeV}$ is sharp, the transition between $25\;{\rm MeV}\lesssim E \lesssim 50\;{\rm MeV}$ occurs in both mass orderings. But when we look at the multi-angle result we see there are no transitions for the normal mass ordering; for the inverted ordering there is a change in the probability beginning around $E\sim 8\;{\rm MeV}$ but it is not as sharp or complete as the single-angle result. For reference, in both calculations the neutrinos which were originally electron flavor become an equal mixture of the $\mu$ and $\tau$ flavors. For this reason, at the present time one is somewhat nervous about declaring which features of systematic surveys of the neutrino spectral parameter space using single-angle calculations, e.g. \cite{2010arXiv1008.0308C}, are robust. 

While an analytical solution of the neutrino flavour evolution seen in figures (\ref{fig:WuMA}), (\ref{fig:multiangleONeMg}) and (\ref{fig:SAvsMA}) is not yet available, in recent years the technique of `stability analysis' \cite{2011PhRvD..84e3013B,2013PhRvD..88f5003V} has been applied to simplified versions of such calculations and found to give predictions which closely match the numerical results. For a review of the stability of the neutrino flavor evolution in supernovae we refer the reader to Chakraborty \textit{et al.} \cite{1475-7516-2016-03-042}. The basic idea of stability analysis is to linearize the Schr\"odinger equation - or the Louiville-von Neumann equation for the evolution for the density matrix - in the off-diagonal elements of $S$ and examine how those off-diagonal elements grow with time or in space. One finds that growth of the off-diagonal elements of $S$ amounts to finding the eigenvalues of a `stability matrix' such that when the eigenvalues are complex, the off-diagonal elements grow exponentially in time/space. While the technique appears to work for simplified models, it is not yet clear if it is able to give good predictions in the more complex supernova neutrino problem. 

Multi-angle calculations of neutrino flavor transformation are still an approximation to the full problem and need to be improved in many ways. Clearly by imposing the symmetry of spherical emission the implications of aspherical neutrino emission - either from hotspots or from a more global asymmetry produced by the LESA - can not be determined. One consequence of aspherical emission may be that there are regions where the antineutrino flux dominates over the neutrino flux. This would cause the self-interaction to be `negative' and it has been seen in the environment of compact object mergers - two merging neutron stars or a black hole/neutron star merger - that the flavor transformation with a negative self-interaction can lead to new flavor transformation phenomenology \cite{2012PhRvD..86h5015M,2014arXiv1403.5797M,Wu:2015fga,Wu:2015fga,2016PhRvD..93d5021M,2017PhRvD..95d3005C}. Even if there are no regions where the overall flux of electron antineutrinos is greater than the electron neutrino flux, there can be locations where there are more electron antineutrinos propagating in particular directions than electron neutrinos. This setup might occur because the electron antineutrinos decouple from the matter deeper within the proto-neutron star than the electron neutrinosphere and thus, at a given radius, the electron antineutrinos are more forward-peaked than the electron neutrinos. If this occurs we can have a case of 'angular crossings' - the electron lepton number current changes sign as a function of  angle relative to the radial direction - and, using stability analysis, such crossings are expected to lead to new flavor transformation effects \cite{2005PhRvD..72d5003S,2012PhRvL.108w1102M,2012PhRvD..85k3002S,2017JCAP...02..019D}. However a recent study by Tamborra \textit{et al.} \cite{2017ApJ...839..132T} of the angular distribution of the different neutrino flavors in three 1D simulations did not find angular crossings: whether crossings occur in 3D simulations which produce a LESA is not yet known. 

Another common assumption of the calculations is that the neutrinos above the neutrinospheres experience no direction-changing scattering events with the matter. This assumption allows the flavor evolution calculations to be posed as an initial value problem: if we define the neutrino states at the neutrinosphere we can solve for the neutrino state at larger radii by integrating the Schr\"odinger equation (\ref{eq:Schrodinger}) along outward trajectories. But if direction-changing scatterings occur then there can be inward moving neutrinos and such neutrinos would not allow us to treat the neutrino evolution as an initial value problem. This effect has been called the `Neutrino Halo' \cite{2012PhRvL.108w1102M} and scatterings are expected to be most important for the flavor evolution during the accretion phase when the density around the proto-neutron star is large and the neutrinos self-interaction not yet strongly suppressed. The number of scattered neutrinos is small but they can contribute significantly to the neutrino self-interaction potential - equation (\ref{eqn:HSI}) - because the angle between the unscattered neutrino trajectories becomes so small suppressing the contribution by the unscattered neutrinos to $H_{SI}$ \cite{2012PhRvL.108z1104C}. At the present time the only calculations which have included this effect are those by Cherry \textit{et al.} \cite{2013PhRvD..87h5037C} for the neutronization burst of a ONeMg supernova and the authors noted significant changes when the halo was included. However Sarikas \textit{et al.}  \cite{PhysRevD.85.113007} argued that the formation of a neutrino halo in a Fe core supernovae does not alter the flavor transformation because the same dense matter that produces the halo also suppresses the flavor transformation during the accretion phase - the so called 'matter suppression' effect \cite{2011PhRvL.107o1101C,2012PhRvL.108f1101S}. 

Finally, laying aside the issues associated with aspherical neutrino emission, the angular distribution of the emission, and the neutrino halo, it has also become apparent that even in the bulb model the self-interaction makes the flavour evolution `unstable' in the Lyapunov sense: small perturbations from the spherical or axial symmetries \cite{Hansen:2014paa,PhysRevLett.111.091101,PhysRevD.90.033004,2015PhLB..747..139D,2015PhRvD..92f5019A} or from stationarity \cite{2015PhLB..751...43A,2015PhRvD..92l5030D,2016JCAP...04..043C}, grow in space and/or time. 
Whether the other sources of asphericity, such as the in neutrino emission and density, make this instability moot is, again, unknown. Many authors have speculated that all the different forms of instability force the neutrino flavor towards the equilibrium of decoherence i.e.\ flavour equality, the same as for strong turbulence. While that is a legitimate opinion, we remind the reader that as shown in the case of turbulence discussed earlier, order in the neutrino flavor evolution can appear out of what appears to be chaos \cite{2014PhRvD..89g3022P}.

\subsection{A timeline of Flavour Changing Effects in Supernovae}
Putting all the different transformation effects together in order to compute the time, energy, mass ordering, progenitor and turbulence dependence of the neutrino flavor composition emerging from the supernova from a given line-of-sight is a huge task. A sample of such calculations is \cite{Kneller:2008PhRvD..77d5023K,2011PhRvD..84h5023R,PhysRevD.91.065016,2009PhRvL.103g1101G} but the landscape of possible flavor transformations is by no means well explored. As mentioned previously, the biggest uncertainty in the flavour evolution is with regard to the self-interactions so while we base our timeline on the results of the multi-angle bulb model calculations, we caution the reader that our statements about the signatures of self-interactions are subject to further review. A rough guide to the expected sequence of changes are:
\begin{itemize}
\item During the presupernova epoch the self-interaction is negligible and the evolution of the neutrino through the H and L resonances is adiabatic. The evolution matrix in the matter basis is diagonal and the matter basis transformation probabilities are very close to $P_{ij}=\bar{P}_{ij}=\delta_{ij}$. 

\item During the collapse epoch there are still no self-interaction effects for Fe core progenitors because they are suppressed by the dense matter close to the proto-neutron star \cite{PhysRevD.91.065016,2011PhRvL.107o1101C,2012PhRvL.108f1101S}. 
For ONeMg supernovae self-interaction effects appear. It is not presently clear how long these continue. For both progenitors the evolution of the neutrino through the H and L resonances is static and adiabatic.

\item Self-interaction effects continue to be suppressed during the accretion phase of Fe core progenitors. Though there is substantial turbulence in the fluid below the stalled shock, it has no effect upon the neutrinos \cite{2017arXiv170206951K}. It is not clear whether self-interactions occur for the ONeMg supernovae during this phase. 

\item Beginning at $t_{pb}\sim 1\;{\rm s}$, i.e.\ soon after the beginning of the cooling phase, self-interaction effects appear in both neutrinos and antineutrinos for Fe core progenitors. 
It is likely these effects are almost independent of the progenitor because of the similarity of the luminosities and mean energies of the neutrinos emitted by the proto-neutron star. The self-interaction effects are dependent upon the mass ordering. 
\begin{itemize}
\item For an IMO a spectral split in the electron neutrinos occurs around an energy of $E\sim 10\;{\rm MeV}$. Below this energy $P_{ee} \sim 1$, above this energy $P_{ee}\sim 0.5\pm 0.15$ for energies up to $E\sim 30\;{\rm MeV}$. For all antineutrinos $\bar{P}_{ee}\sim 0.7$. The self-interaction effects in the electron antineutrinos disappear again by $t_{pb}\sim 1.5\;{\rm s}$. The self-interaction effects in the electron neutrinos above $E\sim 10\;{\rm MeV}$ endure for up to $t_{pb}\sim 2\;{\rm s}$ and thereafter $P_{ee}\sim 0.5$ in a window $10\;{\rm MeV} \lesssim E\lesssim 25\;{\rm MeV}$ . 
\item For an NMO, multi-angle calculations do not show evidence of self-interaction effects. 
\end{itemize}

\item Dynamic matter effects begin to appear during the cooling phase. If the neutrino mass ordering is normal the shock wave affects the neutrinos in the $\nu_2 - \nu_3$ mixing channel, if it is inverted it affects the antineutrinos in the $\bar{\nu}_1 - \bar{\nu}_3$ channel. Later in the cooling phase the shock may affect neutrinos in the $\nu_1 - \nu_2$ channel. The time at which dynamic matter effects appear depends upon the progenitor. The more compact the progenitor the earlier the appearance. For ONeMg supernovae the density profile of the progenitor is so steep that matter effects occur very early and end quickly, well within $\sim 1\;{\rm s}$. For Fe core progenitors the shock arrives after $\sim 1\;{\rm s}$ and continues to affect the neutrinos for several seconds. In all cases the shock feature sweeps through the neutrino spectrum from low energy to high energy accelerating as it does so.

\item Little turbulence is expected for ONeMg progenitors. In contrast, for Fe core progenitors, once dynamic matter and turbulence effects appear even tiny amounts of turbulence have been shown to essentially randomize the transition probabilities in the $\nu_2-\nu_3$ mixing channel if the mass ordering is normal, the $\bar{\nu}_1-\bar{\nu}_3$ mixing channel if the mass ordering is inverted, due to Distorted Phases Effects. Turbulence effects appear first for lower neutrinos energies than higher. As the turbulence moves outwards the Distorted Phase Effects become subordinate to Stimulated Transitions but this may not lead to any noticeable change if the turbulence amplitude is $\lesssim 30\%$. If the turbulence amplitude is greater than $\sim 30\%$, the average transition probabilities within an neutrino energy window temporarily changes from $1/2$ to $1/3$. When the Stimulated Transitions stop and the turbulence affects revert back to 
Distorted Phase Effects, the averages return to $1/2$ before the turbulence fades away. 
The whole process can take several seconds. 

\item The interplay of self-interaction, shock and turbulence effects means that the signature features of one may be compensated by the effects of another or completely obscured. 
\end{itemize}


\section{The flux at Earth}\label{sec:Earth}

The emitted neutrino spectra and the flavor transformation must be combined in order to compute the neutrino flux at Earth. The neutrino flux at Earth can be written as a 3-component vector $F(d) = (F_e(d), F_{\mu}(d), F_{\tau}(d))^{T}$ and is related to the spectra emitted at the proto-neutron star $\Phi(R_{\nu}) = (\Phi_e(R_{\nu}), \Phi_{\mu}(R_{\nu}), \Phi_{\tau}(R_{\nu}))^{T}$ via
\begin{equation}
F(d) = \frac{1}{4\pi d^{2}}\,T\, \Phi(R_{\nu}) \label{eq:fluxesatEarth}
\end{equation}
where $T$ is the transfer matrix. A similar equation exists for antineutrinos. The matrix $T$ is a product $T=D\,P^{(mf)}(R_{\star},R_{\nu})$ where the matrix $P^{(mf)}(R_{\star},R_{\nu})$ is the matrix of transition probabilities relating initial flavor states at the neutrinosphere $R_{\nu}$ to mass states at the edge of the supernovae $R_{\star}$, and $D$ is a matrix accounting for decoherence. For both neutrinos and antineutrinos the elements of $D$ and $\bar{D}$ are simply the square magnitudes of the vacuum mixing matrix, that is $D_{\alpha i}= \bar{D}_{\alpha i} = |U_{V,\alpha i}|^{2}$. In the literature one usually observes particular attention upon the $`ee'$ element of $T$ which is the probability of a pure electron flavor neutrino at the proto-neutron star to be  detected at Earth as an electron flavor neutrino (and similarly for antineutrino states). These probabilities are often denoted simply as $p$ and $\bar{p}$ respectively in the literature and referred to as survival probabilities. The expressions for $p$ and $\bar{p}$ are
\begin{eqnarray}
p & = & \sum_i  D_{e i} P_{ie}(R_{\star},R_{\nu}) \\
\bar{p} & = & \sum_i  D_{e i} \bar{P}_{ie}(R_{\star},R_{\nu}).
\end{eqnarray} 
If one makes the approximation $\Phi_{\mu}(R_{\nu}) = \Phi_{\tau}(R_{\nu}) = \Phi_{x}(R_{\nu})$ and $\bar{\Phi}_{\mu}(R_{\nu}) = \bar{\Phi}_{\tau}(R_{\nu})= \bar{\Phi}_{x}(R_{\nu})$ then the components of the flux vectors at Earth can be written as 
\begin{eqnarray}
& F_e(d) = \frac{1}{4\pi d^2} \left[ p\,\Phi_e(R_{\nu}) + (1-p)\,\Phi_x(R_{\nu}) \right] \label{eq:Fe}\\
& F_{\mu}(d) + F_{\tau}(d) = \frac{1}{4\pi d^2} \left[ (1-p)\,\Phi_e(R_{\nu}) + (1+p)\,\Phi_x(R_{\nu}) \right] \\
& \bar{F}_e(d) = \frac{1}{4\pi d^2} \left[ \bar{p}\,\bar{\Phi}_e(R_{\nu}) + (1-\bar{p})\,\bar{\Phi}_x(R_{\nu}) \right]\\
& \bar{F}_{\mu}(d) + \bar{F}_{\tau}(d) = \frac{1}{4\pi d^2} \left[ (1-\bar{p})\,\bar{\Phi}_e(R_{\nu}) + (1+\bar{p})\,\bar{\Phi}_x(R_{\nu}) \right] \label{eq:Fbarmu}
\end{eqnarray}
so knowing the four spectra $\Phi_{e}$, $\bar{\Phi}_{e}$, $\Phi_{x}$, $\bar{\Phi}_{x}$ and expressions for $p$ and $\bar{p}$ gives the fluxes. 

However what is often reported in the literature is not $P^{(mf)}$ but usually either $P^{(ff)}$ i.e.\ flavor transition probabilities, or $P^{(mm)}$, matter basis transition probabilities. In order to compute $P^{(mf)}$ we have to transform the evolution matrices using $U$ and $\bar{U}$ at the appropriate locations and then square the amplitude of the elements. The mixed basis evolution matrix $S^{(mf)}(R_{\star},R_{\nu})$ is related to $S^{(mm)}(R_{\star},R_{\nu})$ and $S^{(ff)}(R_{\star},R_{\nu})$ by $S^{(mf)}(R_{\star},R_{\nu}) = S^{(mm)}(R_{\star},R_{\nu}) U^{\dagger}(R_{\nu})$ or $S^{(mf)}(R_{\star},R_{\nu}) = U^{\dagger}(R_{\star}) S^{(ff)}(R_{\star},R_{\nu})$. The form of $U(R_{\nu})$ and $\bar{U}(R_{\nu})$ were given previously in equations (\ref{eq:UNMO}) and (\ref{eq:UIMO}). So if $P_{ij}(R_{\star},R_{\nu})$ are the matter basis transition probabilities as calculated from the neutrinosphere to edge of the supernova then $p$ and $\bar{p}$ can be determined. 
For the NMO the expressions for $p$ and $\bar{p}$ are
\begin{eqnarray}
p & = & D_{e1} P_{13}(R_{\star},R_{\nu}) + D_{e2} P_{23}(R_{\star},R_{\nu}) + D_{e3} P_{33}(R_{\star},R_{\nu}) \\
\bar{p} & = &  D_{e1} \bar{P}_{11}(R_{\star},R_{\nu}) + D_{e2} \bar{P}_{21}(R_{\star},R_{\nu}) + D_{e3} \bar{P}_{31}(R_{\star},R_{\nu})
\end{eqnarray}
and in the IMO we find
\begin{eqnarray}
p& = & D_{e1} P_{12}(R_{\star},R_{\nu}) + D_{e2} P_{22}(R_{\star},R_{\nu}) + D_{e3} P_{32}(R_{\star},R_{\nu})\\
\bar{p}  & = &  D_{e1} \bar{P}_{13}(R_{\star},R_{\nu}) + D_{e2} \bar{P}_{23}(R_{\star},R_{\nu}) + D_{e3} \bar{P}_{33}(R_{\star},R_{\nu}).
\end{eqnarray}

\begin{table}
\begin{minipage}{0.5\textwidth}
\begin{tabular}{l}
 {\bf NMO} \\
 1) Adiabatic Evolution \\  \hline
 $p = s_{13}^2$\\
 $\bar{p} = c_{12}^2\,c_{13}^2$\\ 
 \\
 2) $P_{13}=0$, $P_{23}=1$, $P_{33}=0$  \\  \hline 
 $p = s_{12}^2\,c_{13}^2$\\ 
 \\
 3a) $P_{13}=0$, $P_{23}=P_{33}=1/2$  \\ \hline 
 $p = (s_{12}^2\,c_{13}^2 +s_{13}^2)/2$\\ 
 \\
 3b) $P_{13}=P_{23}=P_{33}=1/3$  \\  \hline 
 $p =1/3$ \\
  \\
 4a) $P_{13}=x$, $P_{23}=0$, $P_{33}=1-x$  \\ 
 $\;\;\;\;\; \bar{P}_{11}=0$, $\bar{P}_{21}=\bar{P}_{31}=1/2$  \\ \hline 
 $p = x\, c_{12}^2\,c_{13}^2 +(1-x)\,s_{13}^2 $\\ 
 $\bar{p} = ( s_{12}^2\,c_{13}^2 + s_{13}^2)/2$\\
 \\
 4b) $P_{13}=0$, $P_{23}=x$, $P_{33}=1-x$  \\ 
 $\;\;\;\;\; \bar{P}_{11}=0$, $\bar{P}_{21}=\bar{P}_{31}=1/2$ \\ \hline 
 $p = x\,s_{12}^2\,c_{13}^2 + (1-x)\,s_{13}^2$\\ 
 $\bar{p} = (s_{12}^2\,c_{13}^2 + s_{13}^2)/2$\\
 \\
 \\
 \\
\end{tabular}
\end{minipage} 
\begin{minipage}{0.5\textwidth}
\begin{tabular}{l}
{\bf IMO} \\ 
1) Adiabatic Evolution \\ \hline 
$p =  s_{12}^2\,c_{13}$\\
$\bar{p}   =  s_{13}^2$\\
\\
2a) $\bar{P}_{13}=1$, $\bar{P}_{23}=0$, $\bar{P}_{33}=0$ \\ \hline 
$\bar{p}   = c_{12}^2\,c_{13}^2$\\ 
\\
2b) $P_{12}=1$, $P_{22}=0$, $P_{32}=0$ \\ \hline 
$p  = c_{12}^2\,c_{13}^2$\\ 
\\
3a) $\bar{P}_{13}=1/2$, $\bar{P}_{23}=0$, $\bar{P}_{33}=1/2$ \\ \hline 
$\bar{p}   =  ( c_{12}^2\,c_{13}^2 + s_{13}^2 )/2$\\
\\
3b) $P_{12}=P_{22}=1/2$, $P_{32}=0$ \\ \hline 
$p   =  c_{13}^2 /2$\\
\\
4) $P_{12}=x$, $P_{22}=1-x$, $P_{32}=0$  \\ 
$\;\;\;\; \bar{P}_{13}=0$, $\bar{P}_{23}=\bar{P}_{33}=1/2$  \\ \hline 
$p = x\,c_{12}^2\,c_{13}^2+(1-x)\,s_{12}^2\,c_{13}^2$\\ 
$\bar{p} = (s_{12}^2\,c_{13}^2 + s_{13}^2)/2$\\
\\
5) $P_{12}=x$, $P_{22}=1-x-y$, $P_{32}=y$  \\ 
$\;\;\;\; \bar{P}_{13}=\bar{x}$, $\bar{P}_{23}=\bar{y}$, $\bar{P}_{33}=1-\bar{x}-\bar{y}$  \\ \hline 
$p = x\,c_{12}^2\,c_{13}^2 + y\,s_{13}^2\,+(1-x-y)\,s_{12}^2\,c_{13}^2$ \\ 
$\bar{p} = \bar{x}\,c_{12}^2\,c_{13}^2 + \bar{y}\,s_{12}^2\,c_{13}^2 + (1-\bar{x}-\bar{y})\,s_{13}^2 $\\
\end{tabular}
\end{minipage} 
\caption{The expressions for the electron neutrino and electron antineutrino survival probabilities $p$ and $\bar{p}$ for each mass ordering for different neutrino flavour evolution scenarios. \label{tab:p,pbar}. The quantities $x$ and $y$, $\bar{x}$ and $\bar{y}$ are partial mixing fractions such that $0 \leq x+y, \bar{x}+\bar{y} \leq 1$. Using the Particle Data Group evaluation of the mixing angles \cite{Olive:2016xmw}, $c_{12}^2\,c_{13}^2=0.678$ $s_{12}^2\,c_{13}^2=0.301$, $s_{13}^2= 0.021$}
\end{table}

The expressions given for the electron neutrino and electron antineutrino survival probabilities $p$ and $\bar{p}$ are completely general in the sense that we made no assumptions about which transition probabilities were zero or non-zero. Note that both survival probabilities are bounded not by zero and unity but rather by $s_{13}^2$ and $c_{12}^2\,c_{13}^2$. Using the Particle Data Group evaluation of the mixing angles \cite{Olive:2016xmw}, this translates into $ 0.021 = \leq p, \bar{p} \leq 0.678$. In some older literature which often used two-flavor approximations, one often sees reference to $P_L$ and $P_H$ which are referred to as the crossing probability at the L and H resonance \cite{2000PhRvD..62c3007D}. For a NMO the assignments are $P_{12}=P_L$, $P_{23}=P_H$ and all other transition probabilities are set to zero. For a IMO $P_{12}=P_L$, $\bar{P}_{13}=P_H$ with all other transition probabilities are set to zero.

Using these expressions we can consider a number of different scenarios for the flavor transformation consistent with the timeline of changes given in the previous section. For example: in the case of adiabatic evolution we have $P_{ij}\approx \delta_{ij}$ and $\bar{P}_{ij}\approx \delta_{ij}$. The expressions for $p$ and $\bar{p}$ in the adiabatic limit are shown in table (\ref{tab:p,pbar}) as NMO-1) and IMO-1) using the vacuum mixing matrix elements given in equation (\ref{eq:UV}). Table (\ref{tab:p,pbar}) also includes many other scenarios that the flavor transformation calculations indicate are useful. The expression for case NMO-2) would be relevant for pure matter effects when the shock is in the vicinity of the H resonance when the mass ordering is normal, while the cases IMO-2a) is the same but for the inverted mass ordering and IMO-2b) is for the case when the shock reaches the L resonance. Cases NMO-3a) and IMO-3a) are the median values of $p$ and $\bar{p}$ when two-flavour depolarization occurs in the NMO or IMO - a situation that appears to be occurring in figure (\ref{fig:matterplusturbulence}). Case NMO-3b) is for the situation of three-flavor depolarization while IMO-3b) is the electron survival probability when depolarization occurs in the neutrinos. 
If we introduce partial mixing fractions $x$ and $y$, $\bar{x}$ and $\bar{y}$ such that $0 \leq x+y, \bar{x}+\bar{y} \leq 1$, cases NMO-4a) and NMO-4b) are consistent with the results shown in the upper panel of figure (\ref{fig:multiangleONeMg}) if $x$ is energy dependent and $y=0$, and the expressions for the case IMO-4) correspond to the results shown in the lower panel if $x$ is allowed to be energy dependent and $y=0$. Finally, case IMO-5) is for the case of Fe core supernova when only self interaction effects occur, as seen in figure (\ref{fig:WuMA}) if $x$, $y$, $\bar{x}$ and $\bar{y}$ are energy dependent. 


\section{Neutrino Detection}\label{sec:detection}

Multiple detectors with sensitivity to core-collapse MeV neutrinos are currently in operation. A range of detector technologies and sensitivity to different neutrino flavors are available, and plans for dramatic improvements are also underway. In this section we briefly summarize MeV neutrino detection, focusing on four general classes of detectors: water Cerenkov, liquid argon, liquid scintillator, and other technologies. For each, we first show in parenthesis the number of events in the dominant channel for the largest detector in its class, taken from Table I of \cite{Mirizzi:2015eza}, where a canonical distance to the core collapse of 10 kpc is used. A more comprehensive review can be found in, e.g., Refs.~\cite{2012ARNPS..62...81S,Mirizzi:2015eza}.

In general, detectors are sensitive to one or more of the products of neutrino-target interactions. In the MeV energy range, $\nu_e$ and $\bar{\nu}_e$ undergo both CC and NC interactions, while heavy lepton neutrinos are only accessible via NC interactions. The exact detection method depends on the detector technology as well as on strategies to beat competing backgrounds. Strategic tagging of final state products is a powerful method to distinguish signals from other events and backgrounds.

\subsection{Water Cerenkov} 
(7,000 $\bar{\nu}_e$ at SuperK; 110,000 $\bar{\nu}_e$ at HyperK)\\
The primary neutrino detection channel in water medium is the inverse-beta decay (IBD), $\bar{\nu}_e + p \to e^+ + n$, where $p$ denotes free hydrogen in water. By measuring the Cerenkov rings caused by relativistic positrons, event-by-event reconstruction of the neutrino interaction is possible by low threshold detectors such as the Super-Kamiokande (SuperK; fiducial volume 32 kton, positron detection threshold energy $\sim 3$ MeV) and the proposed Hyper-Kamiokande (HyperK; fiducial volume 374 kton) experiment. Larger megaton class volumes with higher thresholds have been considered by instrumenting ocean water \cite{Suzuki:2001rb,Kistler:2008us}. Since the positron is emitted almost isotropically, there is little directional information of the neutrino that can be obtained. On the other hand, the neutrino energy can be faithfully obtained from the reconstructed positron energy. Both the cross section and kinematics are well understood \cite{Vogel:1999zy,Strumia:2003zx}, making the IBD the golden channel for supernova neutrino detection. 

Single $e^\pm$ searches however have many backgrounds from radioactivities and those induced by atmospheric neutrinos. But by detecting also the neutron in the IBD, a delayed coincident tagging becomes possible. This allows the important separation with background signals that do not produce a neutron final state. In water, the MeV neutron produced by IBD thermalizes by elastic collisions and is eventually captured by a proton, $n+p \to d+\gamma$, after a typical delay of $\sim 200 \, \mu$s. The gamma emitted by the deuteron however has an energy of $\sim 2.2$ MeV, making it challenging to detect in SuperK (tagging efficiency of some 20\% has been achieved \cite{Zhang:2013tua}). To improve on this, an upgrade to SuperK is currently planned: the addition of gadolinium salt to the water. Gadolinium's neutron capture cross section is $>10^5$ that of a proton's, and its subsequent decay generates 3-4 gammas with total energy $\sim 8$ MeV with a delay of $\sim 20 \, \mu$s. Proposed by Beacom \& Vagins \cite{Beacom:2003nk}, the gadolinium upgrade will achieve a neutron tagging efficiency of some 90\%, opening both stronger and more reliable coincidence discrimination.

Neutrino scattering on $e^-$ provides a complementary channel to IBD. Since the recoil electrons are forward-peaked, the direction of the incoming neutrino can be reconstructed. All neutrinos contribute, although $\nu_e$ and $\bar{\nu}_e$ dominate. The cross section is only a few percent of IBD's at supernova neutrino energies \cite{Vogel:1989iv}, making large volumes and efficient IBD tagging particularly powerful for isolating the $\nu_e$ \cite{Laha:2013hva}. Neutrinos can also be detected via interactions with oxygen, including CC captures $\nu_e + {}^{16}{\rm O} \to e^- + {}^{16}{\rm F}^*$ and $\bar{\nu}_e + {}^{16}{\rm O} \to e^+ + {}^{16}{\rm N}^*$, as well as all-flavor NC interactions \cite{Haxton:1987kc}. The former interactions have high thresholds of $\approx 15$ MeV, while the latter have small cross sections, making them unique channels most powerful for nearby core-collapse events.

A special case of water Cerenkov detectors is those intended for higher energies (GeV and above), such as IceCube. These detectors have photomultiplier tubes attached to strings that are periodically placed to instrument large volumes of transparent material, e.g., water or ice. The lower density of photodetectors means they cannot reconstruct the Cerenkov rings of individual MeV neutrino interactions. However, they are able to detect the supernova neutrino burst as a coincident rise of optical hit rates across all photodetectors \cite{Halzen:1994xe,Halzen:1995ex}. IceCube consists of 5160 optical modules, and being submerged in the cold Antarctic ice, has a stable and low background rate, effectively acting as a $\sim 3$ Mton MeV neutrino detector \cite{Abbasi:2011ss}. The proposed IceCube upgrade, PINGU, will introduce a sub-volume of higher photodetector density and improve lower energy capabilities \cite{Aartsen:2014oha}, although not yet to MeV neutrino interaction reconstruction. Other detectors such as ORCA, as part of the KM3NeT project, will implement a similar string-based detector in the Mediterranean Sea \cite{Katz:2014tta}. The ambient detector background is however expected to be larger, in part due to the higher operating temperature but also anticipated bioluminescence.

\subsection{Liquid Argon} 
(3,000 $\nu_e$ at DUNE)\\
Liquid Argon Time Projection Chamber (LAr-TPC) technology provides ample sensitivity to supernova $\nu_e$ through the CC interaction $\nu_e + {}^{40}{\rm Ar} \to e^- + {}^{40}{\rm K}^*$. These detectors drift ionization charge from charged-particle energy loss onto a two-dimensional plane, and with timing providing an additional dimensional probe, makes possible three-dimensional track reconstruction. The technology has been demonstrated in the Icarus experiment utilizing 760 tons of LAr \cite{Amoruso:2003sw}. The Deep Underground Neutrino Experiment (DUNE) will be the next-generation LAr detector providing a clean high-statistics $\nu_e$ detection of supernova neutrinos using four 10 kton LAr TPCs \cite{Acciarri:2016crz,Acciarri:2015uup,Acciarri:2016ooe}. 

To reliably distinguish the $\nu_e$ capture from other interaction channels, most notably $\bar{\nu}_e + {}^{40}{\rm Ar} \to e^+ + {}^{40}{\rm Cl}^*$, there is ongoing photon detector system R\&D \cite{Ankowski:2016lab} that aims to distinguish the $\gamma$-ray emitting de-excitations of ${}^{40}{\rm K}^*$ and ${}^{40}{\rm Cl}^*$. Also, there remains work needed, both theoretical and experimental, for a better understanding of neutrino-argon interactions in the MeV energy regime. At these energies, the impulse approximation is expected to break down as different effects of nuclear dynamics become important. To date, advanced many-body calculations based on realistic models of nuclear dynamics of the responses to electroweak interactions have been successfully carried out up to medium-heavy nuclei \cite{Lovato:2012ux,Lovato:2013dva,Ankowski:2014yfa,Rocco:2015cil}. However, for a nucleus as complex as argon, extensions will be required. Experimentally, there is currently no direct measurement of neutrino-argon cross sections at MeV energies. Although it can be inferred using measurements of other processes \cite{Bhattacharya:1998hc,Bhattacharya:2009zz}, there is currently no consistent framework \cite{Karakoc:2014awa}. Direct measurements are thus highly important. For example, pion decay-at-rest source such as the Spallation Neutron Source (SNS) \cite{Bolozdynya:2012xv} are suitable for its well-understood spectrum. Currently measuring on lead targets \cite{Akimov:2015nza}, it can be adapted to other targets such as argon. 

\subsection{Liquid Scintillator} 
(300 $\bar{\nu}_e$ at KamLAND; 6,000 $\bar{\nu}_e$ at JUNO)\\
Like water, liquid organic hydrocarbons ($\mathrm{C}_n \mathrm{H}_{2n}$) contain large number of free protons that act as neutrino targets. The primary detection channel is IBD, but instead of Cerenkov radiation the scintillator emits photons in response to charged particle energy loss. Since the photon emission is mostly isotropic, there is little directional information. However, since the light yield is high, good event-by-event energy reconstruction and a low detection energy threshold can be achieved. Tagging by observing neutron capture on proton is feasible, although it is still often supplemented by doping (e.g., gadolinium). The low threshold also allows scintillators to have unique sensitivity to NC scattering on protons, which provides a robust and sensitive measure of heavy lepton neutrino spectra \cite{Beacom:2002hs,Dasgupta:2011wg}. Several large underground liquid scintillator detectors are currently online, including Borexino \cite{Monzani:2006jg}, KamLAND \cite{Eguchi:2002dm}, and LVD \cite{Agafonova:2007hn} and target masses of 0.3 kton, 1 kton, and 1 kton, respectively. JUNO \cite{An:2015jdp}, a 20 kton liquid scintillator detector, is currently under construction in Southern China, and LENA is a 50 kton proposal \cite{Wurm:2011zn}.

\subsection{Other technologies}
Heavy nuclei targets, such as iron or lead, have attractive high cross sections for MeV neutrino interactions \cite{Fuller:1998kb,Engel:2002hg}. Detector systems for leptons and ejected nucleons provide signal identification. Since the cross sections for single and double neutron ejection depends strongly on neutrino energy, their relative rates also yield some neutrino spectral information \cite{Vaananen:2011bf}. Lead is especially suitable for its ease of handling and low neutron capture cross section, which allows easier placement of neutron detection systems. The HALO detector \cite{Duba:2008zz}, located at SNO lab, currently utilized 76 tonnes of lead, and will detect neutrinos via $\nu_e + {}^{A}{\rm Pb} \to e^- + {}^{A}{\rm Bi}^*$ and $\nu_x + {}^{A}{\rm Pb} \to \nu_x + {}^{A}{\rm Pb}^*$. The $\bar{\nu}$ are suppressed by Pauli blocking due to lead's neutron excess. A future upgrade to HALO-2, with a kilotonne scale target, is being planned. 

Another anticipated detection channel is coherent elastic neutrino-nucleus scattering, the NC scattering of a neutrino with the entire nucleus \cite{Horowitz:2003cz}. The challenge is to detect the the small recoil energies (tens to hundreds of keV) of the nucleus. Fortunately, direct dark matter detectors provide just the required capabilities. In this case, supernova neutrinos prove a background for dark matter searches (especially the diffuse flux of supernova neutrinos which cannot be rejected based on timing), but next generation tonne-scale direct dark matter detectors are expected to yield useful flavor-blind information of a supernova neutrino burst \cite{Chakraborty:2013zua}. Dual-phase xenon experiments have several advantages over competing direct dark matter detection technologies: the large neutron number of xenon yields a large coherent scattering cross section, they are sensitive down to low (sub-keV) recoils, they have achieved low background rates, they have excellent timing resolution, and finally, there are several plans for scaled-up detectors, e.g., the $\sim 7$ tonnes XENONnT \cite{Aprile:2015uzo} and LZ \cite{Akerib:2015cja} detectors, and the DARWIN consortium \cite{Aalbers:2016jon} is exploring 40 tonnes of instrumented xenon.


\section{Supernova physics we might observe and how to get at them}\label{sec:SNphysics}

Hopefully by now it should be apparent to the reader the physics content of a supernova and its signal is huge, and detector capability is ready for high-statistics neutrino detection. Many fundamental questions about the evolution of the star that led it to collapse, the mechanism by which it subsequently exploded, and the properties of the neutrino could be asked and, hopefully, answered by the signal from the next supernova in the Milky Way. These include:
\begin{itemize}
\item Measuring the energy of the explosion.
\item Measuring the neutrino spectra at Earth.
\item Observing a second collapse due to phase transition during the first \textit{O}(10) seconds.
\item Observing whether a black hole formed during the first \textit{O}(10) seconds.
\item Probing BSM physics from deviations of the neutrino light curve from Standard Model physics.
\item Observing the progenitor structure.
\item Observing the shock stall and the duration of the accretion phase.
\item Observing the accretion of different compositional shells of the progenitor.
\item Determining neutrino properties, e.g., the mass ordering.  
\item Determining if there is any cooling rate evolution. 
\item Determining if the Standing Accretion Shock Instability occurs; if so, measuring its duration, amplitude and frequency; determining whether the Standing Accretion Shock Instability was important for causing the star to explode.
\item Determining the original neutrino spectra during the collapse and accretion phase.
\item Determining an athermal component.
\item Probing BSM physics from deviations of the neutrino spectra from Standard Model physics.
\item Observing the motion of the shock through the mantle. 
\item Determining flavor mixing effects due to self-interactions and turbulence.
\item Determining the original neutrino spectra during the cooling phase.
\item Determining the kind of nucleosynthesis occurred.
\item Observing a Lepton-Emission Self-sustained Asymmetry (LESA).
\item Determining the sphericity of the core collapse. 
\item Determining the angular momentum of the proto-neutron star.
\end{itemize}
The reader can probably think of many more questions that one may also hope to answer. Some of these questions are relatively easy to answer and are robust to uncertainties; others require much more detailed information about the supernova and cannot be tackled until this more fundamental data is provided. The list is ordered by how robust we feel they can be probed given present signal prediction uncertainties. However this is clearly a gross approximation and many can switch positions. 
In what follows we describe some current strategies for answering these questions.

\subsection{Flavour Changing Effects and the Original Spectra}
As equations (\ref{eq:Fe}) to (\ref{eq:Fbarmu}) show, the neutrino signal we receive is the combination of flavour changing effects and the neutrino spectra at the proto-neutron star. Many of the questions we wish to answer depend upon our ability to deconvolve them. Since both the survival probabilities and emitted spectra evolve with time, disentangling them to study each separately means finding periods when one or the other is relatively well understood. Assuming we know the flavor transformation, equations (\ref{eq:Fe}) to (\ref{eq:Fbarmu}) can be inverted to give the neutrino spectra at the neutrinosphere in terms of the observed flux:
\begin{eqnarray}
& \Phi_e  = \frac{4\pi d^2}{1-3p} \left[ \left(1-p\right) \,\left( F_{\mu} + F_{\tau} \right) - \left(1+p\right)\,F_e \right] \label{eq:Phie}\\
& \Phi_x = \frac{4\pi d^2}{1-3p} \left[ \left(1-p\right)\,F_e -p \,\left( F_{\mu} + F_{\tau}\right) \right] \label{eq:Phimu}  \label{eq:Phix}\\
& \bar{\Phi}_e  = \frac{4\pi d^2}{1-3\bar{p}}\left[ \left(1-\bar{p}\right) \,\left( \bar{F}_{\mu} + \bar{F}_{\tau} \right) - \left(1+\bar{p}\right)\,\bar{F}_e \right]  \label{eq:Phibare}\\
& \bar{\Phi}_x = \frac{4\pi d^2}{1-3\bar{p}}\left[ \left(1-\bar{p}\right)\,\bar{F}_e - \bar{p} \,\left( \bar{F}_{\mu} + \bar{F}_{\tau} \right) \right]  \label{eq:Phibarx}.
\end{eqnarray}
The location, energy and time dependence of the spectra, fluxes and survival probabilities have been omitted for the sake of clarity. Note how these equations are singular when $p$ or $\bar{p}$ are equal to $1/3$. Assuming we know the emitted spectra, equations (\ref{eq:Fe}) to (\ref{eq:Fbarmu}) can be rearranged to give the electron neutrino/antineutrino survival probabilities via
\begin{eqnarray}
& p = \frac{  {4\pi d^2} \,\left[ 2\,F_e - F_{\mu} - F_{\tau} \right] }{ 3\left[ \Phi_e - \Phi_x \right] } +\frac{1}{3}\label{eq:pe2} \\
& \bar{p} = \frac{ {4\pi d^2} \,\left[ 2\,\bar{F}_e - \bar{F}_{\mu} - \bar{F}_{\tau} \right] }{ 3\left[ \bar{\Phi}_e- \bar{\Phi}_x \right]} + \frac{1}{3}\label{eq:pebar2}.
\end{eqnarray}
Note how these formulae become singular when the emitted spectra are equal for a given energy. 

With sufficient statistics and detector time/energy resolution one could obtain the emitted spectra and/or survival probabilities in a parameter-independent fashion. Such an approach would be the better choice for the accretion phase when the emitted neutrino spectra are a blend of different spectra across the PNS. In practice, unless the next Galactic supernovae is very close, one will have to integrate over energy and time windows that are substantially larger than both the time and energy resolutions of the detectors, and the time and energy scales for variations of the emitted spectra and survival probabilities. Both integrations generally make it more difficult to obtain the emitted spectra and survival probabilities in a parameter-independent fashion without judicious choices for the time and energy bins used in the analysis. Whatever time and energy windows are chose, we define a time/energy window averaged flux for each flavor, $\langle F \rangle$, and a time/energy window averaged emitted spectra, $\langle \Phi \rangle$ as
\begin{eqnarray}
& \langle F \rangle(t,E)  = \frac{1}{\Delta t\, \Delta E}\int_{t-\Delta t/2}^{t+\Delta t/2}\int_{E-\Delta E/2}^{E+\Delta E/2} F(t,E)\,dt\,dE\\
& \langle \Phi \rangle(t,E)  = \frac{1}{\Delta t\, \Delta E}\int_{t-\Delta t/2}^{t+\Delta t/2}\int_{E-\Delta E/2}^{E+\Delta E/2} \Phi(t,E)\,dt\,dE.
\end{eqnarray}
Up until the cooling epoch it is expected that the self-interaction and turbulence effects are suppressed leaving just the static MSW effect which is purely adiabatic and independent of energy and time. Thus we can use the adiabatic formulae for $p$ and $\bar{p}$ given in table (\ref{tab:p,pbar}) in equations (\ref{eq:Phie}) to (\ref{eq:Phibarx}) and replace the flux and spectra with their time/energy bin averages. While we will lose some fine time and energy structures of the actual spectra---such as the neutronization burst---the procedure is relatively straightforward.

During the cooling phase the emitted spectra become simpler with the expectation that the same neutrino spectra are emitted from all points on the neutrinosphere because the accretion is now absent. One might then be tempted to parameterize the emitted spectra---luminosities, mean energies, and pinch parameters are one obvious set of spectral parameters---and use the data to fit the parameters. This was  done with the SN 1987A data by Loredo and Lamb, and \cite{2002PhRvD..65f3002L} and Costantini, Ianni and Vissani \cite{2004PhRvD..70d3006C}; see Pagliaroli \textit{et al.} \cite{Pagliaroli:2009qy} and Y\"uksel and Beacom \cite{2007PhRvD..76h3007Y} for more recent versions. However one finds the number of parameters to describe the spectra quickly becomes rather large: if we use luminosities, mean energies, and pinch parameters as the spectral parameters and treat the electron neutrinos, electron antineutrinos and heavy lepton neutrinos/antineutrinos independently, then we need nine parameters; if we add a cooling timescale - which could be different for each flavor - we add one to three more; if we try to account for time evolution of the luminosities, mean energies, and pinch parameters then we add at least nine more parameters. One could try to reduce this number by approximating some parameters as equal to each other or constant in time based upon the results of simulations, e.g., as shown in figure (\ref{fig:L,meanE}). Several studies which have used luminosities, mean energies, and pinch parameters as their spectral parameter sets often find significant degeneracies between them and even systematic offsets \cite{2008JCAP...12..006M,2008arXiv0802.1177S,2017arXiv171100008N}. However, applying cuts to the data---for example, a forward angular cut around the arrival direction of the neutrinos---can improve the statistical significance of the parameter derivation \cite{Laha:2013hva}. Only by combining the data from multiple detectors as well as better detector channel separation will we be able break the degeneracies and have reliable measurements of parameters.

Complicating matters further is that during the cooling phase self-interaction and turbulence effects appear and the MSW effect becomes dynamic. Thus the survival probabilities become, in general, energy and time dependent. One can define effective survival probabilities for each time/energy bin by,
\begin{eqnarray}
&  p_{eff}(t,E) \langle \Phi \rangle(t,E)  = \frac{1}{\Delta t\, \Delta E}\int_{t-\Delta t/2}^{t+\Delta t/2}\int_{E-\Delta E/2}^{E+\Delta E/2} p(t,E) \Phi(t,E)\,dt\,dE \\
&  \bar{p}_{eff}(t,E) \langle \Phi \rangle(t,E)  = \frac{1}{\Delta t\, \Delta E}\int_{t-\Delta t/2}^{t+\Delta t/2}\int_{E-\Delta E/2}^{E+\Delta E/2}\bar{p}(t,E) \bar{\Phi}(t,E)\,dt\,dE, 
\end{eqnarray}
so that the measured time/energy window integrated flux for each flavor, $\langle F \rangle$, is related to the time/energy window integrated emitted spectra, $\langle \Phi \rangle$ by, 
\begin{eqnarray}
& \langle F_e\rangle = \frac{1}{4\pi d^2} \left[ p_{eff}\,\langle\Phi_e\rangle + \left(1-p_{eff}\right)\,\langle\Phi_x\rangle \right] \\
& \langle F_{\mu} + F_{\tau}\rangle = \frac{1}{4\pi d^2} \left[ (1-p_{eff})\,\langle\Phi_e\rangle + \left(1+p_{eff}\right)\,\langle\Phi_x\rangle \right] \\
& \langle \bar{F}_e\rangle = \frac{1}{4\pi d^2} \left[ \bar{p}_{eff}\,\langle\bar{\Phi}_e\rangle + \left(1-\bar{p}_{eff}\right)\,\langle\bar{\Phi}_x\rangle \right]\\
& \langle \bar{F}_{\mu} + \bar{F}_{\tau} \rangle = \frac{1}{4\pi d^2} \left[ \left(1-\bar{p}_{eff}\right)\,\langle\bar{\Phi}_e\rangle + \left(1+\bar{p}_{eff}\right)\,\langle\bar{\Phi}_x\rangle \right].
\end{eqnarray}
If the survival probabilities were more-or-less constant across the time/energy bin then $p_{eff}\approx \langle p \rangle$, 
$\bar{p}_{eff}\approx \langle \bar{p} \rangle$ where $\langle p \rangle$ and $\langle \bar{p} \rangle$ are, 
\begin{eqnarray}
& \langle p \rangle(t,E) = \frac{1}{\Delta t\, \Delta E}\int_{t-\Delta t/2}^{t+\Delta t/2}\int_{E-\Delta E/2}^{E+\Delta E/2} p(t,E)\,dt\,dE\\
& \langle \bar{p} \rangle(t,E)  = \frac{1}{\Delta t\, \Delta E}\int_{t-\Delta t/2}^{t+\Delta t/2}\int_{E-\Delta E/2}^{E+\Delta E/2} \bar{p}(t,E)\,dt\,dE.
\end{eqnarray}
This may apply during some time intervals for certain progenitors. For example, if the dynamic MSW and turbulence are absent---see section \S\ref{sec:shock motion} for further discussion on why that can occur---but self-interactions are present such that the flavor survival probabilities looked like those in figure (\ref{fig:WuMA}), one could imagine smoothing the probabilities over $\Delta t \sim 0.5\;{\rm s}$, $\Delta E \sim 10\;{\rm MeV}$ windows and still retaining much of the structure of the flavor transformation in energy and time. 
But if $p$ and $\bar{p}$ vary wildly over the time/energy bin---as seen in figure (\ref{fig:matterplusturbulence})---then one would expect integrating over any $\Delta E \sim 10\;{\rm MeV}$ window would give $p_{eff}$ and/or $\bar{p}_{eff}$ that is close to the median between the limit of the fluctuations. Unfortunately, as we showed earlier, the limits of both $p$ and $\bar{p}$ are $0.021 \lesssim p,\bar{p} \lesssim 0.678$ so one might expect $p_{eff}\approx 0.33$ and/or $\bar{p}_{eff}\approx 0.33$, right on the $p = 1/3$, $\bar{p}=1/3$ limit where inversion is not possible. Extracting the survival probabilities during the cooling phase faces the additional challenge that the emitted spectra become similar at late times, as shown in figure  (\ref{fig:L,meanE}). The task of extracting the original spectra and flavour changes from an observed signal during the cooling phase is a subject that needs much more attention than it has so far received.

\subsection{The progenitor}\label{sec:physics:progenitor}
Knowledge about the progenitor is useful for two reasons: it allows us to test our theories about the late stages of stellar evolution, and it also allows us to determine whether current ideas of the link between pre-collapse structure and successful / failed supernovae are correct \cite{2015arXiv151102820P}. There are multiple pathways to determine the progenitor. The most obvious is pre-supernova images. A high fraction of likely progenitors, some $92$\%, is predicted to have already been observed by 2MASS, a large field of view survey operating in the infrared bands where dust extinction is mild \cite{2013ApJ...778..164A}. The situation is less favorable in the optical bands where the Galactic plane introduces enormous dust extinction. Estimated detection is around 57\% with the USNO-B1.0 catalog assuming a sensitivity limit of $m_{V,lim} = 21.0$ \cite{2013ApJ...778..164A}. Both of these estimates assume no bright nearby object, the presence of which would further limit identification due to confusion. The lack of optical imaging will limit the identification of the progenitor temperature and luminosity. The use of future wide field of view surveys, e.g., LSST, will improve detection prospects to some 66\% detection (assuming observable magnitude of $m_{V,lim}<24.5$). Detailed analyses can be found in Adams \textit{et al.}~\cite{2013ApJ...778..164A}. 

If the star is sufficiently close then the pre-supernova neutrinos themselves provide us with information about the progenitor. Kato \textit{et al.} \cite{Kato:2015faa} predict a total number of events ranging from $\sim 7-61$ depending upon the progenitor mass and neutrino mass ordering for SuperK from an Fe-core pre-supernova star at a distance of 200 pc. Their calculations for the number of events seen in the JUNO detector for a pre-supernova at the same distance ranged from 189 to 864. The KamLAND collaboration also calculated the event rate in their detector and found they could detect pre-supernova neutrinos from a $25~M_{\odot}$ star at 3-$\sigma$ significance if it were closer than 690 pc depending upon the neutrino mass ordering \cite{2016ApJ...818...91A}. The time structure of the pre-supernova emission was considered by Yoshida \textit{et al.} \cite{2016PhRvD..93l3012Y} who showed that the neutrino emission drops during the oxygen and silicon shell burning phases. KamLAND, JUNO, and HyperK would be able to determine the time structure of the nuclear burning if the supernova is closer than 200 pc. The expected number of neutrino events from their $15\;{\rm M_{\odot}}$ model at $d=200$ pc in a Gadolinium doped Hyper-K prior to the explosion is shown in figure (\ref{fig:yoshidapresupernova}). 
\begin{figure*}[t!] 
\includegraphics[width=0.7\linewidth,angle=270]{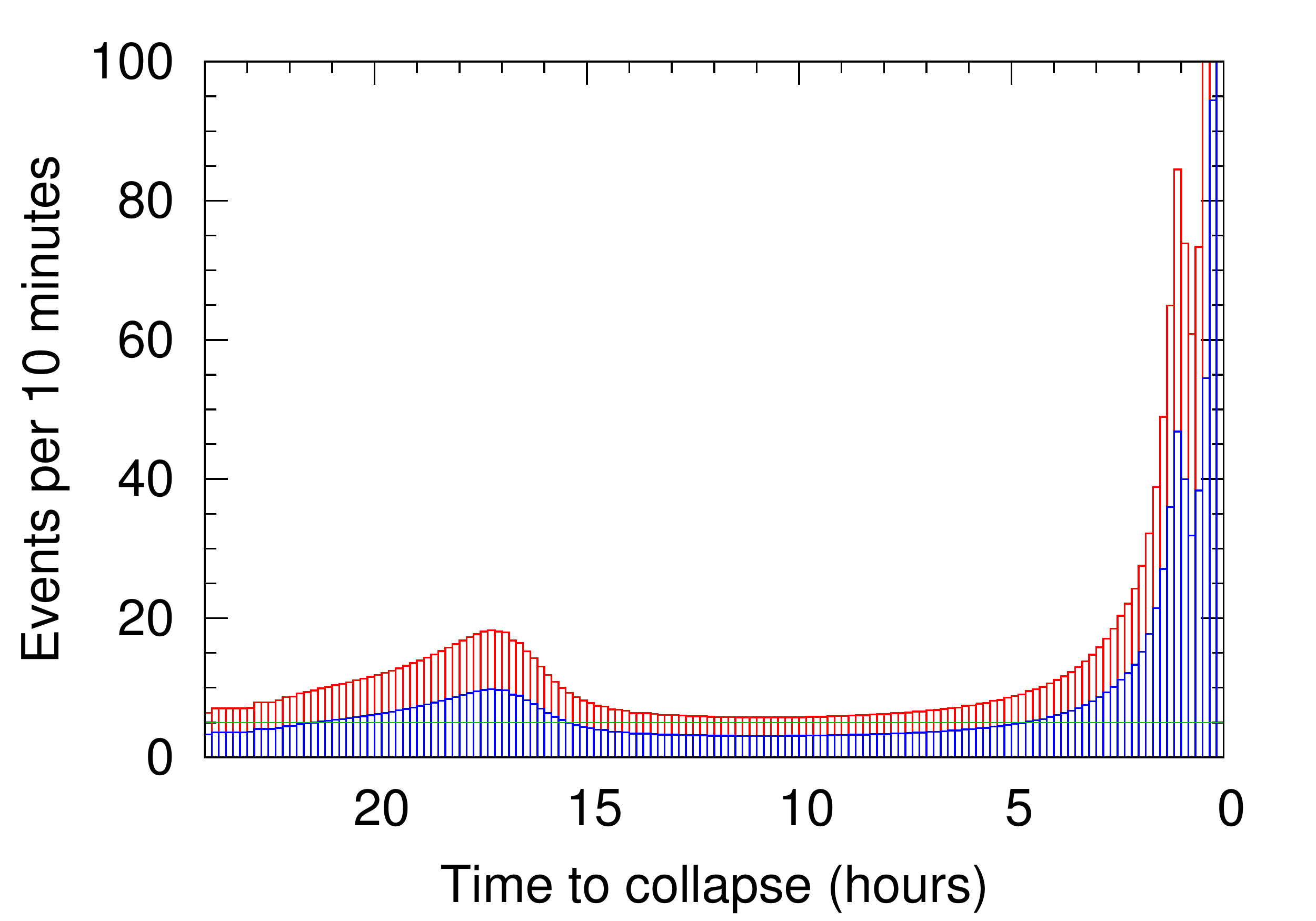}
\caption{Expected neutrino events per ten minutes by Hyper-K with Gd in 24 hours prior to a SN explosion of the $15\;{\rm M_{\odot}}$ model by Yoshida \textit{et al.} \cite{2016PhRvD..93l3012Y}. Red and blue bins indicate the cases of the normal and inverted mass orderings. The green horizontal line is the event number for the unit interval of the significance more than 3$\sigma$. \label{fig:yoshidapresupernova} }
\end{figure*}
As pointed out earlier, the spectrum due to thermal processes peaks at energies around $E_{\nu} \sim 1\;{\rm MeV}$ but the spectrum due to weak processes can extend the spectrum up to $E_{\nu}\sim 10\;{\rm MeV}$. While subordinate in terms of the amount of energy they remove from the core (except in the last few hours of a star's life) it is expected the event rates due to weak processes dominate if detector thresholds are set at $\sim 2\;{\rm MeV}$ \cite{2015arXiv151102820P}.

There is also information about the progenitor from the neutrinos emitted during the collapse phase. As described in Section \ref{sec:accretion:progenitor}, the neutrino emission derives partly from the proto-neutron star cooling and partly from the mass accretion. The mass accretion in turn depends on the progenitor. Thus, the evolution of the neutrino signal holds information of the progenitor's core. Within the delayed neutrino mechanism, the shock revival takes several hundred milliseconds post bounce, after which mass accretion largely halts. Thus, the neutrino signal typically probes the inner several solar masses of the progenitor. This makes it highly complementary to the other probes of the progenitor. 

The neutrino emission during the first tenths of a second should also be able to distinguish a ONeMg supernova from an Iron supernova because the ONeMg supernova does not linger in an accretion phase. A very swift transition to neutrino cooling in the observed signal would strongly imply a ONeMg progenitor. In addition, there are reasons to believe the flavor transformation for ONeMg supernovae may continue for a longer period than in an Fe core case. The neutrino luminosities at late times for ONeMg supernovae are not very different from Fe core supernovae but the matter density outside the proto-neutron star falls away much more rapidly. This allows $H_{SI}$ to dominate over $H_M$ for a longer period and may be another way to distinguish these kinds of supernovae, but this has not been shown. 

Finally, information about the progenitor can also be extracted from the first electromagnetic signal that arises when the supernova shock passes through the stellar photosphere. Neutrinos have a role to play in better securing their detection, and we focus more upon this point in Section \ref{sec:multimessenger}. 

\subsection{Bounce time}\label{sec:physics:bouncetime}
The time of core bounce can be estimated reliably for core collapses inside our Galaxy. The expected distance distribution of Galactic core collapse falls off beyond $\sim 20$ kpc (Figure \ref{fig:SNdistribution}) and even for this upper value of 20 kpc, the largest existing neutrino detectors, Super-K and IceCube, both have sufficient sensitivity to measure the bounce time. By modeling the anti-neutrino flux rise by a parametrized function of the form $\propto (1-e^{-t/\tau_r})$, and fitting at the same time the astrophysical parameters, Pagliaroli \textit{et al.} \cite{Pagliaroli:2009qy} show that Super-K will be able to infer the bounce time to within a few tens of milliseconds. Halzen \& Raffelt \cite{Halzen:2009sm} showed how IceCube will achieve a $\pm 6$--7 msec measurement for a comparable core collapse at 20 kpc, and for a typical distance of 10 kpc, to $\pm 3.5$ msec, as shown in Figure \ref{fig:bouncetime}. As we summarize in Section \ref{sec:multimessenger}, the bounce time opens unique opportunities for coincidence measurements between neutrino and gravitational wave signals from stellar core bounce.
\begin{figure*}[t!] 
\includegraphics[width=\linewidth]{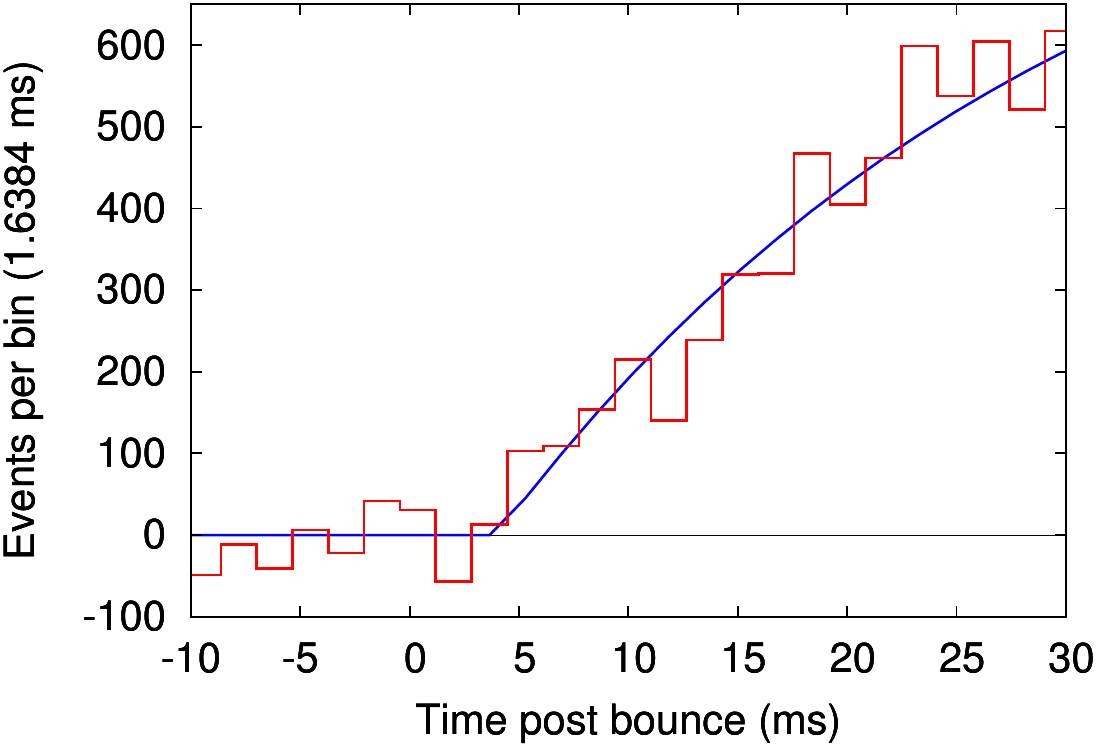}
\caption{Monte Carlo realization (red histogram) and reconstructed fit (blue line) for the photon counts per time bin at IceCube, based on the benchmark core-collapse distance of 10 kpc and using the simulation of the Garching group \cite{2009A&A...496..475M}. The figure is taken from Halzen \& Raffelt \cite{Halzen:2009sm}\label{fig:bouncetime}}
\end{figure*}

\subsection{The SASI}
\begin{figure*}[t] 
\includegraphics[width=0.45\linewidth]{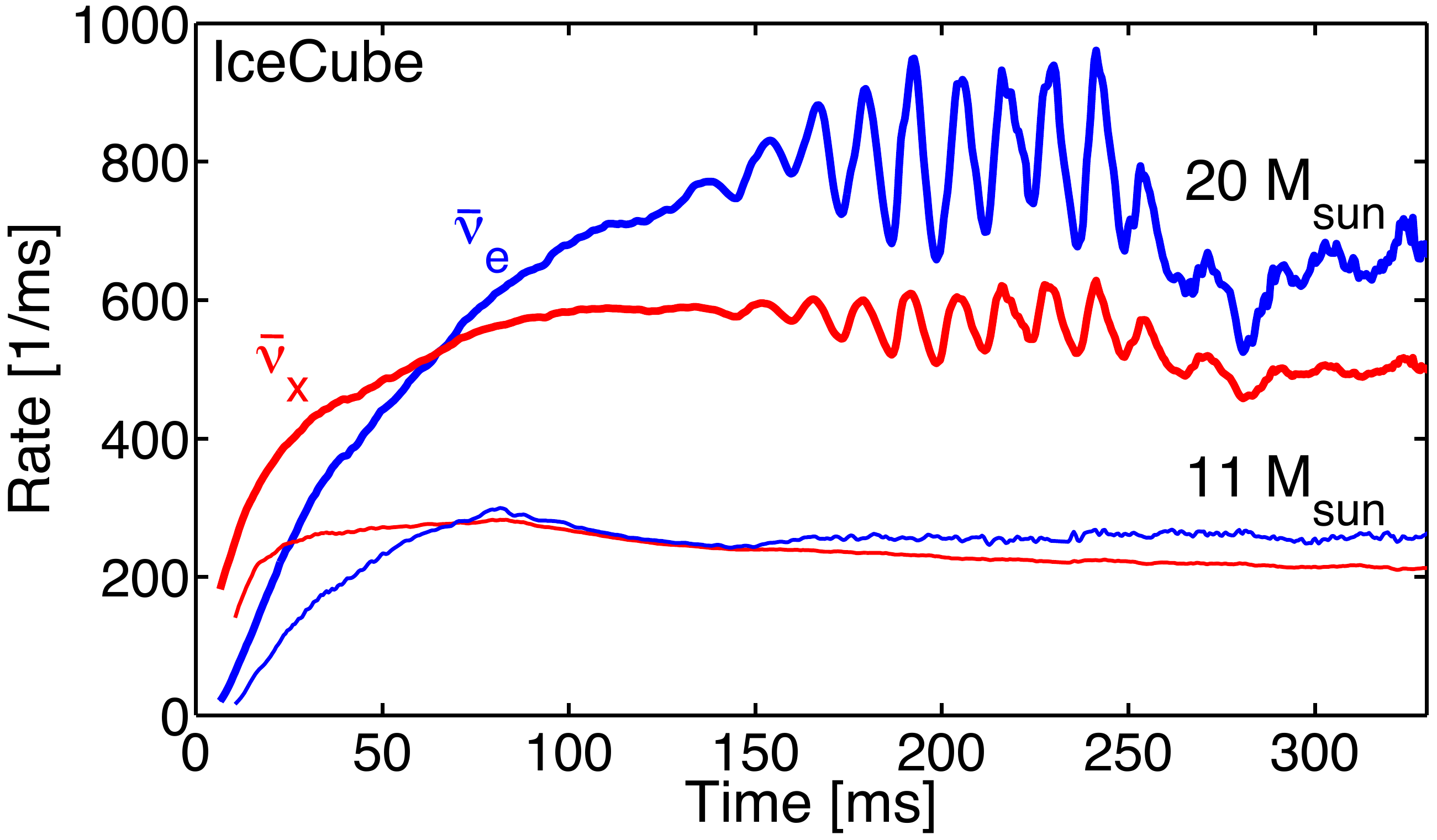}
\includegraphics[width=0.45\linewidth]{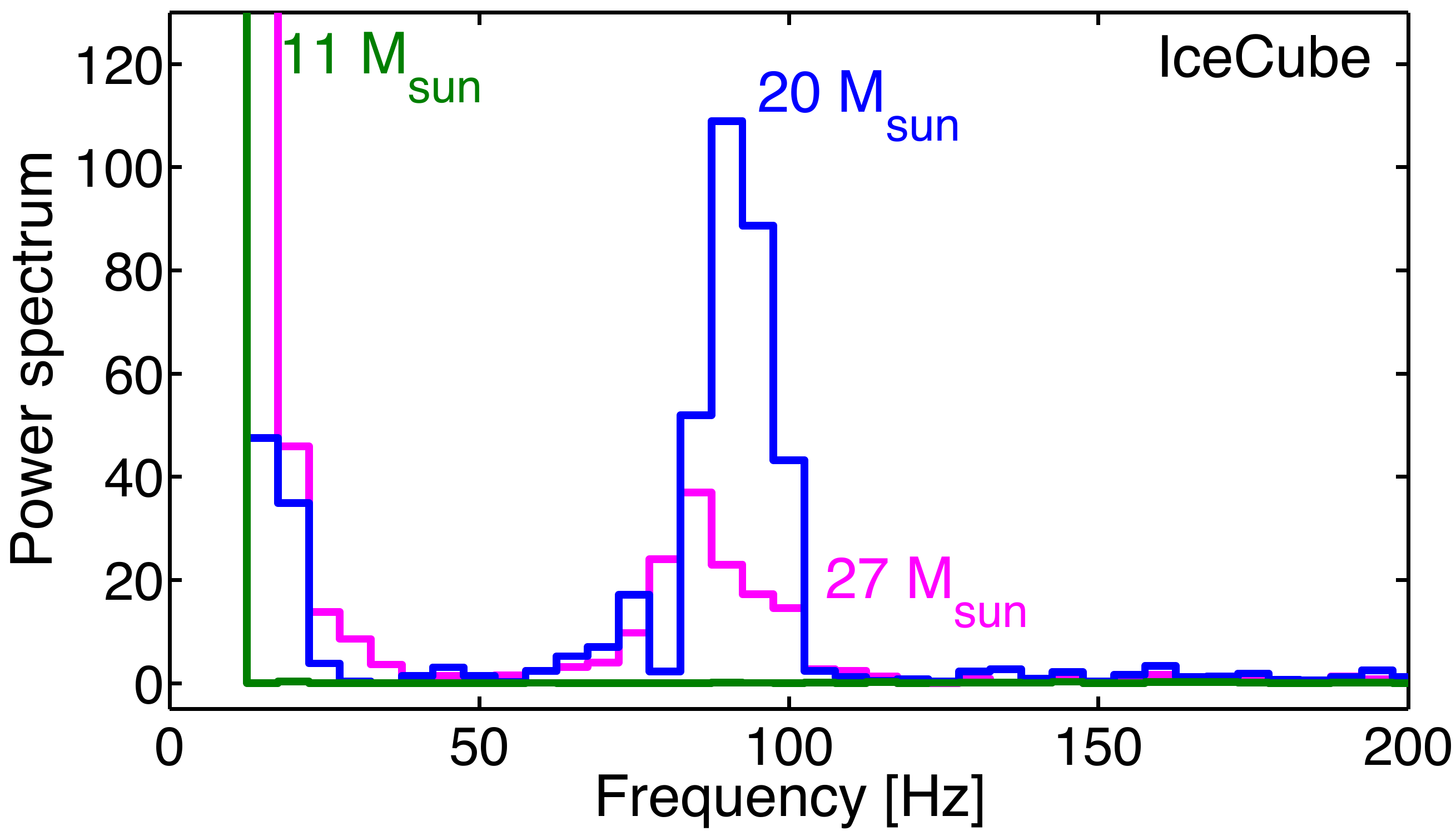}
\caption{ In the left panel is shown the event rate in IceCube for the optimal observing direction to a $11.2\;{\rm M_{\odot}}$ and a $20\;{\rm M_{\odot}}$ model at $10\;{\rm kpc}$ as a function of postbounce time. In the right panel is the power spectrum of the IceCube event rate for the interval 100--€"300 ms for three progenitors, assuming a distance to the supernova of $10\;{\rm kpc}$. The power spectrum is normalized to the shot noise caused by the Ice Cube background rate of $1.48 \times 10^{3} /{\rm ms}$. Both figures are taken from Tamborra \textit{et al.} \cite{Tamborra:2013laa} and reprinted with permission from APS.\label{fig:Tamborra34}}
\end{figure*}
The Standing Accretion Shock Instability should also be relatively straight-forward to detect. Several studies have focused upon this expected feature of the explosion beginning with Lund \textit{et al.} \cite{Lund:2010kh,Lund:2012vm} and more recently Tamborra \textit{et al.} \cite{Tamborra:2013laa}. The frequency of the fluctuations in the event rate in both IceCube and HyperK are directly related to the frequency of the sloshing of the SASI. However the amplitude of the fluctuations is line-of-sight dependent so if we are unlucky they may not lead to detectable fluctuations of the neutrino flux. The existence of a SASI phase is progenitor dependent with higher compactness progenitors producing a SASI and low compactness progenitors none. The detection of a SASI and the progenitor structure would be a good test of our understanding of the dynamics of supernovae. An example of the expected oscillations of the event rate due to SASI in IceCube and the strength of the signal in Fourier space are shown in figure (\ref{fig:Tamborra34}) which are taken from Tamborra \textit{et al.} \cite{Tamborra:2013laa}.

\subsection{The explosion energy}
\begin{figure*}[b!] 
\includegraphics[width=\linewidth]{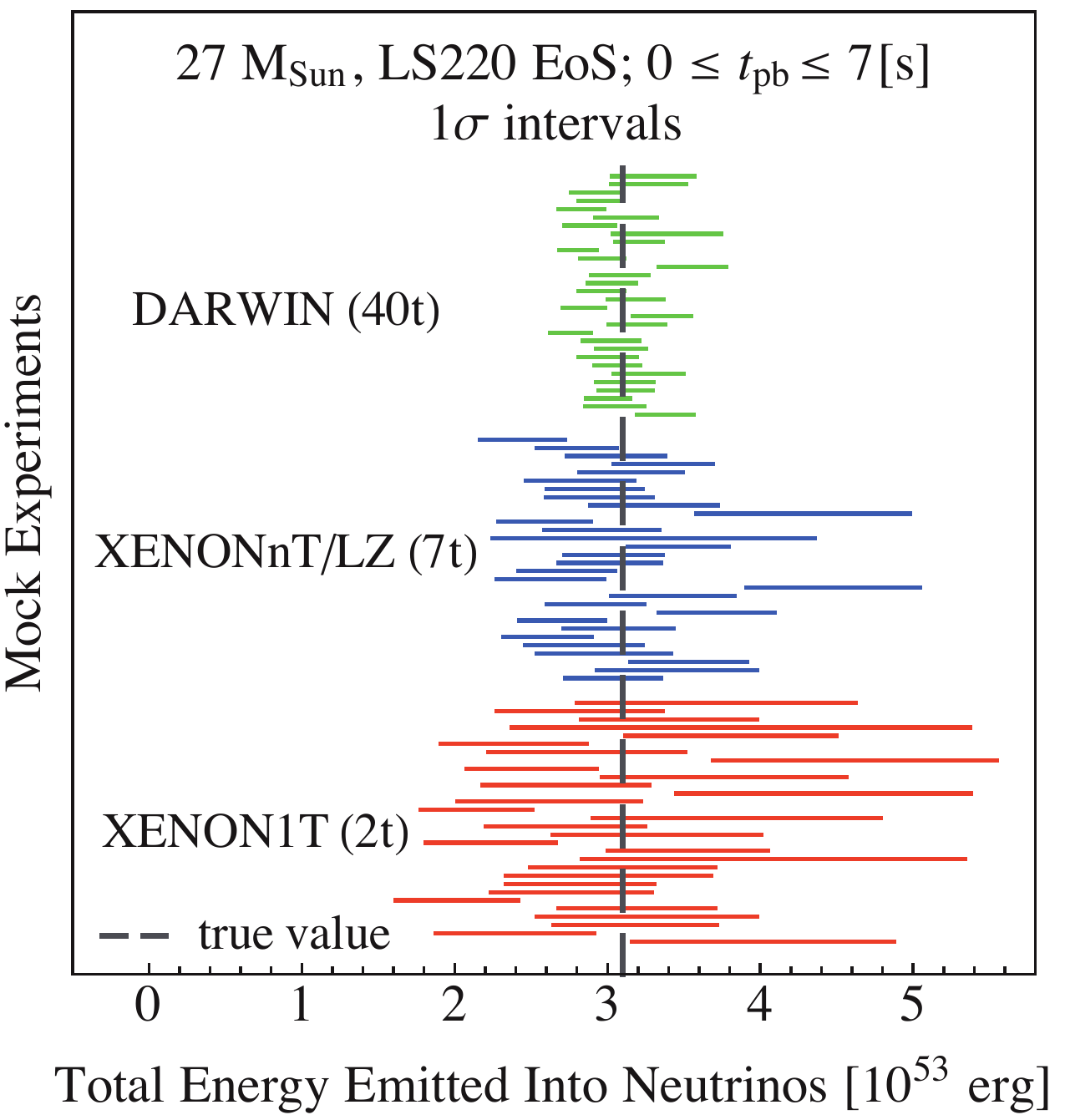}
\caption{The reconstructed 1-$\sigma$ band of the total energy emitted in neutrinos in 30 mock experiments for each of XENON1T (red), XENONnT/LZ (blue) and DARWIN (green).  The true value for the energy emitted in the simulation between $0 \leq t_{pb} \leq 7\;{\rm s}$ of a $27\;{\rm M_{\odot}}$ using the LS220 EOS \cite{1991NuPhA.535..331L} is shown the dashed vertical line. The figure is taken from Lang \textit{et al.} \cite{2016PhRvD..94j3009L} and reprinted with permission from APS.\label{fig:Langetal8}}
\end{figure*}
The explosion energy would be of great interest. The cleanest method for measuring the total energy emitted is to use the neutral current events. The coherent elastic neutrino-nucleus scattering has a large cross section but for an incoming neutrino of energy $E\sim O(10\;{\rm MeV})$, the energy of the recoiling nucleus is only $E\sim O(1\;{\rm keV})$. Nevertheless, tonne scale dark matter detectors can measure it. An example of how well the total energy can be measured is shown in figure (\ref{fig:Langetal8}) which is taken from Lang \textit{et al.} \cite{2016PhRvD..94j3009L}. The figure shows how a 2 tonne liquid Xenon detector can pin down the total emitted energy to a precision of $20\%$ depending upon details such as the progenitor mass and the EOS, which improves to a precision of $5\%$ for a 40 tonne experiment such as DARWIN.

\subsection{Shock motion}\label{sec:shock motion}

The motion of the shock through the envelope of the supernova leads to a time and energy dependent change in the survival probabilities $p$ and $\bar{p}$. The time/energy dependence is always for lower energies to be affected before higher energies for the simple reason that the resonance densities are inversely proportional to the energy, the density profile of the envelope is a monotonically decreasing function of the radius, and the star explodes from the inside out. For a NMO only the the electron neutrino survival probability $p$ is affected: for an IMO both neutrino and antineutrinos are affected by the shock with the antineutrinos affected before the neutrinos because the H resonance, as the name implies, is at high density and the L resonance is at low densities. The expected changes in $p$ and $\bar{p}$ are given by a combination of the cases listed in table (\ref{tab:p,pbar}). At early times the evolution would be adiabatic NMO-1) or IMO-1) but when the shock arrives at the H-resonance density for a given neutrino energy, the survival probability $p$ or $\bar{p}$ would change to that given in NMO-2) or IMO-2a) then IMO-2b) if we ignore self-interactions. If we include self-interactions then we have to apply NMO-4a) / NMO-4b) or IMO-4) / IMO-5) averaged over time and energy windows. 
There are many ways one could look for shock effects in a detector. The most obvious is to look for the change in $\langle p \rangle$ or $\langle \bar{p} \rangle$ in a given energy window as a function of time, or a ratio of event rates in two energy bins \cite{2005JCAP...04..002F}. 
\begin{figure*}[b!] 
\includegraphics[width=\linewidth]{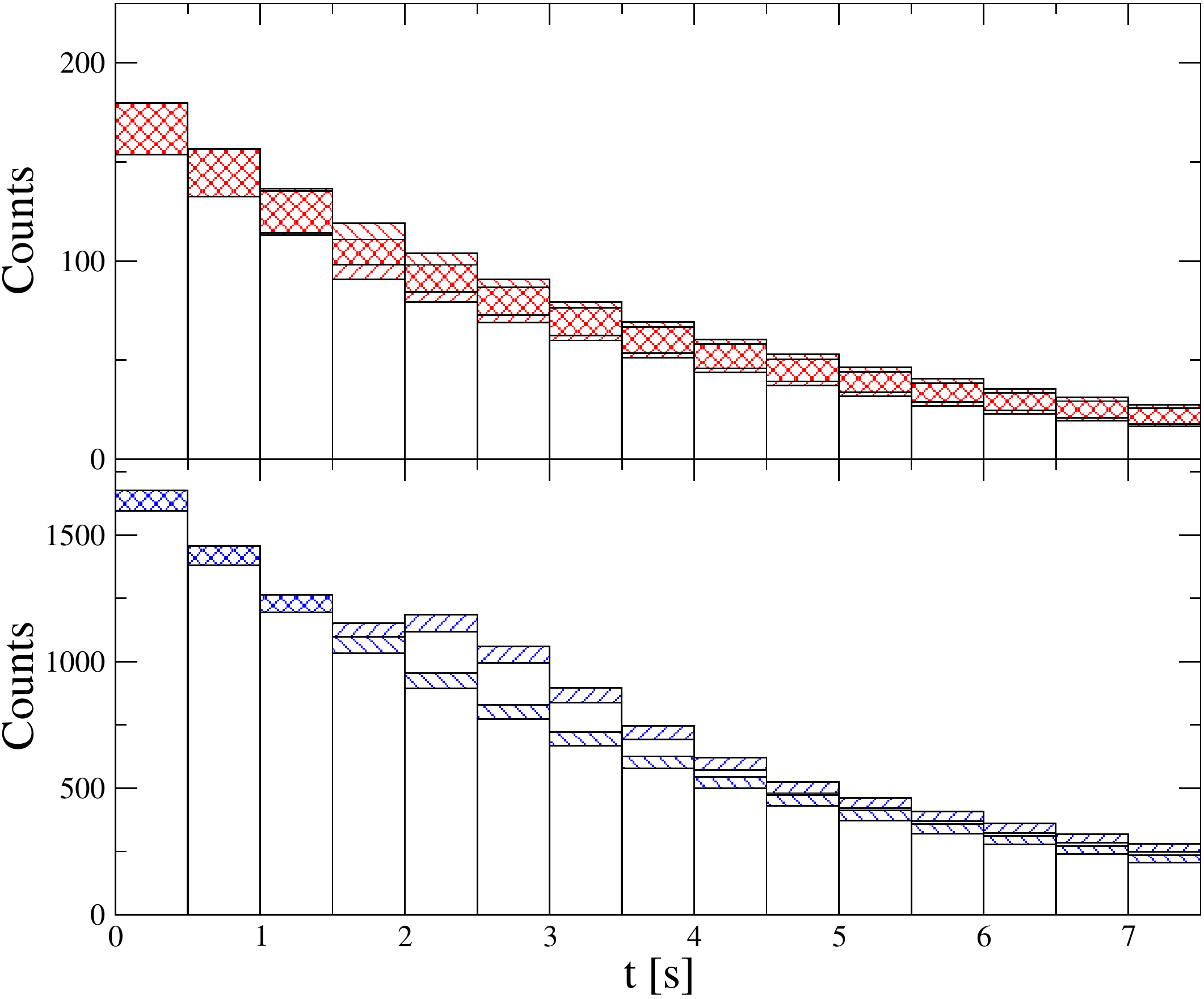}
\caption{The positron event rate in  SuperKamiokande in $0.5\;{\rm s}$ time bins and in two energy bins: $10$ to $19$ MeV (upper panel) and above $25$ MeV (lower panel). An
exponential decay calibrated on the first two time bins is shown for comparison. The passage of the shock through the mantle of the supernova leads to a decrease in the event rate (compared to the pure exponential) for the lower energy bin, and an increase in the event rate for the higher energy bin. The size of the boxes indicate the $1-\sigma$ Poisson error. The figure is taken from Gava \textit{et al.} \cite{2009PhRvL.103g1101G}.\label{fig:Gava4}}
\end{figure*}
An example is shown in figure (\ref{fig:Gava4}) which is taken from Gava \textit{et al.} \cite{2009PhRvL.103g1101G} (some times denoted as GKVM) who computed the positron event rate in SuperK as a function of time in two energy bins - $10\;{\rm MeV} \leq E_{\bar{e}} \leq 19\;{\rm MeV}$ and $E_{\bar{e}} \leq 25\;{\rm MeV}$ - for an IMO, a given set of input luminosities and mean energies, and density profiles taken from a hydro simulation. In their calculation, the arrival of the shock and the assumed spectra means the number of positron events should decrease in the low energy bin and increase in the higher. While turbulence was not included in their calculation, there were Phase Effects which have more-or-less the same effect. As the figure shows, the change from an exponential decay in the event rate starts to appear $\sim 1.5\;{\rm s}$ and lasts for several seconds. The change is significant at several sigma. That the change appears in the positron event rate indicates the mass ordering used must have been inverted.   

As we have already mentioned, the density profile of the progenitor determines how quickly the shock arrives in the H resonance region and how quickly it sweeps through the spectrum. A shock feature that arrives within a second and last no more than a second strongly hints at an ONeMg progenitor. 
If no shock features are seen within the neutrino signal then the envelope of the supernova is very bloated.

\subsection{Phase Transitions and Black hole formation}
\begin{figure*}[b!] 
\includegraphics[width=\linewidth]{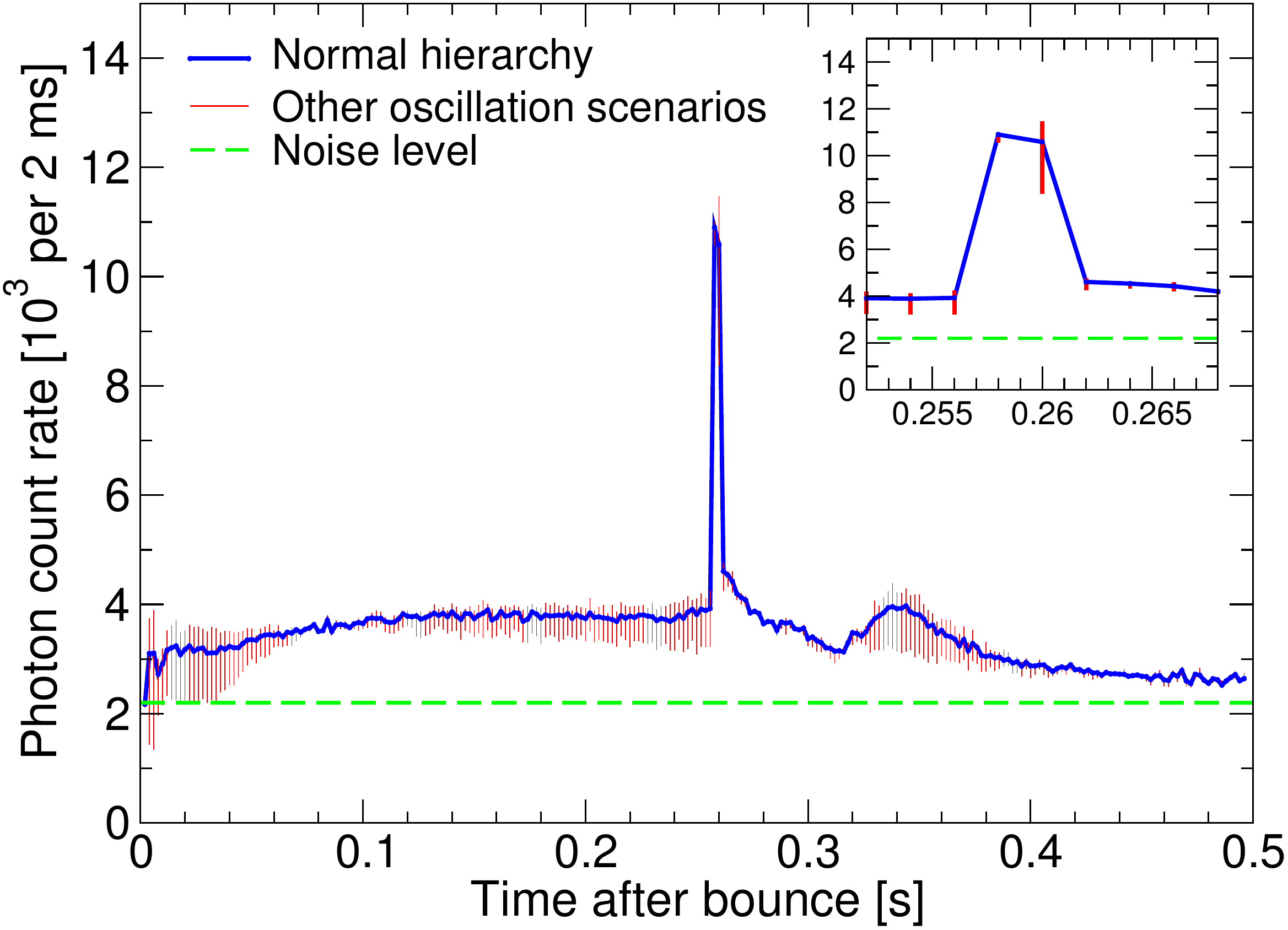}
\caption{Photon count rates at IceCube for the neutrino emission from a core collapse undergoing QCD phase transition at t $\sim 257$--261 ms, based on the numerical calculations of Sarget \textit{et al.} \cite{2009PhRvL.102h1101S}. The green dashed line denotes the ambient noise level at IceCube, and the range shown in red indicate the uncertainty due to beyond MSW oscillation scenarios. The inset shows the second burst blown up, in the same axis units. The figure is taken from Dasgupta \textit{et al.} \cite{Dasgupta:2009yj}.\label{fig:nuebarBurst}}
\end{figure*}
A secondary collapse of the proto-neutron star to a proto-exotic star generically leads to the formation of a second burst of neutrinos similar to the neutronization burst. However this second burst differs from the neutronization burst in that it is a burst in all flavors dominated by electron antineutrinos, while the neutronization burst is dominated by electron neutrinos. In addition to the second burst, the second collapse changes the characteristic of the neutrino spectra. Sarget \textit{et al.} \cite{2009PhRvL.102h1101S} find the mean energies of all flavors increases after the second collapse. The features of the second burst contain information of dense matter: the time width of the burst reveals the time for the shock to cross the hadronic crust, while the decay of neutrino emission after the burst is related with the compressibility of the remnant matter \cite{2009PhRvL.102h1101S,Dasgupta:2009yj}. If the second collapse occurs during the first $O$(10) seconds of the initial collapse, as is predicted in the parameter descriptions adopted in Ref.~\cite{2009PhRvL.102h1101S}, a straightforward method of corroborating the second collapse is to look for a second shock feature \cite{Dasgupta:2009yj}. An example as seen with IceCube is shown in Figure \ref{fig:nuebarBurst}. If the second collapse occurs before the original shock formed by the bounce of the PNS was relaunched or had gotten very far, the two may merge into one shock \cite{1993ApJ...414..701G,2009PhRvL.102h1101S}.

The observation of black hole formation in the neutrino burst signal is a feature that is expected to have a clear signature of a sudden cutoff in the neutrino emission. Secondary features of the neutrino signal such as the increased luminosity and mean energies of the neutrinos would corroborate this conclusion. The increase in luminosity and mean energies just before formation of the black hole mean detectors are able to detect neutrinos from failing supernovae at farther distances than from successful supernovae. Yang and Lunardini \cite{2011PhRvD..84f3002Y} find a megatonne water Cherenkov detector can detect a few events from a failed supernova at distances up to $d\sim 4-5\;{\rm Mpc}$. The occurrence rate of supernovae makes a rapid jump at this distance due to the inhomogeneity of cosmological structure. While the rate of core-collapse in our Milky Way is a few per century \cite{Diehl:2006cf}, the cumulative rate within 5 Mpc is some $\sim 1$ per year \cite{Ando:2005ka,Horiuchi:2013bc} and likely higher given the lack until recently of full-sky high-cadence monitoring for transients in the local Universe. 

\subsection{The mass ordering}
There are multiple ways the mass mass ordering can be determined because all the flavor transformation effects are seen to be sensitive to the mass ordering \cite{2015PTEP.2015f3B01C}. The cleanest signal of the ordering occurs during the pre-supernova phase when the emission does not fluctuate and the neutrino evolution is adiabatic. Unfortunately, the neutrino luminosity is somewhat low and the star would have to be close in order to detect sufficient number of events. The neutrino luminosity increases considerably during the collapse phase without large fluctuations or asphericity and the flavor evolution remains adiabatic. The neutronization burst occurs during this phase and is a nice feature of supernova to exploit because there are few $\mu$ or $\tau$ neutrinos emitted during the burst and because the burst close to being a standard candle for supernovae since it is almost independent of the progenitor \cite{PhysRevD.71.063003}. Using table (\ref{tab:p,pbar}) and the Particle Data Group evaluation of the mixing parameters \cite{Olive:2016xmw}, the probability that an electron neutrino emitted during the burst is detected as an electron neutrino at Earth is $p = \sin^2 \theta_{13}=0.0219$ for a NMO and $p = \sin^2 \theta_{12}\cos^2 \theta_{13} = 0.297$ for an IMO. This is such a big difference even a considerable uncertainty in the distance cannot overwhelm the distinction. While the signal is clean in principle, it requires a detector sufficiently sensitive to electron neutrinos that is capable of detecting enough events to beat Poisson noise. 

An alternative other authors have considered is how the rise time of the electron antineutrino signal can be used as a mass ordering discriminator \cite{2012PhRvD..85h5031S}. Again, using table (\ref{tab:p,pbar}) and the PDG values for the mixing angles, we find the probability that an electron antineutrino emitted during the burst is detected as an electron neutrino at Earth is $\bar{p} = \cos^2 \theta_{12}\cos^2 \theta_{13} = 0.681$ for a NMO and $\bar{p} =\sin^2 \theta_{13}=0.0219$ for an IMO. The difference is larger than for the electron neutrinos but the reader must remember that the emitted flux is smaller and not so different from the $\mu$ and $\tau$ flux - see figure (\ref{fig:L,meanE}). 

In addition there have been other suggestions for signatures of the mass ordering which can be used to cross-check the conclusion from the collapse phase neutrinos. The pattern of spectral splits due to self-interaction effects in either the neutrinos and/or the antineutrinos is very different - see figures (\ref{fig:WuMA}) and (\ref{fig:multiangleONeMg}). This approach would be easiest while matter effects were simple. For a ONeMg supernova this window is late in the signal because the shock races through the mantle of the star very quickly: for an Iron core supernova the window is immediately after the accretion phase and lasts for $\sim 1\;{\rm s}$ for a more compact star up to $\sim 10\;{\rm s}$ for a star with a more extended envelope. However, in order for the technique to work we must be confident we understand the phenomenon fully and that the physics which is missing from the calculations does not radically alter the result. 

The time dependence of the matter effects (the moving shock wave and the turbulence) also distinguishes the mass ordering \cite{2002astro.ph..5390S,2004JCAP...09..015T,PhysRevD.78.023016,Kneller:2008PhRvD..77d5023K,2009PhRvL.103g1101G,2013PhRvD..88b3008L,2016JCAP...02..007V}. The matter effects appear in the neutrinos if the ordering is normal and in the antineutrinos if the ordering is inverted and can be distinguished from self-interaction effects by the sweep from low energy to high. These matter effects are robust because they rely simply on the decreasing density of the progenitor profile and the diabaticity of neutrino propagation through a shock. Where the approach may fail is if the density profile of the progenitor is either too steep or too shallow. In the former - typical of a ONeMg supernova - the sweep through the neutrino spectrum is very rapid \cite{PhysRevD.78.023016} prohibiting the accumulation of lots of events within the time window that it occurs and decreasing the statistical power. In the latter case, shock effects may not appear until very late in the signal also preventing the accumulation of enough events to observe the effect. Another possible source of confusion is the time dependence of the self-interaction effects. 

Finally, Earth matter effects could be exploited to determine the mass ordering \cite{2000PhRvD..62c3007D,1475-7516-2003-06-006,PhysRevD.73.093003,PhysRevLett.101.171801,1475-7516-2004-01-004,PhysRevD.79.113007,PhysRevD.94.113016} because neutrinos crossing through the Earth before reaching a detector are imprinted with an energy dependent oscillation that one would not expect to be there at the source. Combining the signal from two detectors, one shadowed by the Earth and the other not, would be an even more powerful \cite{1475-7516-2003-06-005}. However the feasibility of using the Earth Matter effect was questioned by Borriello \textit{et al.} \cite{2012PhRvD..86h3004B} who found it may be too small to observe unless the supernova was closer than $d\lesssim 200 \;{\rm pc}$. 

\subsection{Neutrino mass}
In addition to the mass ordering, the neutrino signal can also be used to 
constrain the neutrino mass. This can be done by focusing upon a impulse-like emission event such as the the neutronization burst or the sudden termination of the neutrino signal if a black hole forms \cite{2001PhRvD..63g3011B,2001PhRvD..63g3011B}, or from a comparison of high and low energy event timings \cite{PhysRevD.69.103002,2005NuPhB.731..140N}. The sensitivity to the mass arises because a neutrino with a mass $m$ and and energy $E$ traveling a distance $d$ will be delayed by a time $\Delta t$ relative to a massless particle given by \cite{2001PhRvD..63g3011B}
\begin{equation}
\Delta t = 0.515\;\left(\frac{m}{1\;{\rm eV}}\right)\,\left(\frac{10\;{\rm MeV}}{E}\right)\,\left(\frac{d}{10\;{\rm kpc}}\right)\;{\rm s}.
\end{equation}
Upper limits to the neutrino mass are usually of order $\sim 1\;{\rm eV}$ or better depending upon what other information is available. For example, an analysis by Lu \textit{et al.} \cite{2015JCAP...05..044L} indicates the JUNO detector could constrain the neutrino mass to be $m < (0.83 \pm 0.24)\;{\rm eV}$ at the 95\% confidence assuming degenerate masses and a NMO. Pagliaroli, Rossi-Torres and Vissani \cite{2010APh....33..287P} studied the mass sensitivity of a Super Kamiokande like detector and found the similar bound of $m < 0.8$ eV at 95\% confidence. 

Alternatively, rather than comparing neutrinos of different energies, one could compare neutrino arrival times with the arrival time of the gravitational waves \cite{2002PhRvD..65c3010A,2016PhRvD..94e3013L}. The upper limits on the absolute neutrino mass from this comparison are also in the range of $\sim 1\;{\rm eV}$ and are almost independent of distance because the smaller lapse times if the supernova is nearby can be compensated by the increased statistical power of having more events \cite{2002PhRvD..65c3010A}.


\subsection{Probes of Beyond Standard Model physics}

We have not dwelt upon BSM physics in this review for reasons of brevity but one must not forget that core-collapse supernova are an extreme environment and a wide range of BSM physics could be tested. We mention a few examples and their observable consequences. 

\begin{itemize}
\item The prodigious production of axions or any other new particle that interacts weakly with the Standard Model could not only carry away energy that might not be detected at Earth so making the supernova appear under-energetic, they would also decrease the timescale for cooling of the PNS. Even the SN 1987A data is sufficient to provide a very strong constraint on 
the mass of the axion \cite{1988PhRvL..60.1797T,1989PhRvD..39.1020B,1996PhRvL..76.2621J,1997PhRvD..56.2419K}. There has also been consideration of the emission of `unparticles' \cite{2007PhRvL..98v1601G,2007PhRvD..76l3012D,2007JHEP...12..033F} which have a similar effect upon the PNS cooling. 

\item Sterile neutrinos are additional neutrino flavors which do not participate in the weak interaction. This means they cannot be produced by any reaction within the supernova but, if they have a mass and mix with the $e$, $\mu$ and $\tau$ flavours (the `active flavours'), they can be `produced' as a neutrino propagates. Mixing between $N_f$ flavours of neutrino is parameterized by $N_f$ square mass difference, $N_f(N_f-1)/2$ mixing angles and $N_f(N_f-3)/2 +1$ (observable) CP phases. Multiple MSW resonances are created whose location in the supernova and adiabaticity depends upon these parameters. The new MSW resonance of most interest is between the sterile and electron neutrino flavors. 

For $eV$ scale mass differences the new MSW resonances occur at densities that put them in the gain region of ONeMg supernovae. Recently Tamborra \textit{et al.} \cite{2012JCAP...01..013T} and Wu \textit{et al.} \cite{2014PhRvD..89f1303W} both studied the effect of these resonances in the simulations of an ONeMg supernova. Wu \textit{et al.} found the effect of the steriles they introduced reduced the amount of heating in this region. The reduced heating would lengthen the accretion phase of the supernova and make it look more like an Fe core supernova. To determine whether this was the case would require independent knowledge of the progenitor. 

In Fe core supernovae the new MSW resonances are usually located at densities too low to affect the dynamics but they can affect the oscillations and thus the neutrino flux and spectra at Earth. Given current limits on the sterile mixing parameters, there may be only a small window in time during the collapse and accretion phases when the adiabaticity of the new resonances permits substantial flavor transformation thus affecting the observations \cite{2014PhRvD..90c3013E,2012JCAP...01..013T}. Attention has particularly focused upon the neutronization burst. As we previously discussed in relation to the mass ordering, observation of the neutronization burst strongly hints at the IMO but when sterile neutrinos are included, adiabatic conversion of electron to sterile flavor causes the neutronization burst to disappear for the IMO. If the mass ordering were known to be the IMO at the time of the next supernova in the Galaxy, observation of the neutronization burst would eliminate a wide swath of sterile neutrino mixing parameter space \cite{2014PhRvD..90c3013E}.

For $keV$ scale mass differences, the new electron-sterile MSW resonances are located at radii below the shock during the accretion phase of Fe core supernovae and even inside the PNS if the mass difference is high enough. If the mixing angles are sufficiently large, the effect is the transport of high energy neutrinos from the core to gain layer and the very quick revival of the shock \cite{Hidaka:2006sg,2014PhRvD..90j3007W}. Thus the Fe core supernova would be seen to explode very promptly and would not linger in the accretion phase. This would make it look more like a conventional ONeMg supernova in terms of the length of the accretion phase so, again, other discriminators of the progenitor would be necessary.

\item Interactions of neutrinos beyond the standard model can also have observable consequences for supernovae \cite{1987PhLB..199..432V,1996PhRvD..54.4356N,1996NuPhB.482..481N,1998PhRvD..58a3012M,2002PhRvD..66a3009F,PhysRevD.76.053001,2008PhRvD..78k3004B,2010PhRvD..81f3003E,2016PhRvD..94i3007S}. 
These non-standard interactions (NSI) also usually lead to new MSW resonances and thus flavor evolution which differs from the usual case. For example, one can find neutrino self-interaction effects in the NMO when NSI are included \cite{PhysRevD.76.053001,2008PhRvD..78k3004B,2010PhRvD..81f3003E,2016PhRvD..94i3007S}.
As with sterile neutrinos, any change in the flavor transformation deep in the supernova presumably has the potential to change the dynamics of the explosion although this has not yet been shown. 

\end{itemize}


\subsection{Multi-messenger astronomy}\label{sec:multimessenger}

In the core-collapse phenomenon, all the fundamental forces of nature are important, making supernovae naturally a source of multi-messenger signals, from the electromagnetic spectrum, neutrinos and cosmic rays, to gravitational waves. An example multi-messenger time-profile is summarized in figure \ref{fig:multimessenger} from Nakamura \textit{et al.}~\cite{Nakamura:2016kkl}, showing electromagnetic (black), neutrino (red), and gravitational wave (blue) signals for the core collapse of a non-rotating solar metallicity $17 M_\odot$ progenitor from the progenitor suite of Woosley, Heger, and Weaver 2002 \cite{Woosley:2002zz}. Cosmic rays, not shown in the figure, are generated at later epochs. Each messenger provide unique information about the core-collapse phenomenon, but it is also important to emphasize that combining multi-messenger observations goes beyond a simple sum of the parts. Multi-messengers provide complementary information that collectively yield a fuller picture of the collapsing star, some of which we summarize below. Multi-messenger signals also act as alerts and advanced warning for each other. This is especially advantageous for improving detection prospects of signals that are difficult to observe, e.g., because of the nature of the signal and/or detector capabilities. We review how this helps the detection of the first electromagnetic signature, the shock breakout. 

\begin{figure*}[b!] 
\includegraphics[width=0.9\linewidth]{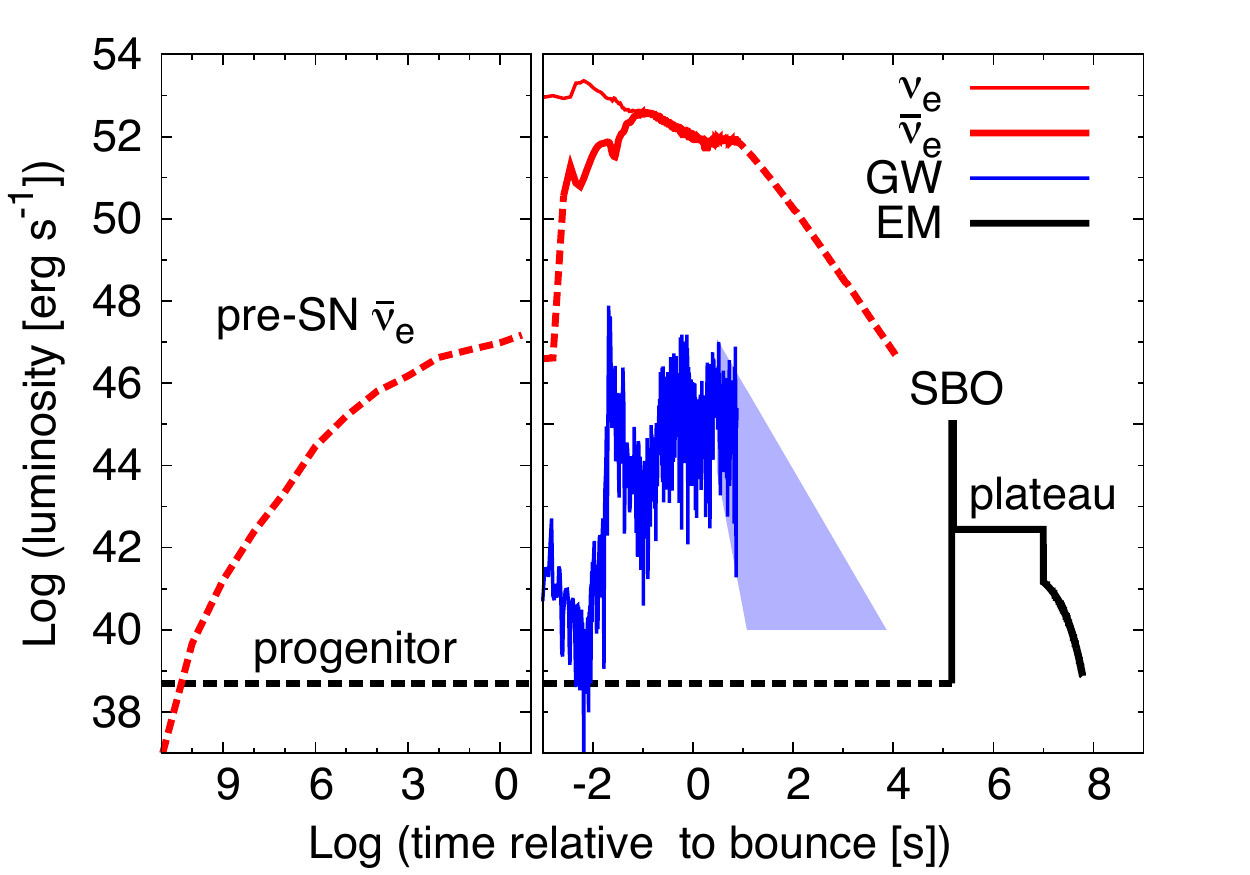}
\caption{Time sequence of multi-messenger signals in a core-collapse supernova, showing electromagnetic (EM; black), neutrinos (red), and gravitational waves (GW; blue) in time relative to core bounce. The left (right) panel shows pre-bounce (post-bounce) evolution. For the electromagnetic signal, ``progenitor'' refers to the star's emission; the SBO refers to shock breakout emission; and ``plateau'' refers to the plateau emission in a supernova. For the neutrino signal, ``pre-SN'' refers to neutrino emission during the last stages of Silicon fusion in the star. The solid phase is based on a two-dimensional core-collapse simulation of the solar metallicity $17 M_\odot$ progenitor of WHW02. The gravitational wave luminosity is similarly calculated from the core-collapse simulation. The total energy liberated after bounce in the form of photons, neutrinos, and gravitational waves are $\sim 4 \times 10^{49}$ erg, $\sim 6 \times 10^{52}$ erg, and $\sim 7 \times 10^{46}$ erg, respectively. Figure from Nakamura \textit{et al.}~\cite{Nakamura:2016kkl}.
\label{fig:multimessenger} }
\end{figure*}

In general, neutrinos and gravitational waves are excellent probes of the stellar interior, while electromagnetic signals are sensitive to the progenitor surface and surrounding interstellar medium.  Prior to core collapse, the progenitor becomes increasingly bright in neutrinos due to advanced stages of silicon burning with detectable consequences \cite{Odrzywolek:2003vn} (labeled ``pre-SN $\bar{\nu}_e$'' in figure \ref{fig:multimessenger}). The rise and luminosity of this signal is dependent on the core structure of the progenitor and can serve to distinguish between progenitor models (see Section \ref{sec:physics:progenitor}). Post core collapse, a wide range of stellar and fundamental physics can be probed by neutrinos, the focus of this review. The gravitational wave signal is highly dependent on the rotation and asphericity of the collapsed core \cite{Ott:2008wt,Kotake:2011yv}, something the neutrinos are more subtly sensitive to \cite{Thompson:2004if}. Studies show the rotation should be reliably measured for nearby ($< 0.2$ kpc) core collapse \cite{Yokozawa:2014tca}. Supernova explosion mechanics such as SASI are predicted to leave their imprint on not only neutrino \cite{Lund:2010kh,Lund:2012vm,Tamborra:2013laa,Tamborra:2014hga} but also gravitational wave \cite{Marek:2007gr,Kotake:2009em,Ott:2012mr,Mueller:2012sv,Yakunin:2015wra,Kuroda:2016bjd} time modulations; thus, cross correlations can yield more sensitive and robust probes. Neutrino-driven convection is also predicted to be robustly imprinted in gravitational waves (e.g., \cite{Mueller:2012sv}), and furthermore can be distinguished from other causes based on a time-frequency analysis \cite{Nakamura:2016kkl}. However, unlike neutrinos, the gravitational wave signal is not guaranteed to be detectable from the entire Milky Way. Using a global network of four detectors (two aLIGO detectors, adVirgo, and KAGRA) in a coherent network analysis \cite{Klimenko:2005xv,Hayama:2015sua}, the horizon for a pessimistic estimate of the gravitational wave signal is a few kpc. For example, starting with a non-rotating progenitor placed at the Galactic Center, Nakamura \textit{et al.}~\cite{Nakamura:2016kkl} showed that the peak gravitational wave signal-to-noise was $\sim 3.5$ using the four detector coherent network analysis. However, precise determination of the core bounce time by neutrino detection (Section \ref{sec:physics:bouncetime}) can be fed into gravitational wave searches to improve sensitivity \cite{Leonor:2010yp,Nakamura:2016kkl}. In the same non-rotating progenitor above, the gravitational wave signal-to-noise increased by a factor of $\sim 2$ when precise core bounce time information from the neutrinos was assumed \cite{Nakamura:2016kkl}. Conversely, information of core rotation obtained from gravitational waves can feed into neutrino observation analyses, allowing improved sensitivity to other physics, e.g., the nuclear EOS. 

The electromagnetic signal on the other hand provides less information about the core, but because it arises from the photosphere, is sensitive the envelope and surrounding medium. The first electromagnetic signal of a supernova is the shock breakout (SBO) signature, which arises when the supernova shock passes the progenitor photosphere \cite{Colgate:1968,Falk:1977,Klein:1978}. Although the SBO is a guaranteed signal of a supernova, it has only been detected in small numbers due to its extremely transient nature. Approximating the progenitor envelope as a polytrope with index $n=3/2$ (suitable for a convective envelope) and $n=3$ (suitable for a radiative envelope), the SBO duration are \cite{Matzner:1998mg},
\begin{eqnarray}
t_{\rm SBO,n=3/2} &\sim& 790 \, {\rm s}
\left( \frac{\kappa}{0.34\, {\rm cm^2 \, g^{-1}}} \right)^{-0.58}
\left( \frac{M_{\rm ej}}{10 \, M_\odot} \right)^{0.21}
\left( \frac{E_{\rm exp}}{10^{51} \, {\rm erg}} \right)^{-0.79}
\left( \frac{R_0}{500 \, R_\odot} \right)^{2.16}, \\
t_{\rm SBO,n=3} &\sim& 40 \, {\rm s}
\left( \frac{\kappa}{0.34\, {\rm cm^2 \, g^{-1}}} \right)^{-0.45}
\left( \frac{M_{\rm ej}}{10 \, M_\odot} \right)^{0.27}
\left( \frac{E_{\rm exp}}{10^{51} \, {\rm erg}} \right)^{-0.72}
\left( \frac{R_0}{50 \, R_\odot} \right)^{1.90},
\end{eqnarray}
where $\kappa$ is the opacity and $\kappa = 0.34 \, {\rm cm^2 \, g^{-1}}$ for Thomson scattering, $M_{\rm ej}$ is the ejecta mass, $E_{\rm exp}$ is the explosion energy, and $R_0$ is the progenitor radius. Thus, the duration can be as short as a few seconds for Wolf-Rayet stars, whose $R_0 \sim 1$-$10 R_\odot$ and $M_{\rm ej} \sim 1$-$10 M_\odot$, posing a real challenge to detect in the absence of an advanced alert. Neutrinos provide a solution, since they arrive before the SBO, the difference being due to the shock crossing time across the progenitor. The time separation has been recently discussed by Kistler et al.~\cite{Kistler:2012as}. For the $n=3/2$ and $n=3$ polytropes these are approximately \cite{Kistler:2012as},
\begin{eqnarray}
\Delta t_{n=3/2} \sim 7226\, {\rm s} &&
\left( \frac{10^{51}\,\mathrm{erg} }{ E_\mathrm{in}} \right)^{1/2} 
\left( \frac{M_\mathrm{ej}}{10\,M_\odot} \right)^{1/2} 
\left( \frac{ R_*}{ 50\,R_\odot} \right) \nonumber \\
&& \times \left[1 -0.738 \left( \frac{M_\mathrm{NS} }{ M_\mathrm{ej}} \right)^{0.80} +0.467 \left(\frac{M_\mathrm{NS} }{ M_\mathrm{ej}}
\right)^{1.20} \right], 
\end{eqnarray}
\begin{eqnarray}
\Delta t_{n=3} \sim 6851\, {\rm s} &&
\left( \frac{10^{51}\,\mathrm{erg} }{ E_\mathrm{in}} \right)^{1/2} 
\left( \frac{M_\mathrm{ej}}{10\,M_\odot} \right)^{1/2} 
\left( \frac{ R_*}{ 50\,R_\odot} \right) \nonumber \\
&& \times \left[1 -0.407 \left( \frac{M_\mathrm{NS} }{ M_\mathrm{ej}} \right)^{0.81} +0.285 \left(\frac{M_\mathrm{NS} }{ M_\mathrm{ej}}
\right)^{1.12} \right].
\end{eqnarray}

\begin{figure*}[b!] 
\includegraphics[width=0.9\linewidth]{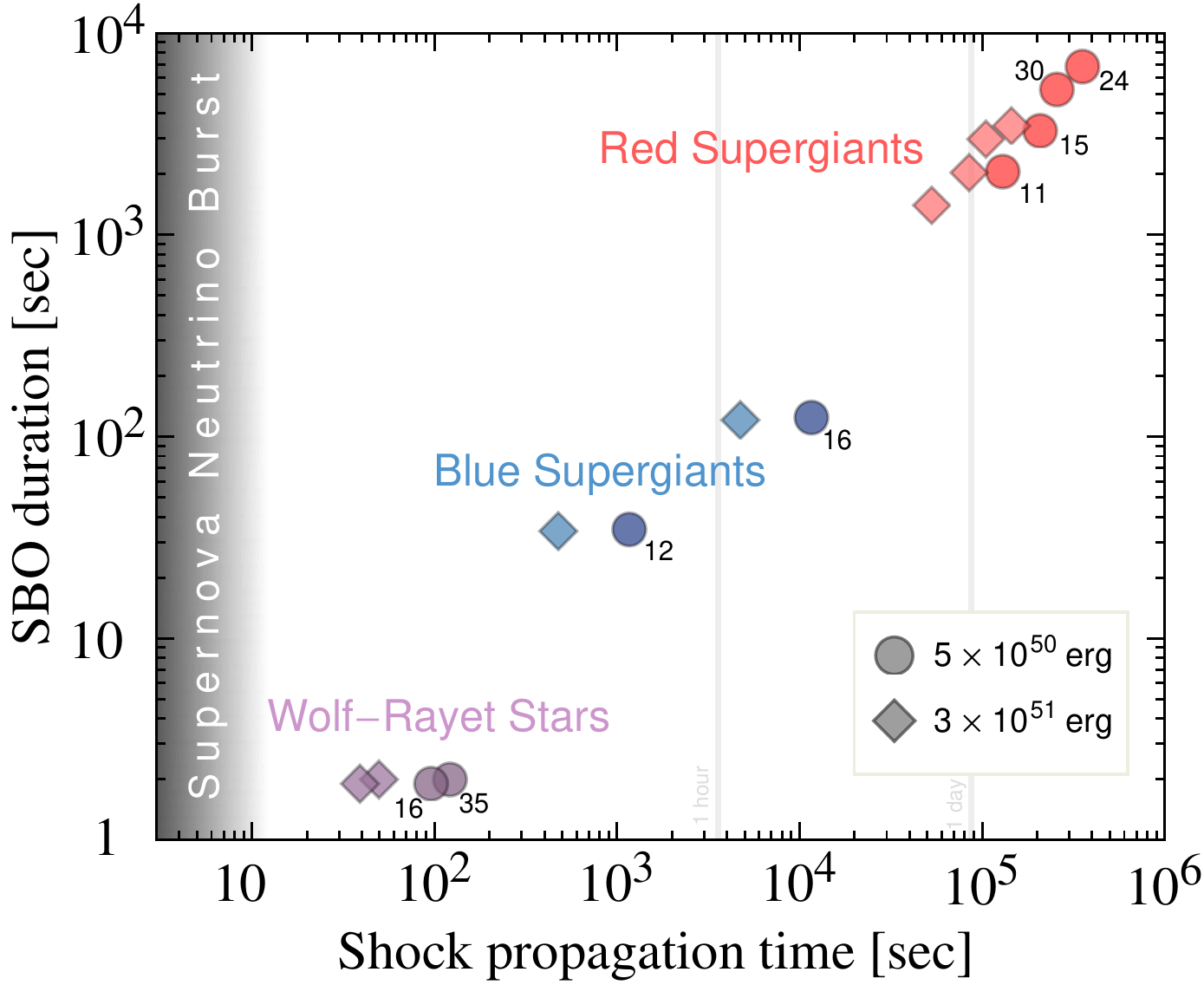}
\caption{SBO duration and shock propagation times in the envelope (which closely equals the time separation between neutrinos and the SBO), for a range of progenitors of initial masses from $11$ to $35 M_\odot$, as labeled. Red supergiants are from Ref.~\cite{Woosley:2002zz}; blue supergiants and Wolf-Rayet stars are from Ref.~\cite{Woosley:2005gy}. Figure from Kislter \textit{et al.}~\cite{Kistler:2012as}.
\label{fig:SBOtimescales} }
\end{figure*}

The typical SBO duration and time separations for three types of progenitors, red supergiants, blue supergiants, and Wolf-Rayet stars, are shown in Figure \ref{fig:SBOtimescales}. Shorter SBO have hotter spectra in the UV/soft-X ray, while longer SBO have peaks in visible band. This necessitates a rapid, pre-determined information transfer from neutrinos to telescopes and satellites across the electromagnetic spectrum. Wider field of view detectors are also beneficial since more sky area can be covered during the limited duration of the SBO. The field of view opening diameter of the largest 8-meter class telescopes are of the order of a few degrees at most. Neutrino alerts to be able to provide comparable angular information about the core collapse also. In principle, neutrino-electron scattering provides the angular information \cite{Vogel:1989iv}. However, they are contaminated by a background of near-isotropic IBD events that overwhelm $e$-scattering in number. Nevertheless, a hotspot of events will be statistically measurable, providing angular resolutions of $\sim 6$ degrees at SuperK for a Galactic Center core collapse \cite{Beacom:1998fj}. With the upgrade of gadolinium, the IBD events will be tagged at some 90\% efficiency, dramatically improving angular resolution to a few degrees \cite{Tomas:2003xn}. Similar performance can also be expected with a DUNE class LqAr detector \cite{Bueno:2003ei}. These are comparable to or better than the fields of view of large optical telescopes, improving the discovery prospect of the first electromagnetic signal of a supernova \cite{Nakamura:2016kkl}. 


\section{Future Prospects}\label{sec:summary}

The understanding of core-collapse supernovae and the neutrino have increased tremendously since the last time we were fortunate enough to detect a neutrino burst from a nearby supernova. While the detection of neutrinos from SN 1987A allowed us to confirm the overall picture of stellar core collapse and the importance of weak-interactions in the collapsed core, the basic paradigm of core-collapse supernovae still remains unanswered. Many possibilities have been developed by multi-dimensional hydrodynamical simulations, and theory is now at a point where we need to test predictions in a more quantitative sense. There is a rich vein of potential physics insights which could be extracted from the next Galactic neutrino burst signal and strategies exist for mining them, even with the present uncertainties in the simulations and flavor transformation. Given the infrequency of core-collapse neutrino bursts in our Milky Way galaxy, the trove of data collected by detectors would be of unprecedented value that would not likely be matched for decades. 

A signal would be a fantastic opportunity to greatly increase our knowledge about multiple frontier areas of physics. In particular, due to the importance of the neutrino in supernovae, a Galactic supernova neutrino burst is a golden opportunity to explore properties of the neutrinos in environments well beyond what we can produce on Earth. Core collapse supernova are one of the few environments where neutrinos interact with each other, not just with the other particle content of the Standard Model. There is also the possibility that information gleaned from the signal could indicate new flavors or new interactions revealing something about what lies beyond the Standard Model. 

Looking forwards, theorists have much to do while they wait. Many more 3D simulations need to be undertaken and there needs to be the development of a method to take those simulations and run them to tens of seconds. Investigations of progenitor asymmetries, improved microphysics, code comparison efforts, and multi-dimensional neutrino transport made possible by next-generation exascale computing, will all contribute to the next-generation suites of predictions. Many holes in the understanding of flavor transformation also exist and need to be filled. There should be no shortage of more robust theoretical results to test against future neutrino datasets. Finally, theorists also need to think more about uncertainty quantification i.e.\ how the signatures of one piece of physics might be confused / degenerate with changes in other pieces of physics or effects. 

On the experimental side, detectors have to be suitably designed for the appropriate energies.  Generally, the better a detector is able to divide events into charged current and neutral current reactions with precise energy deposition and even reaction kinematics so as to determine the incoming neutrino direction, flavor and energy, the more we can learn about what happened in the core of the star. However the criteria for measuring accurately the energy and type of interaction of a supernova neutrino event in a detector are not always the same as measuring the energies and type of interaction of the neutrinos the detectors spend most of their time looking at. Ensuring next-generation detectors have such capabilities in the MeV energies relevant for supernovae needs to be a high priority if we are to realize the rich information content of the signal. 

Finally, the gulf between theory predictions and experiment needs to close substantially and this is best achieved if results---and better yet codes---are shared. The building of more efficient pipelines to be able to take the results of a simulation, process it with a flavor transformation code, and then run the predicted neutrino flux at Earth through detector simulation software such as SNOwGLoBES \cite{SNOwGLoBES}, will be most valuable. First and foremost this would facilitate the quantitative translation of theoretical needs to experimental efforts using a common framework, and would address the task of how the features theorists seek could be detected. This would be particularly fruitful especially in light of the large number of theoretical predictions anticipated in the future. Also, this would be valuable for the process of designing next-generation detectors. 

To finish, we return to the original question we set out to answer: What can be learned from a future supernova neutrino detection? Hopefully our review has revealed that the answer is we will learn a substantial amount about the neutrino, supernovae, nuclear physics and a lot of other fields. Significant strides have already been taken but the ability to exploit the signal still needs to be pursued vigorously by both theorists and experimentalists. Only by combining theoretical, experimental, and bridging efforts, will we see the rich returns on the investments made.


\ack

The authors wish to thank the numerous individuals who provided figures for this paper. 
The list of questions one might hope to answer with a Galactic core-collapse supernova was, in part, taken from a similar list compiled by the attendees at the INT workshop INT-16-61W: ``Flavor Observations with Supernova Neutrinos" held August 15 - 19, 2016.\\

\noindent SH is supported by the U.S.~Department of Energy under award number DE-SC0018327. JPK is supported at NC State by the U.S.~Department of Energy award DE-FG02-10ER41577. \\


\bibliographystyle{IEEEtran}
\bibliography{main}

\end{document}